%% file: main.tex
\documentclass[10pt, journal]{IEEEtran}
\input{preamble_new.tex}

\title{Traffic State Estimation for Connected Vehicles using the Second-Order Aw-Rascle-Zhang Traffic Model}

\author{Suyash C. Vishno$\text{i}^{\dagger, \ddagger}$, Sebastian A. Nugroh$\text{o}^{\star}$, Ahmad F. Tah$\text{a}^{\dagger\dagger}$, and Christian G. Claude$\text{l}^{\mathsection}$ \vspace{-0.5cm}
	\thanks{
        $^{\dagger}$School of Civil and Environmental Engineering, Georgia Institute of Technology, 790 Atlantic Dr NW, Atlanta, GA 30332.
		$^{\star}$Cummins Technical Center, Cummins Inc., Columbus, IN 47201.
		$^{\dagger\dagger}$Department of Civil and Environmental Engineering, Vanderbilt University, 2201 West End Ave, Nashville, TN 37235.
        $^\mathsection$Department of Civil, Architectural, and Environmental Engineering, The University of Texas at Austin, 301 E. Dean Keeton St. Stop C1700, Austin, TX 78712.
		$^\ddagger$Corresponding author. 
		
		Emails: scvishnoi@gatech.edu, sebastian.nugroho@cummins.com, ahmad.taha@vanderbilt.edu, christian.claudel@utexas.edu.
		
		This work is partially supported by the Valero Energy Corporation and National Science Foundation (NSF) under Grants 1636154, 1728629, 2152928, and 1917056, and 2152450, and USDOT CAMMSE.}}

\begin{document}

\maketitle

\setlength{\abovedisplayskip}{3.5pt}
\setlength{\belowdisplayskip}{3.5pt}
\setlength{\abovedisplayshortskip}{3.2pt}
\setlength{\belowdisplayshortskip}{3.2pt}

\newdimen\origiwspc%
\newdimen\origiwstr%
\origiwspc=\fontdimen2\font
\origiwstr=\fontdimen3\font

\fontdimen2\font=0.64ex

\begin{abstract} 
 This paper addresses the problem of traffic state estimation (TSE) in the presence of heterogeneous sensors which include both fixed and moving sensors. Traditional fixed sensors are expensive and cannot be installed throughout the highway. Moving sensors such as Connected Vehicles (CVs) offer a relatively cheap alternative to measure traffic states across the network. Moving forward it is thus important to develop such models that effectively use the data from CVs. One such model is the nonlinear second-order Aw-Rascle-Zhang (ARZ) model which is a realistic traffic model, reliable for TSE and control. A state-space formulation is presented for the ARZ model considering junctions in the formulation which is important to model real highways with ramps. A Moving Horizon Estimation (MHE) implementation is presented for TSE using a linearized ARZ model. Various state-estimation methods used for TSE in the literature along with the presented approach are compared with regard to accuracy and computational tractability with the help of a numerical study using the VISSIM traffic simulation software. The impact of various strategies for querying CV data on the estimation performance is also considered. Several research questions are posed and addressed with a thorough analysis of the results.
\end{abstract}


\section{Motivation and Paper Contributions}\label{sec:Introduction}
 \IEEEPARstart{W}{ith}   the large number of vehicles overloading the transportation systems across the world, problems like congestion, accidents, and pollution have become common. As a remedy to such circumstances, control methods such as variable speed limits, ramp metering, route control and their combinations have become quite popular for instance see~\cite{gomes2006optimal,carlson2010optimal,han2017resolving,van2018efficient}. 

These methods require the knowledge of the system at all times to make them work effectively. A popular method for real-time monitoring of traffic systems is by means of traffic state estimation (TSE) using dynamic traffic models which provide a high-fidelity picture of the traffic spatio-temporally while utilizing data from sensors available throughout the highway. In general, more data results in better estimates of the system states. However, since fixed sensors like inductive loop detectors are quite expensive, they cannot be placed at short intervals throughout the highway. Connected vehicles (CVs) offer a potential solution to this problem by providing additional sources of data relatively free of cost~\cite{lu2014connected}. Here, we assume that most of the communication between the CVs and the network operator will take place via existing cellular networks so there will be no additional costs of building connected highway infrastructure everywhere. As the proportion of CVs in the traffic rises, CVs will be able to provide useful data from across the system including both traffic density and speed. Thus, moving forward, it is imperative to develop such models that can utilize well different types of data from both fixed sensors and CVs to perform state estimation and control.

Traditionally, TSE is performed using first-order traffic models such as the Lighthill-Whitham-Richards (LWR) model~\cite{Lighthill1955b,Richards1956}. First-order models are simple to implement as they only have a single equation which is the conservation of vehicles to describe the traffic dynamics. They also have very few calibration parameters, making them a popular choice for state estimation.  However, they only consider equilibrium traffic conditions, that is, the traffic density ({number of vehicles per unit space expressed in vehicles per unit length for example, veh/km~\cite{HARROU202215}}) and traffic flux ({number of vehicles that cross a given point per unit of time expressed in vehicles per unit time for example, veh/hr~\cite{HARROU202215}}) are assumed to follow a predefined relationship known as the \textit{fundamental diagram}. This makes them unable to represent certain non-equilibrium traffic phenomena like capacity drop which are essential for the purpose of traffic control~\cite{daganzo1995requiem}. Thus, the use of these models in traffic control is considered less effective. Second-order traffic models such as the Payne-Whitham (PW) model~\cite{payne1971model,whitham1974linear} and the Aw-Rascle-Zhang  (ARZ) model~\cite{aw2000resurrection,zhang2002non}, on the other hand, can represent non-equilibrium traffic phenomena with the help of an additional equation to describe the traffic dynamics. They are, therefore, considered more realistic than the first-order models. As a result, these models are not only good for state estimation but are also reliable for control. Additionally, second-order models provide a natural way to incorporate multiple sources of data as they consider both density and speed to be independent variables. In first-order models that only consider either the density or the speed as a variable at a time, any deviation of the speed from its equilibrium relationship must be considered a part of the modeling error. Thus, second-order models become a natural choice for state estimation using CVs. Note that while Lagrangian (vehicle-based) models of traffic exist~\cite{leclercq2007lagrangian} which are arguably more compatible with trajectory-based data from CVs, here we are using road density and average vehicle speed information obtained using both CVs and fixed detectors and not just relying on trajectory based information. Therefore, we have chosen an Eulerian (location-based) second-order model over a Lagrangian model besides the above reasons.

In light of the aforementioned discussion, the objective of this work is to develop a state-space representation of a reliable second-order traffic model and show the potential of CV data for TSE under different scenarios. Given this objective, in the following discussion, we present a literature review of traffic models used for TSE and studies utilizing heterogeneous sources of data, followed by a brief discussion on the estimation methods used.

The most popular model for TSE in the literature is the first-order LWR model. The simple form of the model with a minimal number of calibration parameters makes it an attractive option for large-scale implementation. Some works that implement a first-order model to perform state estimation using heterogeneous sensors include~\cite{work2008ensemble,wright2016fusing,nantes2016real}. Readers can also refer to \cite{seo2017survey} for a comprehensive review of TSE literature involving first-order models. Due to the known limitations of first-order models, several studies have also undertaken state estimation using second-order models such as in~\cite{wang2005real,zhao2020real,agalliadis2020traffic} and the references therein. Most of these studies use the second-order PW model implemented in the METANET~\cite{kotsialos2002,messner1990metanet} framework. The PW model has well-known limitations~\cite{daganzo1995requiem} such as physical inconsistency under certain heterogeneous traffic conditions which make it unreliable. A significantly better model is the ARZ model which retains the benefits of second-order models without sacrificing the physical consistency of the first-order models. Despite this, there are very few studies in the literature that use the ARZ model for state estimation. The work in \cite{Seo2017} develops a state-space formulation for the nonlinear ARZ model and performs state estimation using Extended Kalman Filter (EKF) considering both fixed and moving sensors. In \cite{yu2019boundary}, the authors propose a boundary observer for state estimation using a linearized ARZ model. The study in \cite{wang2017comparing} uses Particle Filter (PF) for the estimation of traffic states using a modified ARZ model. However, it is worth noting that none of these papers considers junctions in the modeling. Modeling the traffic dynamics at junctions is essential to the modeling of traffic on real highways which consist of on-ramp and off-ramp connections. Therefore, unlike past studies, we formulate herein a state-space model for the nonlinear ARZ model considering junctions.

Note that the aforementioned studies using second-order models as well as the present work are different from studies like \cite{bekiaris2016highway} which while do consider the speed to be an independent variable like the second-order models but consider it to be known everywhere and at all times using CV data. These have been categorized as \textit{data-driven} methods by \cite{seo2017survey}. In the current work, unlike \cite{bekiaris2016highway}, we assume a bandwidth restriction on the data that can be transferred from the CVs to the network operator which forces data to be available only from a subset of all segments for estimation while the traffic data on other segments is considered unknown. Further, the impact of various strategies associated with the selection of the subset of segments to query CV data for estimation is also investigated. Besides \cite{bekiaris2016highway}, several other studies utilize heterogeneous data sources for TSE. Detailed reviews of the related literature can be found in \cite{xing2022traffic, seo2017survey, kashinath2021review}. Most studies focus on data fusion methods to combine fixed and moving sensor data to achieve improved estimation performance. Some studies such as \cite{Seo2017} focus on the impact of the penetration rate of CVs on TSE. To the best of the authors' knowledge, none of the studies investigate the impact of different approaches to query subsets of segments for estimation in a moving sensor setting.

A majority of the model-driven TSE literature either uses one of the Kalman Filter (KF) variants from among EKF, Unscented Kalman Filter (UKF), and Ensemble Kalman Filter (EnKF), or other methods like PF, and observers to perform state estimation, for instance, see \cite{seo2017survey,nugroho2019control,vishnoi2020asymmetric}. While these methods are computationally attractive, they have certain limitations with respect to TSE. The primary limitation is that they do not have an inherent way to deal with state constraints. Thus, it is possible that the estimates generated from these methods contain nonphysical values of certain states which can further cause the process model to collapse.

An estimation method that handles this limitation naturally, due to its optimization-based structure, is MHE. MHE has been explored extensively in the general state estimation literature, for instance in ~\cite{rao2001constrained, rao2003constrained,wang2017resilient,alessandri2017fast}, but not so much in the TSE literature. In \cite{sirmatel2019nonlinear} and \cite{sirmatel2020model}, the authors propose an MHE formulation for the estimation and control of large-scale highway networks using the Macroscopic fundamental diagram (MFD). MFD is a network-level traffic model and does not consider the variation in traffic density on individual stretches of the highway. Unlike \cite{sirmatel2019nonlinear} and \cite{sirmatel2020model}, we investigate estimating the density throughout the highway stretch. The study in \cite{timotheou2015moving} presents an MHE formulation for traffic density estimation using the Asymmetric Cell Transmission model (ACTM). ACTM is based on the LWR model and therefore, has the drawbacks previously mentioned for first-order models. Moreover, the work in \cite{timotheou2015moving} does not consider moving sensors from CVs. 

{Besides the above approaches, a recent paradigm of TSE explores physics-informed deep learning (DL)~\cite{huang2020physics, zhao2022integrating, shi2021physics}. These approaches aim to guide the training of DL-based models for TSE through physics-based traffic laws, such as those governing the first and second-order models discussed above. The traffic model parameters are automatically tuned as the DL model is trained, thus offering the determination of accurate traffic flow laws for a given scenario. However, these approaches suffer from several limitations, including a lack of robustness to noisy data and the need for extensive tuning of training algorithm parameters for individual scenarios, limiting the models' applicability to real-world use cases. Interested readers are referred to \cite{di2023physics} for a comprehensive survey on this paradigm of TSE. Compared to these, model-driven approaches such as the one presented in this work are favored for real-world applications due to their interpretability and computational advantage.}

Given that, the main research gaps on this topic are a) the absence of a state-space formulation for a reliable second-order traffic model with junctions, b) the lack of exploration of MHE in the context of TSE and comparison with other state-estimation methods, and c) the absence of an investigative study on the impact of different strategies for querying data from CVs for TSE as opposed to fixed sensors. In what follows, we highlight the main contributions of this paper:

\begin{itemize}
    \item We derive a nonlinear state-space formulation for the second-order ARZ model with junctions in the form of ramp connections. In that, we present the detailed dynamic equations of the model. This is a development over~\cite{Seo2017} which does not consider junctions in the formulation. The inclusion of junctions adds additional complexity to the model in terms of the nonlinearity which now comprises of minimum and piecewise functions in the model. Second-order traffic models are more realistic than first-order models like the LWR model as they can capture certain phenomena like capacity drop which are essential to control applications. The obtained state-space formulation can thus be used for state estimation as well as control purposes.
    \item We consider heterogeneous sensors including both fixed and moving sensors. The former consists of sensors like inductive loop detectors while the latter includes CVs. The state-space description is appended to include the measurement model which is also nonlinear thus resulting in a nonlinear input-output mapping of the system dynamics.
    \item We investigate the performance of various state estimation methods in terms of accuracy and computational tractability using the VISSIM traffic simulation software. As a departure from estimation based on KFs, PF, observers, and so on, we investigate MHE for TSE. MHE, unlike the other methods, naturally allows us to include constraints on the state variables making the problem more practical.
    \item The impact of moving sensors including CVs on the performance of TSE is studied under various scenarios including different frequencies of change in sensor positions, different sensor placement configurations, and different levels of {measurement errors from varying penetration rate of CVs and sensor noise}. The estimated states are examined qualitatively to understand the implication of moving sensors on TSE.
\end{itemize}
\noindent {\textbf{Paper's Notation:}} \; Let $\mathbb{N}$, $\mathbb{R}$, $\mathbb{R}^n$, and $\mathbb{R}^{p\times q}$ denote the set of natural numbers, real numbers, and real-valued {column} vectors with size $n$, and $p$-by-$q$ real matrices respectively. $\mathbb{S}^{m}_{++}$ denotes the set of positive definite matrices. For any vector $z \in \mathbb{R}^{n}$, $\Vert z\Vert_2$ denotes its Euclidean norm, i.e. 
$\Vert z\Vert_2 = \sqrt{z^{\top}z} $, where $z^{\top}$ is the transpose of $z$. 
Tab. \ref{tab:notation} provides the nomenclature utilized in this paper.

\setlength{\textfloatsep}{10pt}

\begin{table}
	\footnotesize	\renewcommand{\arraystretch}{1.0}
	\caption{Paper nomenclature: parameter, variable, and set definitions.}
	\label{tab:notation}
	\centering
	\begin{tabular}{||l|l||}
		\hline
		\textbf{Notation} & \textbf{Description}\\
		\hline
		\hline
		\hspace{-0.1cm}$\Omega$ & \hspace{-0.1cm}the set of highway segments on the stretched highway \\
		\hspace{-0.1cm} & \hspace{-0.1cm}$\Omega = \{ 1,2,\hdots,N \}$ , $N := \abs{\Omega}$ \\
		\hline
		\hspace{-0.1cm}$\Omega_I$ & \hspace{-0.1cm}the set of highway segments with on-ramps \\
		\hspace{-0.1cm}	&  \hspace{-0.1cm}$\Omega_I = \{ 1,2,\hdots,N_I \}$ , $N_I := \abs{\Omega_I}$ \\
		\hline
		\hspace{-0.1cm}$\Omega_O$ & \hspace{-0.1cm}the set of highway segments with off-ramps \\
		\hspace{-0.1cm}& \hspace{-0.1cm}$\Omega_O = \{ 1,2,\hdots,N_O \}$, $N_O := \abs{\Omega_O}$ \\ 
		\hline
		\hspace{-0.1cm}$\hat{\Omega}$ & \hspace{-0.1cm}the set of on-ramps, $\hat{\Omega} = \{ 1,2,\hdots,N_I \}$ , $N_I = |\hat{\Omega}|$\\
		\hline
		\hspace{-0.1cm}$\check{\Omega}$ & \hspace{-0.1cm}the set of off-ramps, $\check{\Omega} = \{ 1,2,\hdots,N_O \}$ , $N_O = |\check{\Omega}| $\\
		\hline
		\hspace{-0.1cm}$T$ & \hspace{-0.1cm}duration of each time-step\\
		\hline		
		\hspace{-0.1cm}$l$ & \hspace{-0.1cm}length of each segment, on-ramp, and off-ramp\\
		\hline	
		
		\hspace{-0.1cm}$\rho_i [k],\psi_i[k],w_i[k]$ & \hspace{-0.1cm}traffic density, relative flow and driver characteristic for\\
		\hspace{-0.1cm}	&  \hspace{-0.1cm}Segment $i \in \Omega$ at time $kT$, $k \in \mathbb{N}$ \\
		\hline
    	\hspace{-0.1cm}$q_i[k],\phi_i[k]$ & \hspace{-0.1cm}traffic flow and relative flux from Segment $i \in \Omega $ into the \\
    	\hspace{-0.1cm}	&  \hspace{-0.1cm}next segment\\ 
		\hline
    	\hspace{-0.1cm}$D_i[k], S_i[k]$ & \hspace{-0.1cm}demand and supply functions for Segment $i \in \Omega$\\
		\hline
		
		\hspace{-0.1cm}$\hat{\rho}_i [k],\hat{\psi}_i[k],\hat{w}_i[k]$ & \hspace{-0.1cm}traffic density, relative flow and driver characteristic for\\
		\hspace{-0.1cm}	&  \hspace{-0.1cm}On-ramp $i \in \hat{\Omega}$ at time $kT$, $k \in \mathbb{N}$ \\
		\hline
    	\hspace{-0.1cm}$\hat{q}_i[k],\hat{\phi}_i[k]$ & \hspace{-0.1cm}traffic flow and relative flux from On-ramp $i \in \hat{\Omega} $ into the \\
    	\hspace{-0.1cm}	&  \hspace{-0.1cm}attached highway segment\\ 
		\hline
    	\hspace{-0.1cm}$\hat{D}_i[k], \hat{S}_i[k]$ & \hspace{-0.1cm}demand and supply functions for On-ramp $i \in \hat{\Omega}$\\
		\hline		
		
		\hspace{-0.1cm}$\check{\rho}_i [k],\check{\psi}_i[k],\check{w}_i[k]$ & \hspace{-0.1cm}traffic density, relative flow and driver characteristic for\\
		\hspace{-0.1cm}	&  \hspace{-0.1cm}Off-ramp $i \in \check{\Omega}$ at time $kT$, $k \in \mathbb{N}$ \\
		\hline
    	\hspace{-0.1cm}$\check{q}_i[k],\check{\phi}_i[k]$ & \hspace{-0.1cm}traffic flow and relative flux from Off-ramp $i \in \check{\Omega} $\\
    	\hline
    	\hspace{-0.1cm}$\check{D}_i[k], \check{S}_i[k]$ & \hspace{-0.1cm}demand and supply functions for Off-ramp $i \in \check{\Omega}$\\
		\hline

        \hspace{-0.1cm}$\bar{q}_i[k],\bar{\phi}_i[k]$ & \hspace{-0.1cm}incoming traffic flow and relative flux for Segment $i \in \Omega $\\
    	\hline
    	\hspace{-0.1cm}$\bar{\check{q}}_i[k],\bar{\check{\phi}}_i[k]$ & \hspace{-0.1cm}incoming  traffic flow and traffic flux for Off-ramp $i \in \check{\Omega} $\\
		\hline
		
    	\hspace{-0.1cm}$D_{in}[k],w_{in}[k]$ & \hspace{-0.1cm}demand and driver characteristic of traffic wanting to\\
    	\hspace{-0.1cm}	&  \hspace{-0.1cm}enter Segment 1 of the highway\\
		\hline
    	\hspace{-0.1cm}$\rho_{out}[k]$ & \hspace{-0.1cm}traffic density downstream of Segment $N$ of the highway\\
		\hline
    	\hspace{-0.1cm}$\hat{D}_{in,i}[k],\hat{w}_{in,i}[k]$ & \hspace{-0.1cm}demand and driver characteristic of traffic wanting to\\
    	\hspace{-0.1cm}	&  \hspace{-0.1cm}enter On-ramp $i \in \hat{\Omega}$\\
		\hline
    	\hspace{-0.1cm}$\check{\rho}_{out,i}[k]$ & \hspace{-0.1cm}traffic density downstream of Off-ramp $i \in \check{\Omega} $\\
		\hline

    	\hspace{-0.1cm}$\beta_i[k]$ & \hspace{-0.1cm}proportion of traffic entering from Segment $i\in \Omega$ into the\\
    	\hspace{-0.1cm}	&  \hspace{-0.1cm}next segment at an on-ramp junction,
    	where $\beta_i[k] \in [0,1]$\\
        \hline
    	\hspace{-0.1cm}$\alpha_i[k]$ & \hspace{-0.1cm}split ratio for the off-ramp attached to Segment $i \in \Omega$,\\
    	\hspace{-0.1cm}	&  \hspace{-0.1cm}where $\alpha_i[k] \in [0,1]$\\
    	
    	\hline              
		\hspace{-0.1cm}$v_f$ & \hspace{-0.1cm}free-flow speed \\
		\hline
		\hspace{-0.1cm}$\rho_m$ & \hspace{-0.1cm}maximum density  \\
		\hline
		
		\hspace{-0.1cm}$\alpha$ & \hspace{-0.1cm}model parameter called relaxation time, where $\alpha\in\mathbb{R}_+$ \\
		\hline
		\hspace{-0.1cm}$\gamma$ & \hspace{-0.1cm}fundamental diagram parameter, where $\gamma\in\mathbb{R}_+$  \\
		\hline
		\hspace{-0.1cm}$p(\rho)$ & \hspace{-0.1cm}pressure function which takes traffic density $\rho$ as input\\
		\hline
		\hspace{-0.1cm}$V_e(\rho)$ & \hspace{-0.1cm}equilibrium traffic speed at traffic density $\rho$ \\
		\hline
	\end{tabular}
	\vspace{-0.0cm}
\end{table}
\setlength{\floatsep}{10pt}

\section{Nonlinear Discrete-Time Modeling of Traffic Networks with ramps}\label{sec:nonlinearmodel}

The objective of this section is to develop a state-space formulation for the nonlinear second-order ARZ model describing the evolution of traffic density on highways with ramps. The developed formulation is useful for several control theoretic purposes including state estimation and control of highway traffic.
\subsection{The Aw-Rascle-Zhang model}\label{s:arz}
In this section, we present the modeling of traffic dynamics for a stretched highway connected with ramps. To that end, we use the second-order ARZ Model \cite{aw2000resurrection,zhang2002non} given by the following partial differential equations:
\begin{subequations} \label{e:ARZ_model}
\begin{align}
    \frac{\partial\rho}{\partial t}+\frac{\partial \rho v}{\partial d}&=0,\\
    \frac{\partial \rho\left(v+p\left(\rho\right)\right)}{\partial t}+\frac{\partial\rho\left(v+p\left(\rho\right)\right)v}{\partial d}&=-\frac{\rho\left(v-V_e\left(\rho\right)\right)}{\tau},
    \end{align}
\end{subequations}
where $t$ and $d$ denote the time and distance; $\rho$ is shorthand for $\rho(t,d)$ which denotes the traffic density (vehicles/distance), and $v$ is shorthand for $v(t,d)$ which denotes the traffic speed (distance/time). Here, $p(\rho)$ is given by
\begin{align}\label{e:pressure}
    p\left(\rho\right) = v_f\left(\frac{\rho}{\rho_m}\right)^\gamma,
\end{align}
 and $V_e(\rho)$ is given by
\begin{align}\label{e:pressure_equilibrium_relationship}
    V_e(\rho)=v_f\left(1-\left(\frac{\rho}{\rho_m}\right)^\gamma\right).
\end{align}
In traffic literature, relationships like \eqref{e:pressure_equilibrium_relationship} are commonly called the \textit{fundamental diagram}.
The first PDE in the ARZ model ensures the conservation of vehicles which is also present in the first-order traffic models. The second PDE which ensures conservation of traffic momentum is unique to second-order models and accounts for the deviation of traffic from an equilibrium position. This equation makes the second-order models more realistic than the first-order models as it allows them to represent some non-equilibrium traffic phenomena such as capacity drop. As second-order models allow traffic flow to deviate from equilibrium, they also inherently allow traffic speed to deviate from the equilibrium speed which allows speed data to be incorporated independent of the density. With first-order models, any deviation of the speed from the equilibrium speed would have to be considered a part of the modeling error. Therefore, second-order models are more naturally suited to perform estimation using both density and speed data provided by the fixed sensors and CVs. The quantity $v+p(\rho)$ is also called the \textit{driver characteristic} and is denoted by the variable $w(t,d)$. The expression $\rho(v+p(\rho))$ is also called the \emph{relative flow} denoted by $\psi(t,d)$ which is essentially the difference between the actual flow and the equilibrium flow at any $\rho$. Notice that in \eqref{e:ARZ_model}, $\rho v$ is the flux of traffic (vehicles/time) which will be denoted by $q(t,d)$, while $\rho(v+p(\rho))v$ is the flux of relative flow (vehicles/time$^2$), also called the relative flux, which will be denoted by $\phi(t,d)$. Using the relative flow and the two flux, the ARZ model can simply be rewritten as
\begin{subequations} \label{e:ARZ_model_flux}
\begin{align}
    \frac{\partial\rho(t,d)}{\partial t}+\frac{{\partial} q(t,d)}{\partial d}&=0,\\
    \frac{\partial \psi(t,d)}{\partial t}+\frac{\partial \phi(t,d)}{\partial d}&=-\frac{\psi(t,d)}{\tau}+\frac{v_f\rho(t,d)}{\tau},
    \end{align}
\end{subequations}
which can be converted to a state-space equation with $\rho$ and {$\psi$} as the states.

To represent this model as a series of difference, state-space equations, we discretize the ARZ Model \eqref{e:ARZ_model_flux} with respect to both space and time, also referred to as the Godunov scheme~\cite{Godunov1959}. This allows us to divide the highway of length $L$ into segments of equal length $l$ and the traffic networks model to be represented by discrete-time equations. These segments form both the highway and the attached ramps. Throughout the paper, the segments forming the highway are referred to as mainline segments. We assume the highway is split into $N$ mainline segments.

To ensure computational stability, the Courant-Friedrichs-Lewy condition (CFL)~\cite{courant1967partial} given as ${v_f T}l^{-1}\leq 1$ has to be satisfied. Since each segment is of the same length $l$, then we have {$\rho(t, d) = \rho(kT, il)$, where $i = 1, 2, \dots, N$ represents the segment index}, and $k\in\mathbb{N}$ represents the discrete-time index. For simplicity, we define {$\rho(kT,il):=\rho_i[k]$}. The other variables are also defined in the same way, namely {$w_i[k],\psi_i[k],q_i[k]$, and $\phi_i[k]$}. The expressions for the flux function {$q_i[k]$ and $\phi_i[k]$} for Segment $i$ depend on the arrangement of the segments before and after that segment. {Mathematical expressions for the flux across different types of segment junctions and those for the traffic demand and supply functions needed to define the flux are omitted for brevity. Interested readers are referred to Appendix A for the same. Here, the demand of a segment denotes the traffic flux that wants to leave that segment while the supply of a segment denotes the traffic flux that can enter that segment.}

\subsection{State-space equations}
The discrete-time traffic flow and relative flow conservation equations for any Segment $i\in\Omega$ can be written as
\begin{subequations}\label{e:Godunov_update_equations1}
\begin{align}
    \rho_i[k\hspace{-0.5mm}+\hspace{-0.5mm}1]&\hspace{-1mm}=\hspace{-0.5mm}\rho_i[k]\hspace{-0.5mm}+\hspace{-0.5mm}\dfrac{T}{l}(q_{i-1}[k]\hspace{-0.5mm}-\hspace{-0.5mm}q_i[k]),\\
    \psi_i[k\hspace{-0.5mm}+\hspace{-0.5mm}1]&\hspace{-1mm}=\hspace{-1.5mm}\left(\hspace{-1mm}1\hspace{-0.5mm}-\hspace{-0.5mm}\dfrac{1}{\tau}\hspace{-0.5mm}\right)\hspace{-0.5mm}\psi_i[k]\hspace{-0.5mm}+\hspace{-0.5mm}\dfrac{T}{l}(\phi_{i-1}[k]\hspace{-0.5mm}-\hspace{-0.5mm}\phi_i[k])\hspace{-0.5mm}+\hspace{-0.5mm}\dfrac{v_f}{\tau}\rho_i[k]
\end{align}
\end{subequations}
Similar equations can be written for ramp segments as well. Here, $q_i[k]$ and $\phi_i[k]$ take the expressions presented in Appendix A depending upon the arrangement of Segment $i$ with respect to other segments.
The state vector for this system can be defined as 
\begin{align*}
\m x[k] := &[\rho_i[k]\hspace{1mm} \psi_i[k]\hspace{1mm} \ldots\hspace{1mm}\hat{\rho_j}[k]\hspace{1mm} \hat{\psi_j}[k]\hspace{1mm} \ldots\hspace{1mm}\check{\rho_l}[k]\hspace{1mm} \check{\psi_l}[k]\hspace{1mm} \ldots]^{\top}\\ \nonumber
&\in \mbb{R}^{2(N+N_I+N_O)},
\end{align*}
for which $i\in\Omega$, $j\in\hat{\Omega}$ and {$l\in\check{\Omega}$}. In this work, we assume that the demand and the driver characteristic upstream of the first mainline segment are known, that is, $D_0[k]\hspace{-1mm}=\hspace{-1mm}D_{in}[k]$ and $w_0[k]\hspace{-1mm}=\hspace{-1mm}w_{in}[k]$ and the density downstream of the last mainline segment is also assumed to be known, that is $\rho_{N+1}[k]\hspace{-1mm}=\hspace{-1mm}\rho_{out}[k]$. Similarly, the demand and driver characteristic upstream of the on-ramps and the density downstream of the off-ramps is also considered to be known. These values can be obtained using conventional detectors like the inductive loop detectors placed upstream of the input segments and downstream of the output segments of the highway. An approximate value of the demand can also be obtained using Origin-Destination flow matrices~\cite{yang2019novel} if available for the given region. Then,
\begin{align*}
    \m u[k] := [&D_{in}[k] \hspace{1mm}w_{in}[k]\hspace{1mm}\rho_{out}[k] \hspace{1mm} \ldots\hspace{1mm} \hat{D}_{in,j}[k]  \hspace{1mm} \hat{w}_{in,j}[k]
     \ldots\hspace{1mm}\hspace{1mm} \\ \nonumber
     &\check{\rho}_{out,l}[k]\hspace{1mm}\ldots]^{\top}\in \mbb{R}^{3+2N_I+N_O},
\end{align*}
where $j\in\hat{\Omega}$ and $l\in\check{\Omega}$.

The evolution of traffic density and relative flow 
 described in \eqref{e:Godunov_update_equations1} can be written in a compact state-space form as follows 
\begin{empheq}[box=\fbox]{align}
    \m x[k+1]=\m A\m x[k]+\m G\m f(\m x,\m u),
\label{eq:state_space_gen}
\end{empheq}
\noindent where $\m A \in \mbb{R}^{n_x\times n_x}$ for $n_x := 2(N+N_I+N_O)$ represents the linear dynamics of the system, $\m f:  \mbb{R}^{n_x}\times  \mbb{R}^{n_u}\rightarrow  \mbb{R}^{n_x}$ for $n_u=3+2N_I +N_O$ is a vector-valued function representing nonlinearities in the state-space equation and $\m G \in \mbb{R}^{n_x\times n_x}$ is a matrix representing the distribution of nonlinearities.

The nonlinearities in $\m f$ are in the form of a minimum of weighted nonlinear functions of the states and inputs. The structure of the above-mentioned matrices and functions is provided in Appendix B. Next, we discuss the measurement model for the ARZ model which is also nonlinear in nature.

\subsection{Sensor data and measurement model}
We consider two types of sensors in this work, the first is fixed sensors like the inductive loop detectors, and the second is moving sensors which include CVs. This study assumes that it is possible to retrieve density and speed data from both types of sensors. Two loop detectors installed at opposite ends of a segment can be used to obtain the traffic density (using an approach similar to \cite{lee2011density}) as well as the average speed of vehicles on the segment \cite{klein2006traffic}. CVs are known to provide the current position and speed data for individual vehicles directly. The average speed of a segment can be assumed to be the average of the speed data provided by all the queried CVs in that segment similar to ~\cite{bekiaris2017highway}. To obtain density data from CVs, we assume additional functionality including either spacing measurement equipment which is available as part of advanced driver assistance systems~\cite{seo2015estimation} or availability of vehicular ad-hoc networks (VANETs) which allow vehicles to communicate with each other in a neighborhood around the queried CV~\cite{panichpapiboon2008evaluation}. When assuming the latter it is important to note the limitation imposed by the communication range of the vehicles on the maximum cell length for traffic modeling. In the case of the former, while a cell length limitation may not be required, sufficient penetration of CVs is necessary on the segments that are queried for data. The data from the CVs is sent via cellular network to a network operator who performs any prior computation if necessary to convert the received information like the spacing data or neighborhood counts into density measurements before using them for state estimation. A measurement error can also be associated with the data at this point based on the available information on penetration rate and other factors.

{Note that in this setting, both fixed sensors and CVs are assumed to provide similar data on the density and speed of traffic on segments. Traditionally, CVs are considered akin to floating cars which provide only trajectory information at high sampling rates and with a broader spatial coverage as compared to fixed sensors. However, with the increasing number of vehicles and devices capable of sending and receiving data over the internet allowing vehicles and objects to communicate with each other such as in the case of VANETS, it is reasonable to expect that CVs could provide data comparable to fixed sensors in quality and type but superior in spatial coverage allowing similar data retrieval over the entire road stretch rather than a few fixed segments. Also, CVs being multi-functional and mobile require a lower commitment than fixed sensors.
}

\begin{figure}
    \centering
    \includegraphics[ width=0.4\textwidth]{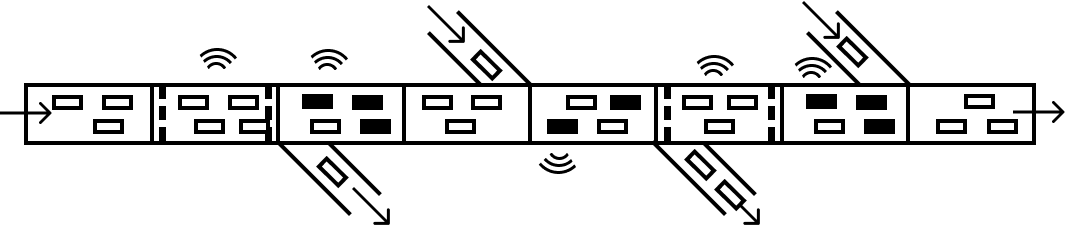}
    \caption{Heterogenous sensors on the highway: fixed sensors represented by dashed lines across the highway and CVs represented by the solid black rectangles.}
    \label{f:heterogenous_sensors}
\end{figure}

Figure \ref{f:heterogenous_sensors} presents a schematic of the sensors' placement on the highway.
Among the measurements, density $\rho_i[k]$ for any mainline segment $i\in\Omega$, and similarly for the ramps, is directly a state and is used as it is, while the velocity $v_i[k]$ can be written in terms of the states as follows:
\begin{align*}
    v_i[k] = \frac{\psi_i[k]}{\rho_i[k]} - p(\rho_i[k]).
\end{align*}
{We define a nonlinear vector-valued measurement function $\m h(\m x[k])$ which maps the state vector to a corresponding vector of all possible measurements in the system such that $\m h(\m x[k])\in\mathbb{R}^{n_x}$. Note, that here the number of possible measurements is equal to the number of states in the system. For any mainline segment $i\in\Omega$, the corresponding measurements are denoted by $h_{2i-1}$ and $h_{2i}$, where the subscripts represent the position of the element in the measurement function vector. These represent the density and speed of traffic on the segment, respectively. As mentioned above, these can be computed using the states corresponding to Segment $i$ that is $x_{2i-1}$ and $x_{2i}$ representing the density and relative flow, respectively, and are defined as follows:}
\begin{subequations}\label{e:measurement_function}
\begin{align}
    h_{2i-1}(\m x[k]) &= x_{2i-1}[k],\\
    h_{2i}(\m x[k]) &= \frac{x_{2i}[k]}{x_{2i-1}[k]} - p(x_{2i-1}[k]).
\end{align}
\end{subequations}
The mapping corresponding to the ramp segments can also be defined similarly. Now, we can define the measurement vector $\m y[k]\in n_p[k]$, where $n_p[k]$ is the number of available measurements from sensors at time $k$, as follows:
\begin{empheq}[box=\fbox]{align*}
    \m y[k] = \m C[k]\m h(\m x[k])+\m \nu[k],
\end{empheq}
where $\m C[k]\in\mathbb{R}^{n_p[k]\times n_x}$ is the observation matrix at time $k$ describing the availability of measurements from sensors. Note, that the observation matrix here is variable in time because of the measurements from CVs which allow data to be measured from different numbers and positions of segments with time. Here, $\m \nu[k]\in\mathbb{R}^{n_{\nu}[k]},n_{\nu}[k]=n_p[k]$ lumps all the measurement errors including the sensor noise into a single vector.

The above results are important as they allow us to perform state estimation for traffic systems using the second-order ARZ model. The state-space equation \eqref{eq:state_space_gen} can also be used for control purposes using control theoretic approaches from the literature. In the following section, we discuss a method for linearization of nonlinear functions which allows us to apply some linear state estimation methods to the otherwise nonlinear ARZ model. 

{\subsection{Linear Model Approximation}}\label{sec:model_linearization}
{The ARZ model specified in Section \ref{s:arz} is nonlinear due to the presence of the piecewise linear and nonlinear expressions in the traffic flux and relative flux terms. This prevents directly using some of the well-known and efficient linear state estimation methods from the literature. However, it is still possible to apply linear state estimation methods to a linearized version of the ARZ model. Methods such as Taylor series expansion~\cite{moreno1996vector}
can be used to obtain a good linear approximation of nonlinear functions about a suitable operating point. The detailed equations for linearization are omitted for brevity. The same can be found in Appendix C.}

{Note that linearization is usually associated with reduced model accuracy and hence worse estimation performance as compared to using the nonlinear model when the model is an exact representation of the system. However, when the process is not exactly governed by the model dynamics, linearization may not necessarily result in a degradation of the estimation results. The latter is particularly relevant when using traffic flow models such as the ARZ model for estimation as they only focus on the aggregate behavior of traffic and do not capture the nuances of vehicle-to-vehicle interaction. A validation study against real-world traffic data similar to \cite{yu2020pde} is required to quantify the trade-off between any loss of accuracy due to linearization versus the reduced computational load of using linear state estimation compared to nonlinear estimation. Such an investigation is considered out of the scope of the present work which mainly focuses on the theory and examples of using CV data for estimation from the perspective of linear state estimation.}

\section{State Estimation methods}\label{s:state_estimation}
In this section, we briefly discuss the different methods implemented in this work for TSE using the ARZ model. 

\subsection{Moving Horizon Estimation}\label{sec:state_estimation_mhe}
MHE is an optimization-based state estimation method that uses measurement data in batches from the most recent time horizon along with a process model to determine the states of the system. It involves solving an optimization problem at every time step of the process with the objective of minimizing the deviation of the estimated states from the modeled states as well as from the measurement data. Being an optimization problem, it is possible to include additional constraints in the problem such as bounds on the state variables. Depending upon whether the model is linear or nonlinear, MHE is divided into linear MHE and nonlinear MHE, both of which have been well explored in the literature. While linear MHE only requires solving a linear program or a quadratic program (QP) and is generally fast and easy to solve using available solvers, nonlinear MHE involves solving a nonlinear optimization problem which is both time-consuming and difficult. Since TSE for control is required to be done in real-time, in practice it is not always possible to spend enough time in solving a nonlinear optimization problem. Therefore, in this paper, we implement a linear MHE approach on a linearized version of the process model.

Throughout the paper, $N$ is used to denote the size of the horizon for optimization. For time steps up to $N$, that is, near the start of the process, the horizon size is kept equal to the number of time steps from the initial time up to that time. {The decision variables for the MHE optimization problem at any time step $k$ are the state vectors from step $k-N$ to $k$ out of which the vector at step $k$ is considered the final output for that step. The MHE algorithm implemented in this work has a similar objective function to \cite{rao2001constrained} with three components. The first component is known as the \textit{arrival cost} which serves to connect the decision variables of the current optimization problem with the estimates up to the previous time step. This effectively allows us to consider the impact of data prior to the current horizon in the estimation process. The second and third components are penalties on the deviation of the estimates from the measurement data and the modeled dynamics respectively. The notations $\mu, w_1$ and $w_2$ are used to denote the weights specifying our relative confidence on the past data and past estimates, the current measurement data, and the process model, and can be set by the modeler accordingly. The goal of the problem is to minimize these errors over the decision vectors. The remaining implementation including a thorough description of the decision variables, the objective function, and the constraints is omitted from the main body of this article as it does not contribute directly to the results of this paper. Interested readers are referred to Appendix D and Appendix E for detailed implementation and notes on comparison with other existing MHE algorithms. Algebraic transformations allow us to write the problem as a convex QP which can be solved using readily available QP solvers like CPLEX or MATLAB's \texttt{quadprog} function. Next, we present a brief discussion on the usage of KFs for TSE.}

\subsection{Kalman Filter variants and limitations}\label{sec:state_estimation_kf}
KFs are quite popular when it comes to TSE. Since the traffic process models are nonlinear we cannot use the ordinary KF, instead, most works use variants of KF designed for nonlinear systems namely the EKF, UKF, and EnKF. There is ample literature available on the design of these filters and their application in TSE, see \cite{seo2017survey} for references. A common limitation of the KF variants is that they do not inherently allow bounds on the state estimates. Since traffic states can only take values from a particular range, this makes it difficult to apply the KF variants directly. Instead, some modifications are required such as manually restricting the states to within their bounds after the state estimate for any time step is obtained. Another limitation of the KF variants is that they assume all errors to be Gaussian. This assumption is not necessarily true in many cases including the traffic system which can result in potential errors in state estimation. MHE naturally overcomes both of these limitations. 

In the following section, we discuss the implementation and results obtained by applying the above-mentioned estimation methods with the help of a numerical example.

\section{Numerical Study using VISSIM}\label{s:numerical_study}
In this section, we apply the state estimation methods discussed above namely EKF, UKF, EnKF, and MHE, on a traffic simulation example generated in VISSIM micro-simulation software under both fixed and moving sensors to highlight their advantages and limitations with respect to TSE and investigate the performance of moving sensors as compared to fixed sensors. 

All the simulations are carried out using MATLAB R2020a running on a 64-bit Windows 10 with 2.2GHz Intel$^\textrm{R}$ Core$^\textrm{TM}$ i7-8750H CPU and 16GB of RAM. We use the \texttt{quadprog} function in MATLAB to solve the MHE optimization problem. 

\subsection{Numerical study objectives}\label{s:case_study}
The primary goal of this study is twofold- to test the performance of the state estimation methods discussed in Section \ref{s:state_estimation} and to investigate the performance of moving sensors under various scenarios. In particular, we are interested in knowing the answers to the following questions:
\begin{itemize}
    \item \textit{Q1:} How does the number of fixed sensors on the highway impact the performance of the various estimation methods? Which method has the best estimation performance across different numbers of fixed segments?
    \item \textit{Q2:} How does the state estimation performance of various methods vary with moving sensors? What is the impact of different frequencies of change in measurement positions on the estimation performance?
    \item \textit{Q3:} Does the positional configuration of moving sensors impact state estimation performance?
    \item \textit{Q4:} Which state estimation method is more robust to {measurement errors}? Do moving sensors impact estimation performance with different levels of {data quality due to factors such as sensor noise and CV penetration rate}?
\end{itemize}
Following is a description of the highway structure used for this study.

\subsection{Highway setup and VISSIM simulation}\label{s:highway_structure}

\begin{figure}
    \centering
    \includegraphics[width=0.4\textwidth]{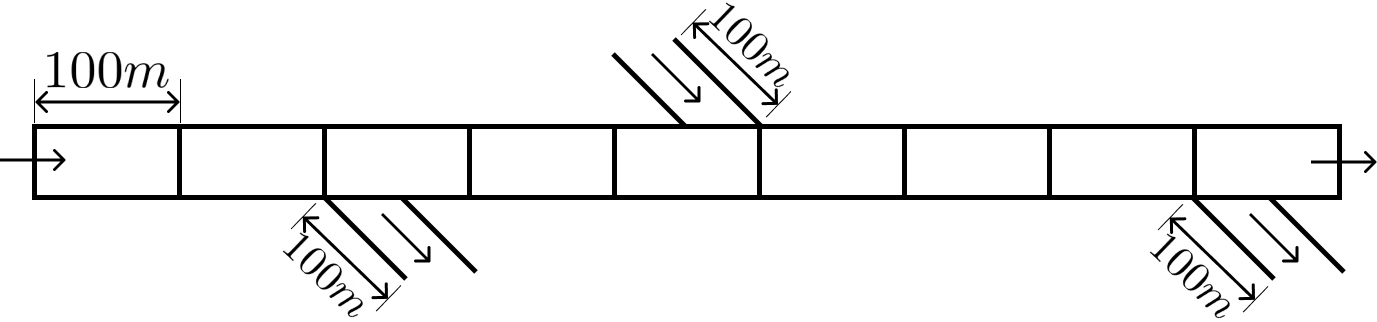}
    \caption{Schematic diagram of the highway considered in this study.}
    \label{f:highway_schematic}
\end{figure}

In this study, we model the highway stretch as shown in Figure \ref{f:highway_schematic} consisting of one on-ramp and two off-ramps. An additional 100 m of highway stretch is modeled in VISSIM preceding the shown stretch. While we only perform state estimation on the latter 900 m and the attached ramps, this additional stretch of highway modeled in VISSIM provides us with the system inputs namely the demand upstream of Segment 1 and the upstream density and speed which are used to calculate the upstream driver characteristic. A similar 100 m stretch is modeled upstream of the on-ramp as well and serves the same purpose of providing the exact inputs. We set the following parameters for the Weidemann 99 car-following model in VISSIM: CC0 $1.50$ m, CC1 $0.9$ s, CC2 $4.00$ m, CC3 $-8.00$, CC4 $-0.50$, CC5 $0.60$, CC6 $6.00$, CC7 $0.25$ m/s$^2$, CC8 $1.00$ m/s$^2$, and CC9 $1.50$ m/s$^2$. The speed limit is set to 102 km/hr. Under the Godunov scheme, the highway and ramps are divided into segments of length 100 m each with a time-step value of 1 s, which satisfies the CFL condition. Thus, there are a total of 24 states in this highway system.  

A traffic jam is introduced in the middle of the highway stretch to replicate a congested scenario which is more interesting for studying state estimation performance. In VISSIM, the jam is created with the help of a reduced speed decision area implemented on Segment 7 of the stretch. The reduction in traffic speed causes a reduction in flow creating a traffic jam that travels upstream on the highway. The jam dissipates once the speed of the reduced speed area is restored. The simulation scenario replicates the formation and dissipation of a traffic jam similar to that caused by an incident in the middle of the stretch. Note that the scenario considered in this work differs from those considered in previous studies such as in \cite{Seo2017} where congestion travels upstream from the downstream end of the road stretch where estimation is performed. In that case, the source of the jam would be captured in the downstream supply conditions which are input to the system. In the scenario considered in this paper, the jam originates in the middle of the stretch. It is therefore not directly captured by any of the inputs and therefore the process model. The given scenario is arguably more difficult to estimate due to the absence of informative inputs to guide the process model.

The ARZ model parameters are selected to keep the simulated state trajectories from the macroscopic model as close to the VISSIM simulation as possible. The selected values are: $v = 102$ km/hr, $\rho_m = 345$ veh/km, $\tau = 20,$ and $\gamma = 1.75$. As mentioned before, in this work we do not track individual vehicles, instead, we consider VANETs formed from CVs capable of measuring the density and speed of segments apart from fixed sensors. We consider a high penetration rate of CVs on the network such that we can query any desired segment for data. The only constraint we impose is bandwidth constraint on data transfer which limits the number of segments from which data can be obtained simultaneously.

\subsection{Observability of the system}\label{s:observability}
To determine the required minimum number and the corresponding placement of sensors, we perform a test of observability for our system using the concept of \textit{Observability Gramian} for discrete-time systems~\cite{chen1984linear}. The method is originally meant to determine the observability of linear systems. In this case, we use it to check the observability of the linearized ARZ model. The observability Gramian is defined as
\begin{align*}
    \m W_k = \sum_{m=0}^{\infty} (\tilde{\m A}_k^{\top})^m \tilde{\m C}[k]^{\top} \tilde{\m C}[k] \tilde{\m A}_k^m,
\end{align*}
where $\tilde{\m A}_k$ is the coefficient matrix of the linearized state-space model and $\tilde{\m C}[k]$ is the observation matrix of the linearized measurement model at time $k$ around a suitable operating point. The system is considered observable if $\m W_k$ is positive definite. In this case, since the model parameters change with time due to changing operating points of linearization, the Gramian changes with time as well. This can result in a change in the observability properties. To check if the system is observable for a given sensor placement, we calculate the Gramian for each time step over the duration of the simulation. 

From this study, we find that to make the system observable, we need to at least sense the states on the last mainline segment and on all the off-ramps. Therefore, throughout the study, we keep fixed sensors on these segments. Any additional sensors are placed after these segments are populated with sensors. This is similar to the observations in \cite{bekiaris2016highway} with respect to the observability of the model used in that paper. It appears to be a common property of traffic models that the states of the output segments of the network (last mainline segment and off-ramps) need to be measured to ensure full-observability of the system. This is not surprising as traffic models share similar state-update equations and therefore have a similar structure of the state-space parameters which form the observability matrix. While the concept of observability can also be used to determine the optimal sensor placement for state estimation for any given number of sensors under certain conditions~\cite{9585066}, here we only use it to determine a minimum number of sensors and their placement. In the following section, we discuss some nuances of implementing the aforementioned estimation methods in the current study.

\subsection{Implementation of estimation methods}
\subsubsection{Evaluation metrics}
We use the \textit{root mean squared error} ($\textrm{RMSE}$) and {the \textit{symmetric mean absolute percentage error} ($\textrm{SMAPE}$)}~\cite{6739119} between the estimated and ground truth (simulated on VISSIM) density denoted by $\textrm{RMSE}_{\rho}$ and {$\textrm{SMAPE}_{\rho}$}, respectively, and those between the estimated and ground truth speed denoted by $\textrm{RMSE}_{v}$ and {$\textrm{SMAPE}_{v}$}, respectively to evaluate the performance of different methods. These metrics are defined as follows:
\begin{align}
    \textrm{RMSE}_{\rho} & = \sqrt{\dfrac{1}{n_xt_f}\sum_{i=1}^{n_x}{\sum_{k=1}^{t_f}{(e^{\rho}_i[k])^2}}},\\
    \textrm{{SMAPE}}_{\rho} & = \dfrac{100}{n_xt_f}\sum_{i=1}^{n_x}{\sum_{k=1}^{t_f}\dfrac{|e^{\rho}_i[k]|}{\Xi^{\rho}_i[k]}},\\
    \textrm{RMSE}_{v} & = \sqrt{\dfrac{1}{n_xt_f}\sum_{i=1}^{n_x}{\sum_{k=1}^{t_f}{(e^v_i[k])^2}}},\\
    \textrm{{SMAPE}}_{v} & = \dfrac{100}{n_xt_f}\sum_{i=1}^{n_x}{\sum_{k=1}^{t_f}\dfrac{|e^v_i[k]|}{\Xi^v_i[k]}},
\end{align}
where $t_f=500$ sec is the total time of simulation, $n_x=N+N_I+N_O$ is the total number of segments in the system,  $e^{\rho}_i[k]$ and $e^{v}_i[k]$ denote the difference between the actual and estimated density and speed, respectively for the $i^{th}$ segment at time-step $k$, {and $\Xi^{\rho}_i[k]$ and $\Xi^{v}_i[k]$ denote the sum of absolute values of the actual and estimated density and speed, respectively for the $i^{th}$ segment at time-step $k$}.
We do not consider the error in the relative flow states for evaluation since it is not directly relevant for traffic operators as compared to density and speed which are fundamental quantities in traffic. {Note that, in this work, SMAPE is selected over the \textit{mean absolute percentage error} (MAPE)~\cite{Seo2017} as MAPE has no upper bound and can give infinitely large values when the actual value is close to zero which is possible with traffic densities and speeds. The SMAPE on the other hand is bounded and can only assume values from $0\%$ to $100\%$. While MAPE is easier to interpret than SMAPE, it is highly unstable in the present scenario and hence not preferable.}

\subsubsection{Parameter tuning}
\label{s:parameter_tuning}
In implementing KFs, three parameters need to be set in advance namely the estimate error covariance matrix ($\m P$), the process noise covariance matrix ($\m Q$), and the measurement noise covariance matrix ($\m R$). In practical applications, these matrices are not known in advance or are difficult to get. In this paper, for all the KF variants, we use a process noise covariance matrix of the form $\m Q = q\m I_{n_x}$ where $q\in\mathbb{R}_+$ and $\m I_{n_x}$ is an identity matrix of dimension $n_x$. Similarly, the measurement noise covariance matrix is set as $\m R = r\m I_{n_p[k]}$ with $r\in\mathbb{R}_+$ and $n_p[k]$ is the number of measured states at time $k$. The initial guess for the estimate noise covariance matrix is taken as $\m P=10^{-3}\m I_{n_x}$. We set the values of $r$ and $q$ to $1$ which is found to be sufficient. Marginally better results for the KFs can be obtained in each case by fine-tuning these matrices but it is avoided as in reality the real states are not known in advance.
In general, algorithms requiring minimal tuning to achieve a reasonable quality of state estimates are desirable. For an estimation algorithm, it is important to determine a set of parameters robust to both traffic conditions and sensor placement. Here we focus on testing the algorithms with parameter values that yield reasonable state estimates across all scenarios, rather than fine-tuning parameters for individual cases.

 Besides these values, there are also some method-specific parameters such as in UKF and EnKF. We find that fine-tuning the values of these parameters does not influence the performance of the methods considerably. For UKF, we set the following values: $\alpha=0.1, \kappa=-4$, and $\beta=2$, and for EnKF, we set the number of ensemble points to $100$. These values are found to be sufficient for the respective methods. Interested readers can refer to \cite{wan2000unscented} and \cite{evensen2003ensemble} for interpretation of parameters and more detail on implementation of UKF and EnKF respectively.

 For MHE, we set the values of the weights $\mu=1, w_1=1$, $w_2=1$, and the horizon length $N=4$. Just as with the KF parameters, fine-tuning MHE parameters is not a focus of this study, and the same parameter values are used in all tested scenarios without further tuning. In general, a large $N$ is considered ideal as it allows the algorithm to track the system dynamics for a longer duration and also considers more data. However, this is not necessarily beneficial to estimation if the process model does not closely follow the real system states. In that case, particularly with a large weight $w_2$ on the process model error, the error in estimates can increase with increasing $N$ as the error due to incorrect dynamics is amplified. To roughly tune $N$, we vary the horizon length from 1 to 10 with different numbers of fixed measurement segments. It is seen that the best $N$ becomes smaller with an increasing number of measured segments. Also, a larger weight on the measurement error improves the results when there are more sensors. Both these observations are reasonable since the measurement data in this case is more accurate than the modeled states and so with sufficient data, increasing $N$ only deteriorates the estimates by increasing the influence of the process model. Given a combination of these reasons, the aforementioned values are found reasonable for MHE.

\subsubsection{Re-scaling to avoid numerical issues}
The large difference in the order of magnitude of the two states, density, and relative flow, results in numerical issues in both the KFs as well as in MHE. This is handled by re-scaling the objective and constraints of the optimization problem in the case of MHE and by re-scaling the state vector in the case of KFs.

\subsubsection{Applying external bounds on states}
The KFs sometimes run into the problem of producing non-physical states such as negative or extremely large densities and relative flows. This is an issue for the process model which includes terms like density raised to fractional power as in \eqref{e:pressure}, which results in numerical issues and forces the estimation to stop. Therefore, it is important to bind the estimates from KFs to only the physical values of the states. In that, we project the obtained estimates in the case of EKF to a range with a lower bound of zero on all states, and an upper bound of $\rho_m$ on the traffic densities and $\rho_m v_f$ on the relative flows. In the case of UKF, the sigma points are projected first followed by the obtained estimate. In the case of EnKF, the ensemble points are projected within specified bounds. This method of projecting vectors for EKF and UKF has been shown to fit in the KF theory mathematically and is among popular methods mentioned in \cite{simon2010kalman}.

We present the results of the study in the following section.

\subsection{Results and discussion}

\subsubsection{Comparison under fixed sensor positions}
\label{s:number_of_fixed_sensors}
\begin{figure}
    \centering
    \includegraphics[ width=0.35\textwidth]{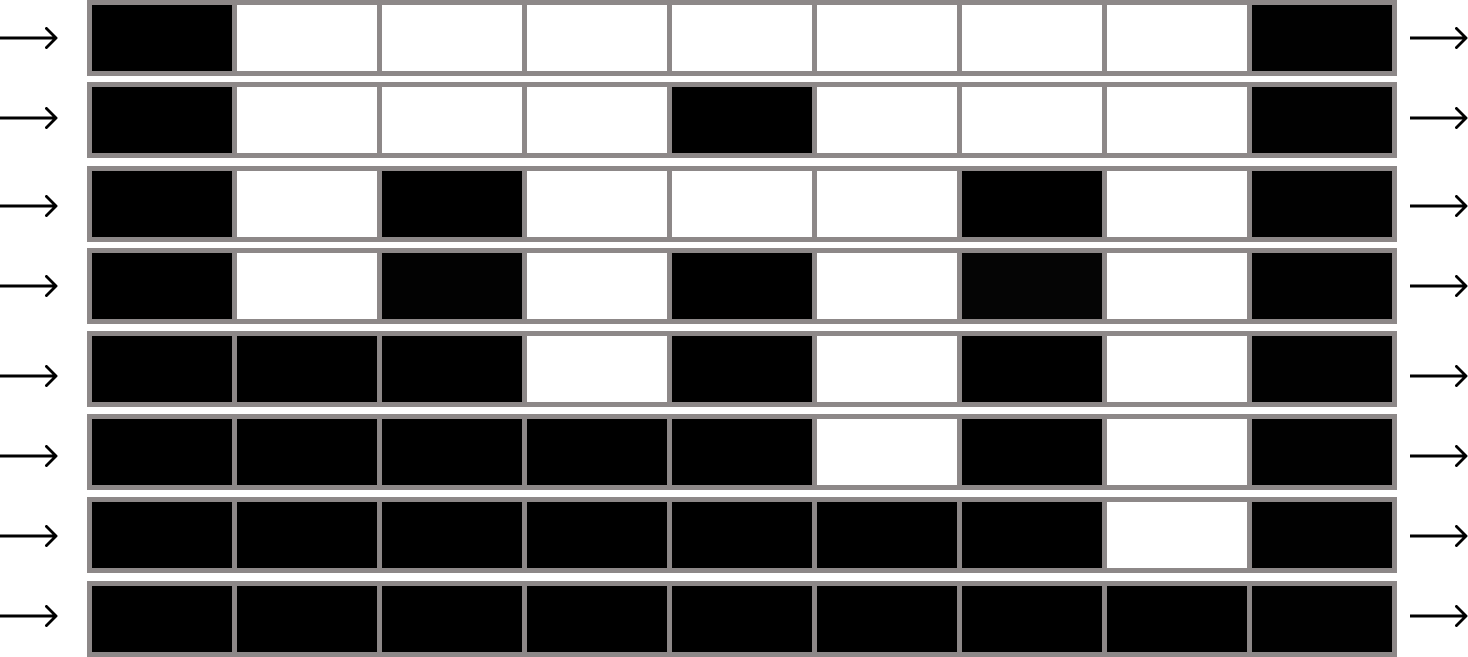}
    \caption{Configurations for fixed sensor placement on the mainline segments. Black boxes depict segments with sensors and white boxes depict otherwise. Arrows indicate the direction of traffic. The ramp segments containing an additional 3 sensors are not presented in this figure. The top and bottom rows present the configuration with a total of 5 and 12 sensors in the system, respectively.}
    \label{f:sensor_config}
\end{figure}

As sensors are indeed costly, it is imperative to determine which state estimation methods perform better with less number of sensors, and how the performance varies with the changing number of sensors. Herein, we test the effect of increasing the number of fixed sensors on the performance of the four estimation methods. We do not consider any CVs in this case. As discussed in Section \ref{s:observability}, we have a minimum of three sensors, one on the last mainline segment and one each on the off-ramps. We also assume that there is always a sensor on the on-ramp. As we add more sensors we try to keep them well-distributed across the highway. The placement of the mainline sensors is depicted in Figure \ref{f:sensor_config}.
No additional process noise is added to the state values generated from VISSIM while a zero mean uniform random noise with a standard deviation of $1$ is added to the sensor measurements.

\begin{figure}
    \centering
    \includegraphics[ width=0.23\textwidth]{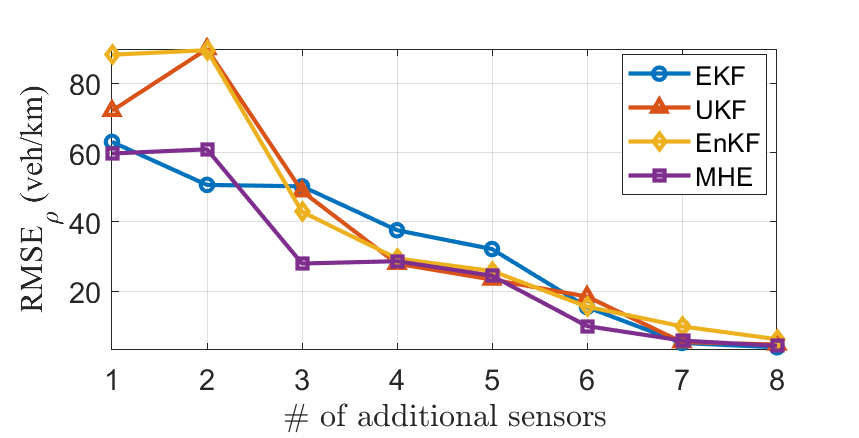}
    \includegraphics[ width=0.23\textwidth]{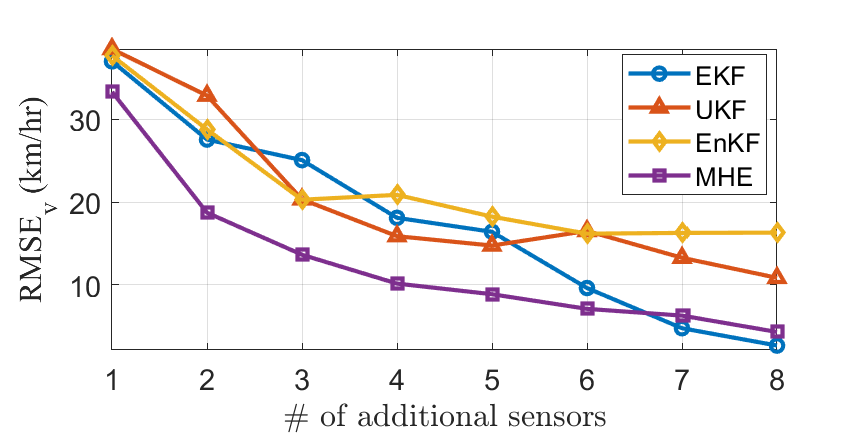}
    \includegraphics[ width=0.23\textwidth]{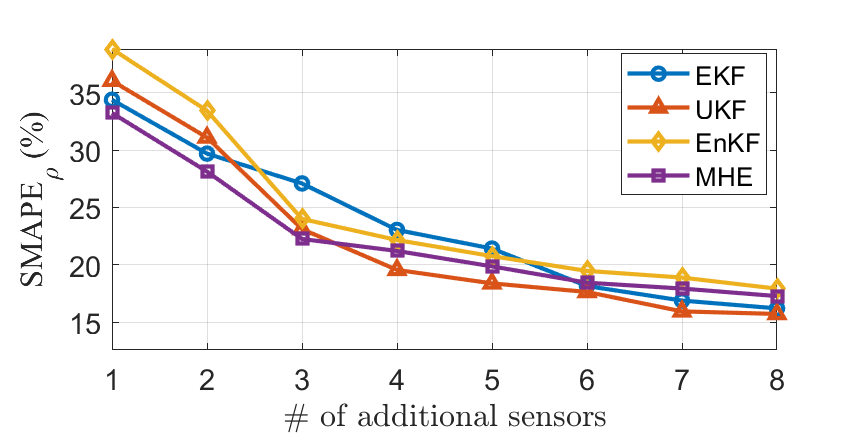}
    \includegraphics[ width=0.23\textwidth]{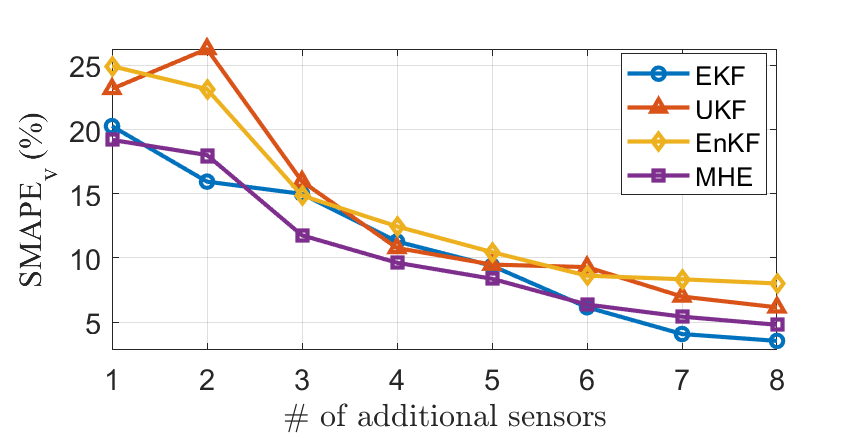}
    \caption{{$\mathrm{RMSE}_{\rho}$ [top left], $\mathrm{RMSE}_{v}$ [top right], $\mathrm{SMAPE}_{\rho}$ [bottom left], and $\mathrm{SMAPE}_{v}$ [bottom right] with different numbers of fixed sensors.}}
    \label{f:number_of_sensors}
\end{figure}

\begin{figure}
  \subfigure[]{\includegraphics[width=0.24\textwidth]{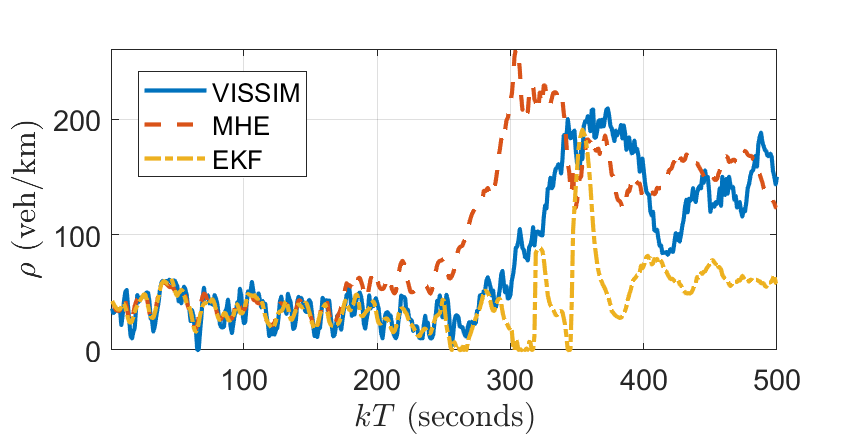}}
  \vspace{-1.5mm}
  \subfigure[]{\includegraphics[width=0.24\textwidth]{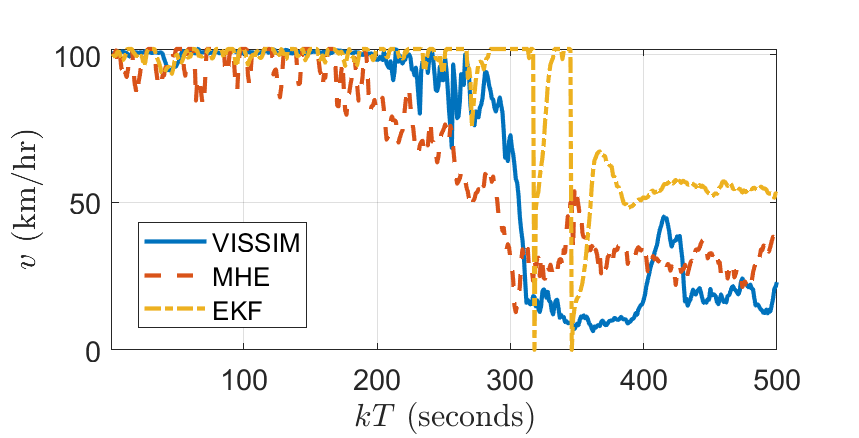}}
  \vspace{-1.5mm}
    \subfigure[]{\includegraphics[width=0.24\textwidth]{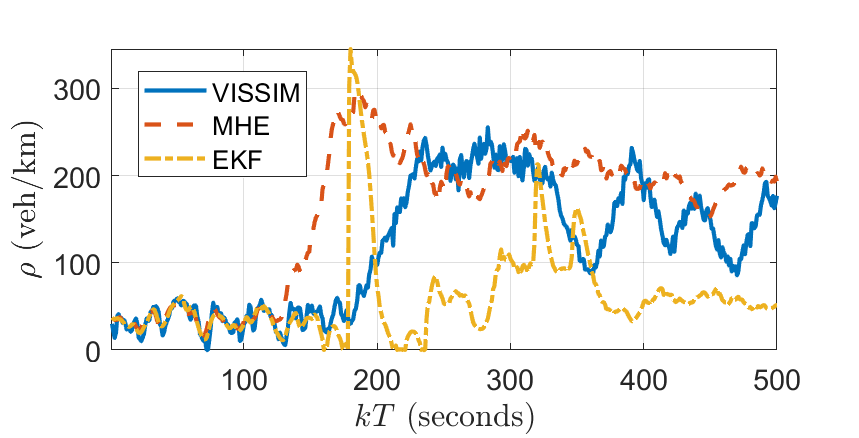}}
    \vspace{-1.5mm}
    \subfigure[]{\includegraphics[width=0.24\textwidth]{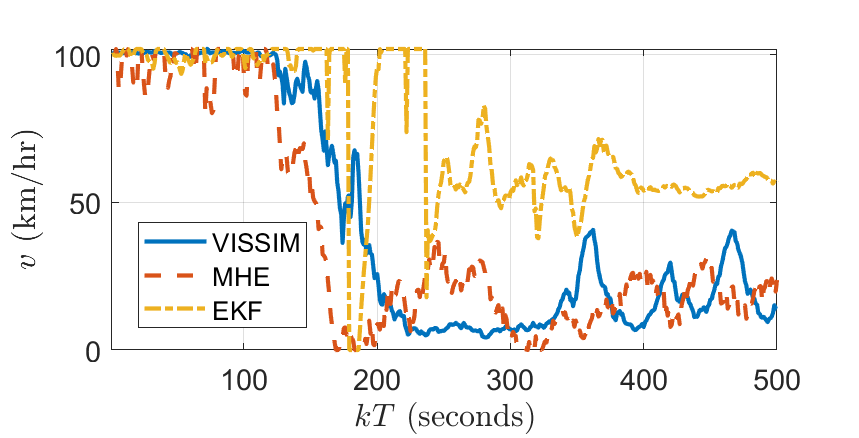}}
    \vspace{-1.5mm}
      \subfigure[]{\includegraphics[width=0.24\textwidth]{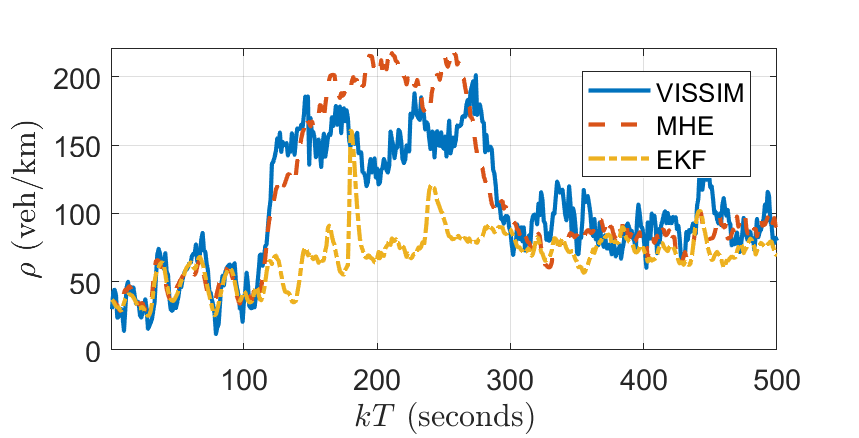}}
      \vspace{-1.5mm}
    \subfigure[]{\includegraphics[width=0.24\textwidth]{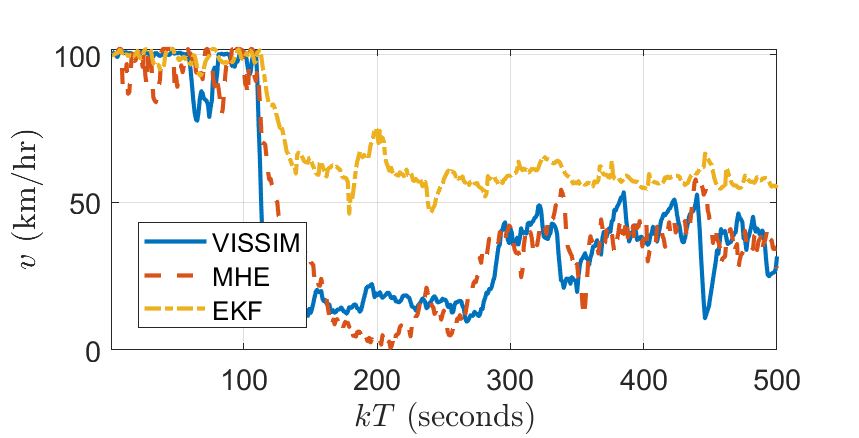}}
    \vspace{-1.5mm}
   \subfigure[]{\includegraphics[width=0.24\textwidth]{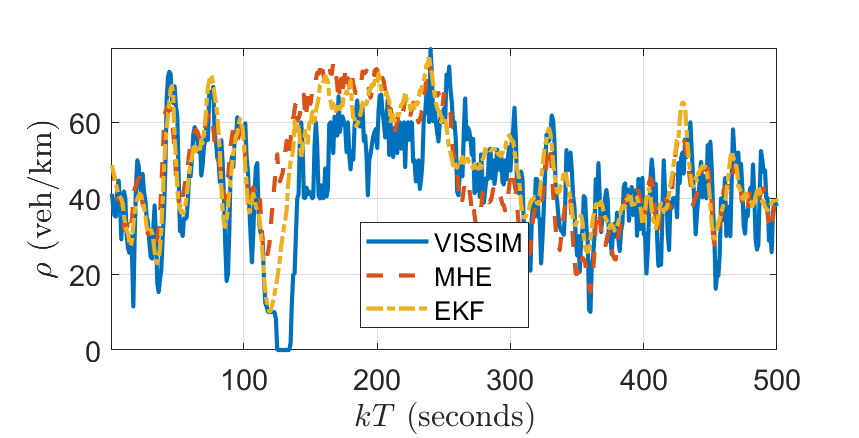}}
    \subfigure[]{\includegraphics[width=0.24\textwidth]{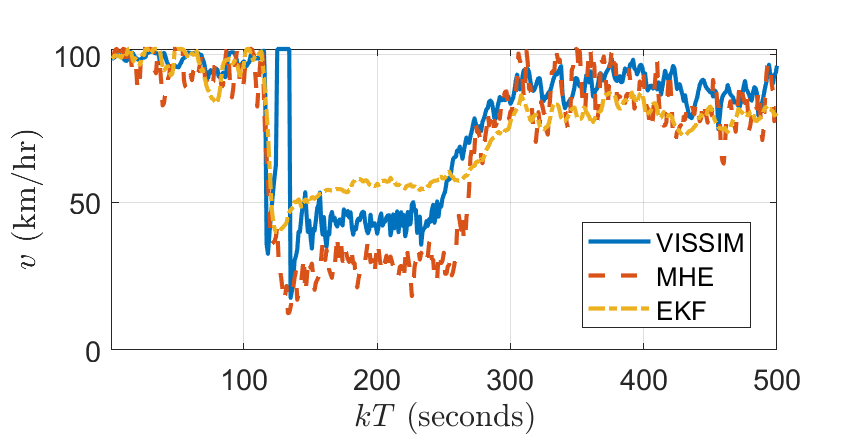}}
  \caption{Plots of simulated and estimated trajectories for densities [left] (a, c, e, g) and speeds [right] (b, d, f, h) in the presence of 4 additional fixed sensors. Rows of figures correspond to the unmeasured Segments 2, 4, 6, and 8 respectively.}\label{f:trajectories}
\end{figure}

\begin{figure}
    \includegraphics[width=0.23\textwidth]{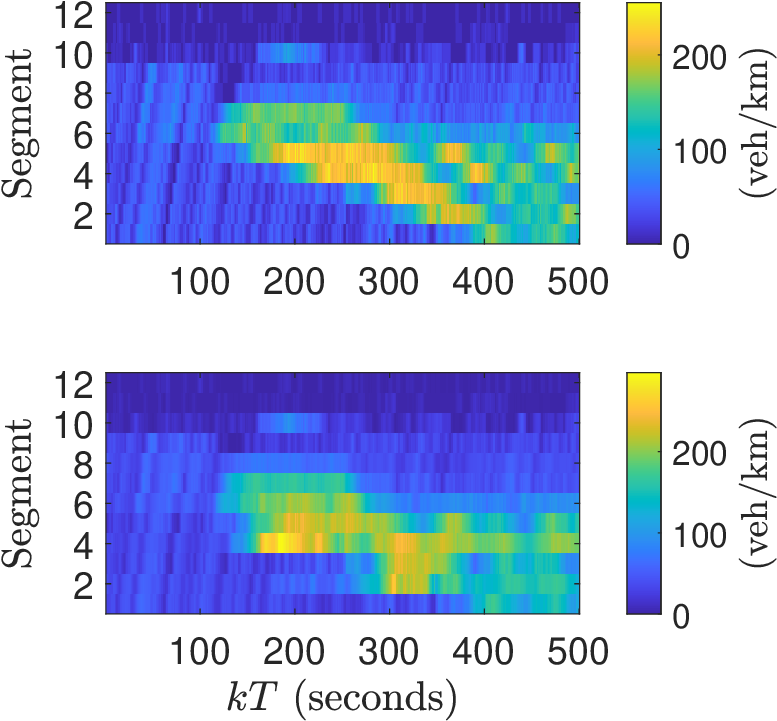}
    \includegraphics[width=0.23\textwidth]{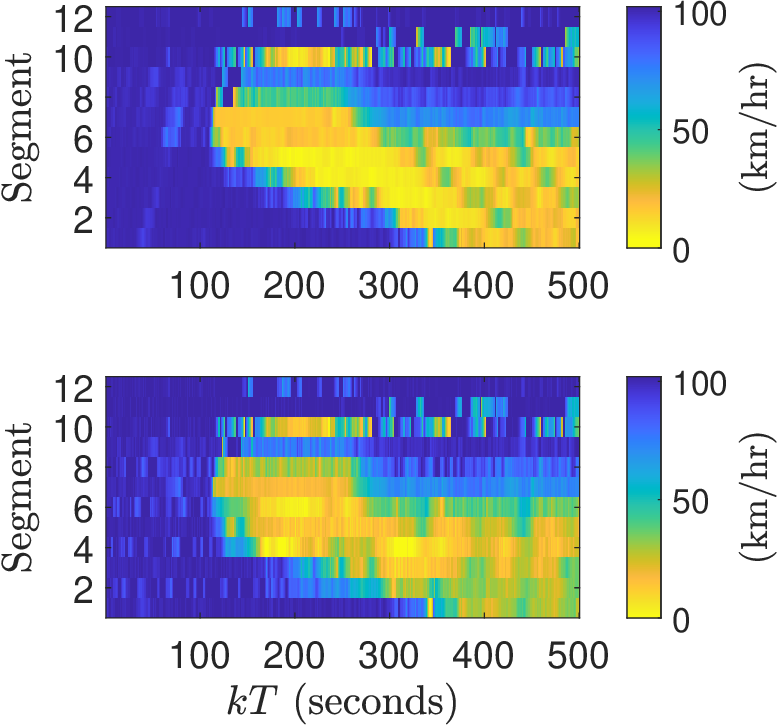} 
    \caption{Comparison of simulated and estimated densities [left] and speeds [right] obtained from MHE for the mainline segments (1 to 9), on-ramp segment (10), and off-ramps segments (11 \& 12).}
    \label{f:estimates2}
\end{figure}

Figure \ref{f:number_of_sensors} presents the plots of the evaluation metrics for each state estimation method. The x-axis presents the number of additional fixed sensors considered on the highway other than the sensors on the last mainline segment and ramps.
Figure \ref{f:number_of_sensors} shows that the performance of all the state estimation methods in terms of density and speed estimation improves with more additional sensors. {Between methods, the $\mathrm{RMSE}_{\rho}$ and $\mathrm{SAMPE}_{\rho}$ at different numbers of segments appear comparable. A difference in performance in favor of MHE and EKF is observed in terms of $\mathrm{RMSE}_{\rho}$ at a small number of additional sensors but the difference diminishes as the number of sensors is increased. In terms of $\mathrm{RMSE}_{v}$, MHE outperforms other methods at all numbers of additional sensors and is marginally outperformed by EKF at the two highest numbers of additional sensors. In terms of $\mathrm{SAMPE}_{v}$, MHE and EKF perform comparably and better than other methods at a smaller number of additional sensors while all methods show a comparable performance at a higher number of sensors.} While RMSE values are suitable for quantitative comparison between estimation methods, we need to compare the trajectories of the estimated states for a qualitative comparison. Figure \ref{f:estimates2} presents a 2-dimensional plot of the simulated and estimated density and speed evolution using MHE on all segments for the discussed scenario to provide a complete picture of the traffic evolution for the reader's reference. Figure \ref{f:trajectories} presents the simulated and estimated trajectories for the unmeasured segments for the case with 4 additional measured segments using MHE and EKF. The estimated trajectories obtained using UKF and EnKF are omitted from the plots in the main text to ensure clarity. The latter is presented in Appendix \ref{a:trajectories} for interested readers. The trajectories for the other cases of additional sensors are also omitted for brevity as they do not add value to the discussion provided in the context of the presented plots.

While EKF and MHE perform similarly for Segment 8 which does not have congestion, Figure \ref{f:trajectories}e shows that EKF is not able to estimate the congested density on Segment 6. It does show a few spikes and a slight gradual increase in density but overall there is no significant congestion depicted by EKF. On the other hand, MHE follows the congested density more closely and also returns to the less congested ground truth condition once the congestion ends. In Figures \ref{f:trajectories}a and \ref{f:trajectories}c, MHE estimates congestion on the respective segments although the congestion is shown to start earlier than when it actually occurs. This is because the considered model inputs do not force congestion on any part of the stretch but one is observed in the measured data for middle segments from about 100 seconds into the simulation. The correction applied to the modeled states that are otherwise free-flowing to replicate the congestion in the measurement data causes congestion to be depicted earlier in the upstream segments corresponding to when it first occurs in the downstream segments rather than when it actually occurs on the upstream segments. Except for the time of the start of congestion in the estimated states on upstream segments, MHE is able to replicate well the magnitude of the congestion in terms of both density and speed which EKF fails to do. The trajectory plots also explain the closeness of EKF and MHE in terms of $\text{RMSE}_{\rho}$ and the significantly larger difference in $\text{RMSE}_{v}$. Both MHE and EKF observe a similar deviation from the actual density, with MHE depicting the congestion to start earlier while EKF not depicting or only partially depicting the congestion. As a result, both methods show a close $\text{RMSE}_{\rho}$. However, since MHE replicates the congestion while EKF does not, the former results in reduced speeds which are closer to the actual speeds than the higher ones estimated by EKF as a result of estimating lower densities. This results in a significantly smaller $\text{RMSE}_{v}$ for MHE as compared to EKF. {On the other hand, since EKF constantly overestimates the speeds, the denominator of $\mathrm{SAMPE}_{v}$ becomes large causing this metric to be close to its value for MHE despite a larger absolute error.}

The average run times per time step of simulation for the methods are given in Table \ref{t:run_times}.
\begin{table}
\renewcommand{\arraystretch}{1.4}
\caption{Computational time for state estimation per time step (1 sec) of simulation.}
    \label{t:run_times}
\centering
    \begin{tabular}{|c||c|c|c|c|}
    \hline
         \textbf{Method} &  \textbf{EKF} &  \textbf{UKF} & \textbf{EnKF} & \textbf{MHE}\\
         \hline
         \textbf{Computation Time (sec)} & 0.002 & 0.006 & 0.016 & 0.075\\

         \hline
    \end{tabular}
\end{table}
The run times include the time from when the data is received along with the information about the current observation matrix $\m C[k]$ to when an estimate is produced.
While the computation time of MHE is significantly higher than the KF variants, it is still only a fraction of the second and useful for real-time control. Moreover, the increased compute time can be justified by the improved estimation performance offered by MHE.

\subsubsection{Effect of moving sensors}\label{s:speed_impact}
\begin{figure}
    \centering
    \includegraphics[width=0.24\textwidth]{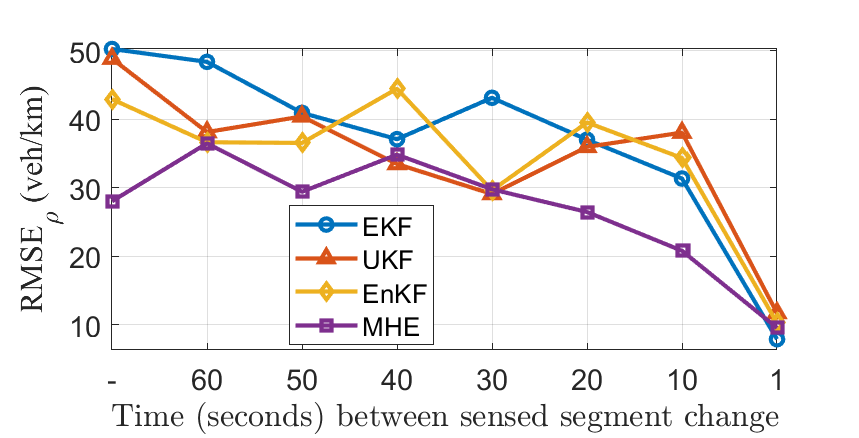}
    \includegraphics[width=0.24\textwidth]{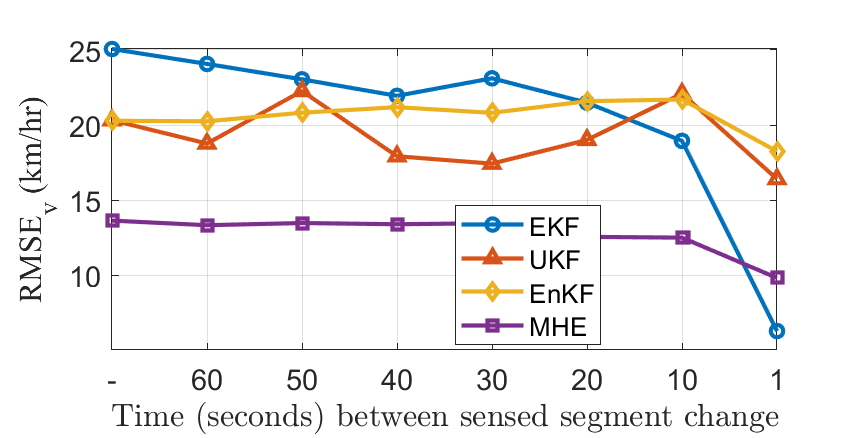}
    \includegraphics[width=0.24\textwidth]{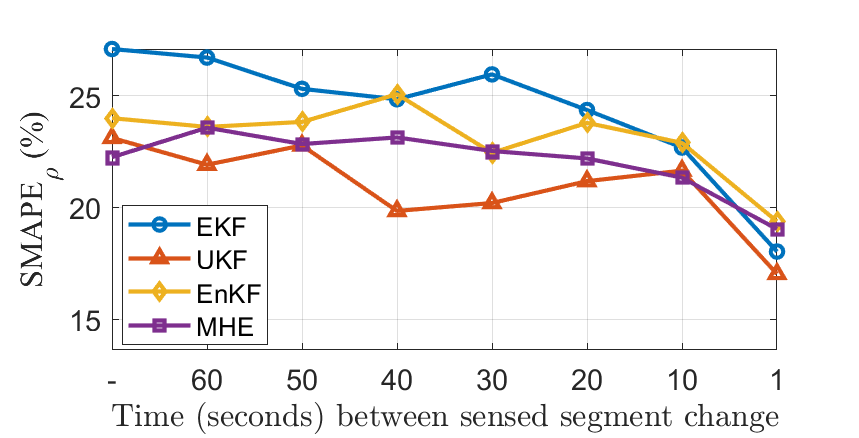}
    \includegraphics[width=0.24\textwidth]{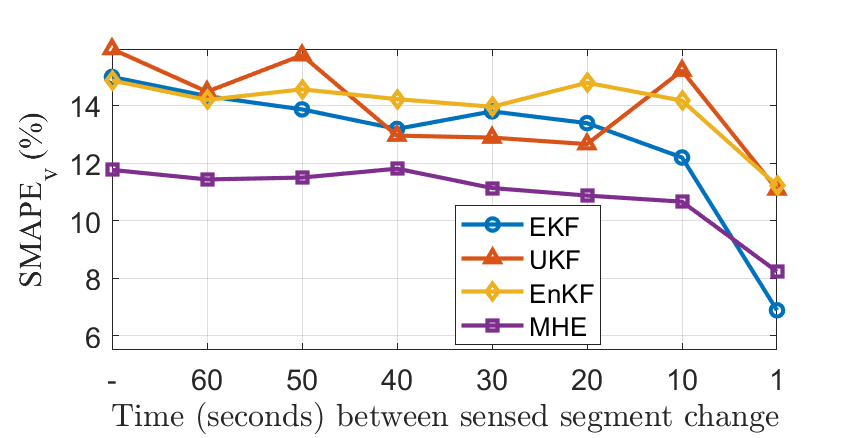}
    \caption{{$\mathrm{RMSE}_{\rho}$ [top left], $\mathrm{RMSE}_{v}$ [top right], $\mathrm{SAMPE}_{\rho}$ [bottom left], $\mathrm{SAMPE}_{v}$ [bottom right]  with different duration between changes in sensor positions. The symbol `-' at the beginning of the x-axis represents the scenario with fixed sensor locations throughout the simulation.}}
    \label{f:cav_speeds}
\end{figure}

CVs can be used to measure traffic data from different segments over time giving more flexibility in terms of data collection than fixed sensors. Here, we test the impact of changing the segments from which measurements are obtained over time on the estimation performance of the considered state estimation methods. The frequency of change in measured segments is also varied and its impact on the estimation performance is analyzed. The last mainline segment and all ramp segments are assumed to have fixed sensors and CVs are used to get data from other segments. We assume that there is a sufficient penetration of CVs on the roadway to allow data collection from any segment on the stretch. However, we assume a restriction on the bandwidth for data transfer such that density and speed data collected using CVs can only be transferred from 3 segments at a time. A fixed bandwidth for data transfer in real-time is a realistic assumption however a stringent one of 3 segments is considered here in particular to clearly observe any benefit of covering different segments over time than fixed sensors which is difficult to observe if data is collected from several segments at all times. Hereafter, in the context of moving sensors, the term measured segment is used to refer to segments from which data is transferred and used for estimation rather than where data is collected (which is assumed to be all segments). Similarly, sensor position is used to refer to the position of a measured segment ignoring segments where sensors are present but data is not used for estimation. The initial sensor positions are the same as the third row from the top in Figure \ref{f:sensor_config} with data being obtained from Segments $\{1, 3, 7\}$. The sensor positions are changed after a fixed duration of time. For this analysis, the duration is varied from indefinite (equivalent to fixed sensors) to 1 second (collecting data from a different set of segments every time step). A systematic update of sensor positions is utilized such that at every change the segments immediately following the current segments are selected. For instance, after Segments $\{1, 3, 7\}$, the positions are changed to Segments $\{2, 4, 8\}$. From Segment 8, the position is changed directly to Segment 1 skipping Segment 9 since Segment 9 already has a fixed sensor. So from Segments $\{2, 4, 8\}$, the positions are changed to Segments $\{3, 5, 1\}$, and so on after the duration of change in each case. Figure \ref{f:cav_speeds} presents the plots of the evaluation metrics with increasing frequencies of changing the position of sensors for the four state estimation methods.

\begin{figure}
  \subfigure[]{\includegraphics[width=0.24\textwidth]{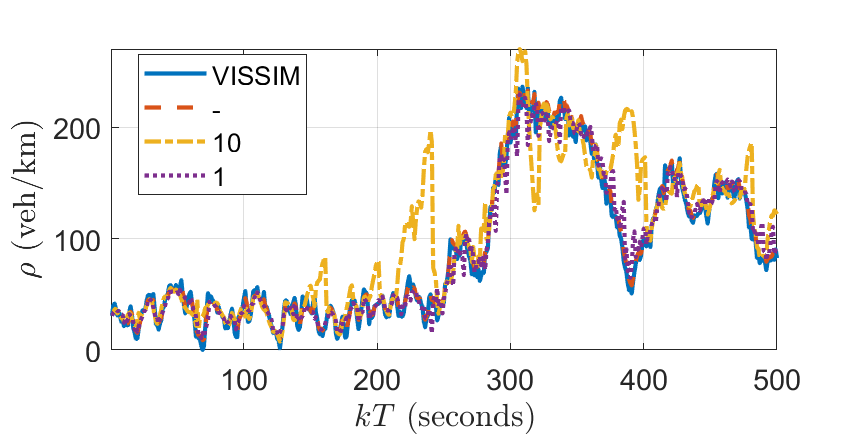}}
  \vspace{-1.5mm}
  \subfigure[]{\includegraphics[width=0.24\textwidth]{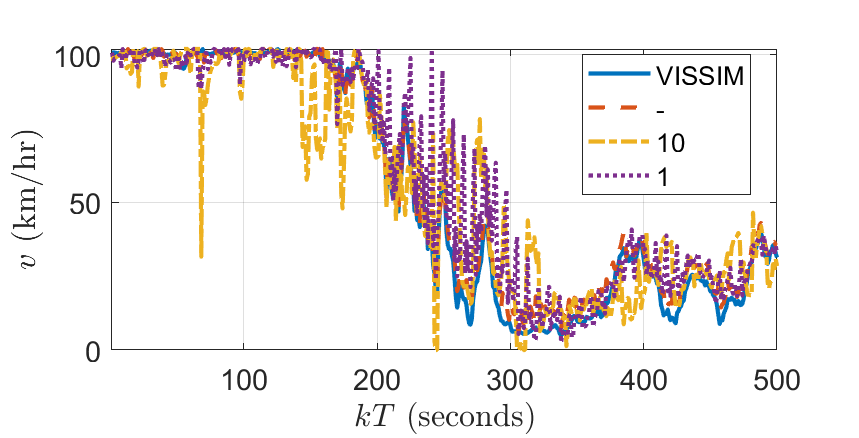}}
  \vspace{-1.5mm}
    \subfigure[]{\includegraphics[width=0.24\textwidth]{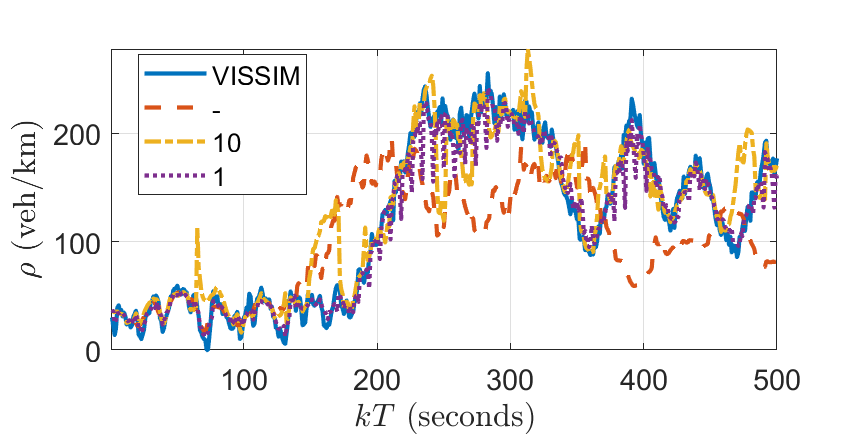}}
    \vspace{-1.5mm}
    \subfigure[]{\includegraphics[width=0.24\textwidth]{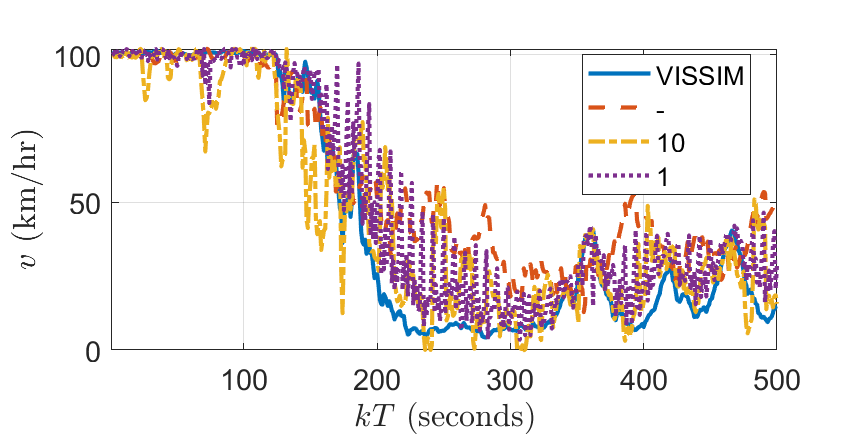}}
    \vspace{-1.5mm}
    \subfigure[]{\includegraphics[width=0.24\textwidth]{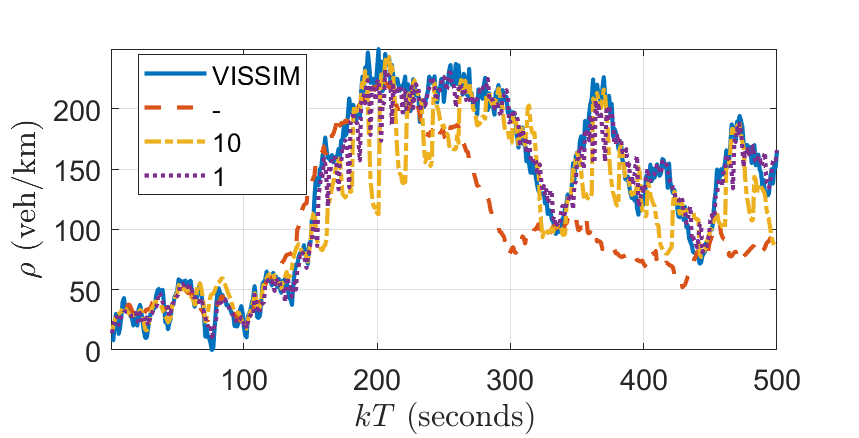}}
    \vspace{-1.5mm}
    \subfigure[]{\includegraphics[width=0.24\textwidth]{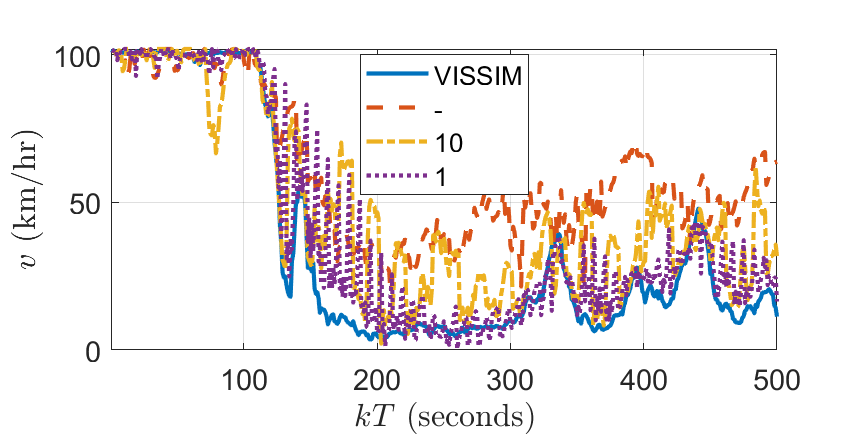}}
    \vspace{-1.5mm}
      \subfigure[]{\includegraphics[width=0.24\textwidth]{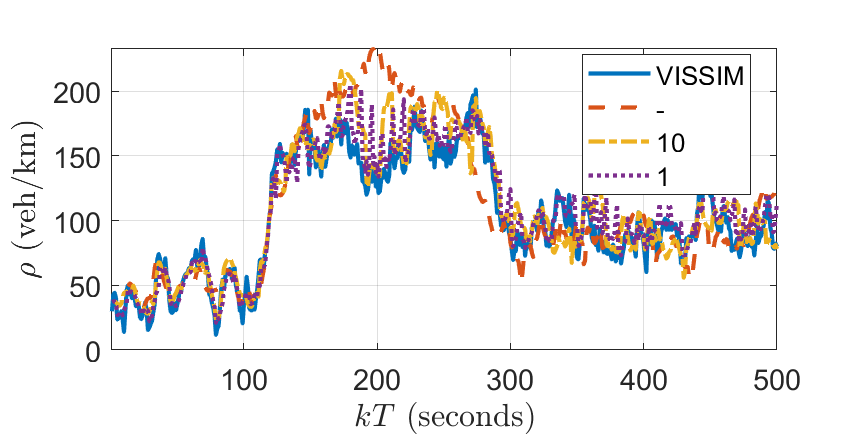}}
    \subfigure[]{\includegraphics[width=0.24\textwidth]{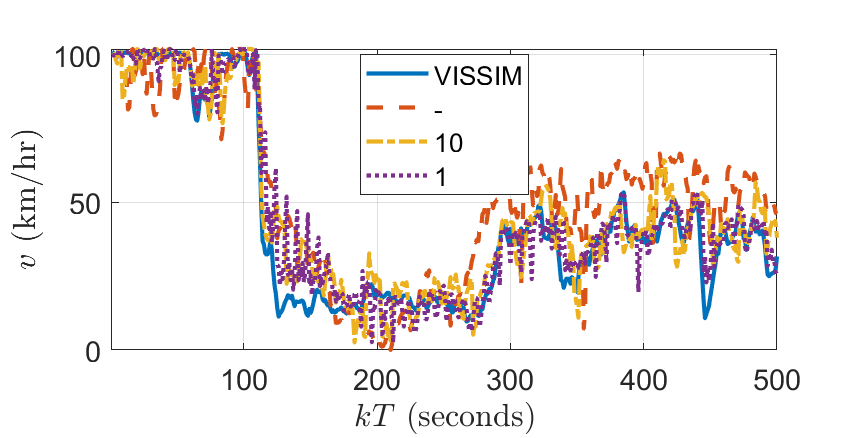}}
  \caption{Plots of simulated and estimated trajectories for densities [left] (a, c, e, g) and speeds [right] (b, d, f, h) in the presence of 3 additional measured segments with changing positions over time. Rows of figures correspond to Segments 3, 4, 5, and 6 respectively.}\label{f:trajectories_moving}
\end{figure}

{It is observed that overall for all estimation methods the value of both $\text{RMSE}_{\rho}$ and $\text{SAMPE}_{\rho}$ decreases with a decrease in the duration between consecutive changes in sensor positions.} A majority of the improvement for all methods occurs with a change duration of fewer than 20 seconds and all methods converge in $\text{RMSE}_{\rho}$ at a change duration of 1 second. While overall there is an improvement with decreased duration between changes in sensor position, the trend is not completely monotonic as the error increases at few values of change duration. Since on some occasions, the process model may not be able to capture the traffic dynamics as well as on other occasions, the estimation error increases if the queried sensors end up not being at the location where the worse modeled behavior occurs at a given time. Given a total simulation duration of 500 seconds, certain values of change duration only result in a handful of position changes during the estimation period. In this case, the time in which data is not collected from segments whose dynamics are not captured well by the model may also be increased causing the error to increase, although marginally. With a higher frequency of sensor position changes, the error decreases monotonically, as the sensors send data from all the segments more frequently. {The two metrics show a similar comparative trend between MHE, EKF, and EnKF while UKF shows a better performance than other methods in terms of $\text{SAMPE}_{\rho}$. This is primarily because UKF makes errors at higher actual density values compared to other methods and also relatively overestimates the densities (similar to the observations with EKF for $\text{SAMPE}_{v}$ in the case of fixed sensors) both of which lead to larger denominator values and a smaller percentage error. The corresponding trajectory plots are presented in Appendix F.} As compared to density, the values of $\text{RMSE}_{v}$ and $\text{SAMPE}_{v}$ are less affected by the variation in frequency of change in sensor positions. Also, both metrics show a similar trend. In terms of speed, UKF, EnKF, and MHE only improve marginally compared to values with fixed sensor positions with the improvement observed at a change duration of 1 second.
EKF, on the other hand, shows a bigger improvement outperforming other methods at the same duration between changes. This is similar to the observation in Figure \ref{f:number_of_sensors} where EKF performs marginally better than MHE when all segments are measured. Overall, MHE is observed to outperform other methods at all different durations between changes in sensor positions except when the positions are changed every 1 second. The simulated and estimated trajectories for density and speed using MHE for the cases with fixed sensors, sensor positions changing every 10 seconds and every 1 second are presented in Figure \ref{f:trajectories_moving} to observe the qualitative improvement in estimation from using CVs as sensors.

Out of the plotted segments, Segment 3 is measured in the fixed sensor case therefore the estimated trajectory from fixed sensors overlaps well with the real trajectory. The other three plotted segments are unmeasured in the fixed sensor case. As expected, the moving sensors result in estimated densities and speeds that follow the real states more closely than the fixed sensor case except for Segment 3. Note that the trajectories estimated using the moving sensors show oscillations in both density and speed which are a result of the segment measurements becoming unavailable over regular intervals. A duration of 10 seconds between changes results in less frequent but larger oscillation as measurements are unavailable for more time steps allowing for larger deviations from the real state. The oscillations in speed estimates are more profound than those in density plots which can be attributed to speeds being obtained by a division of the density and relative flow states and thus being sensitive to changes in both. Since such oscillations in estimated states are generally undesirable, a smoothing filter such as a moving average filter may be applied to the estimated states from the moving sensors to make them more realistic and usable. Plots obtained by applying a moving average filter with values averaged over 15 time steps are presented in Appendix \ref{a:trajectories}.

\subsubsection{Impact of segment selection for sensing} \label{s:impact_fixed_sensor_placement}
\begin{figure}
    \centering
    \includegraphics[width=0.23\textwidth]{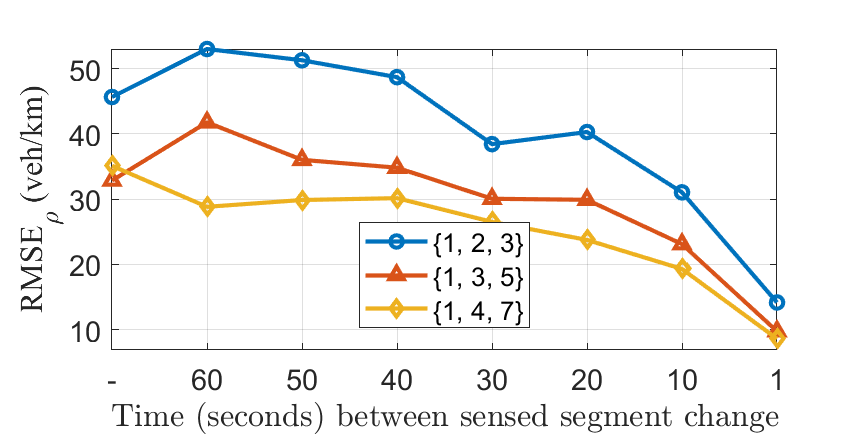}
    \includegraphics[width=0.23\textwidth]{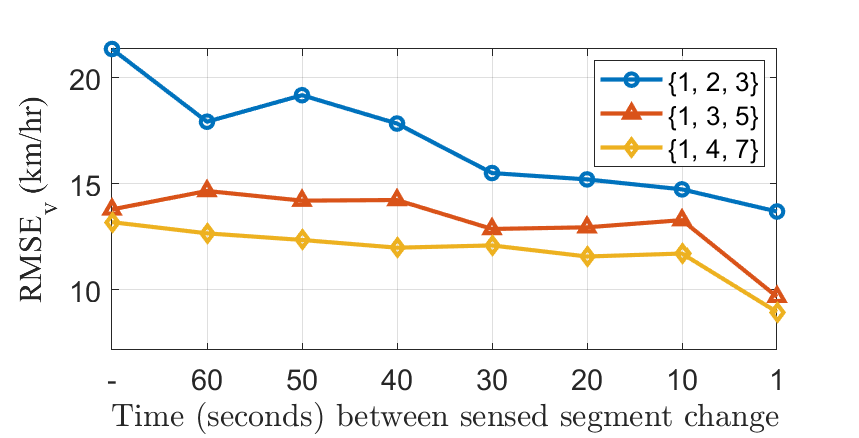}
    \includegraphics[width=0.23\textwidth]{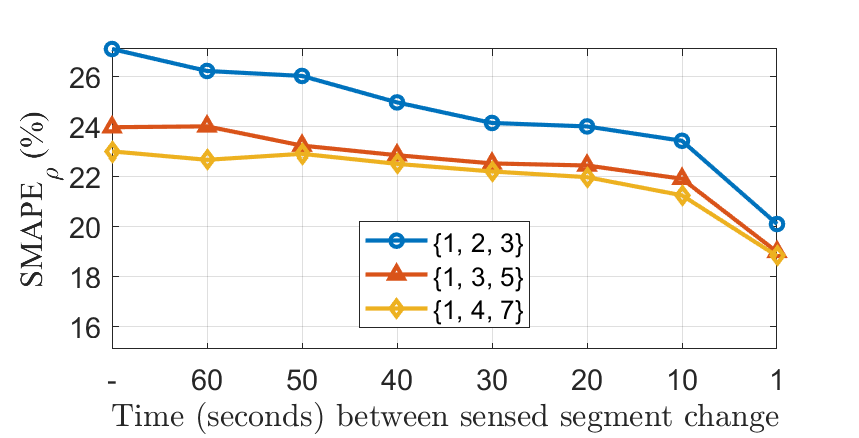}
    \includegraphics[width=0.23\textwidth]{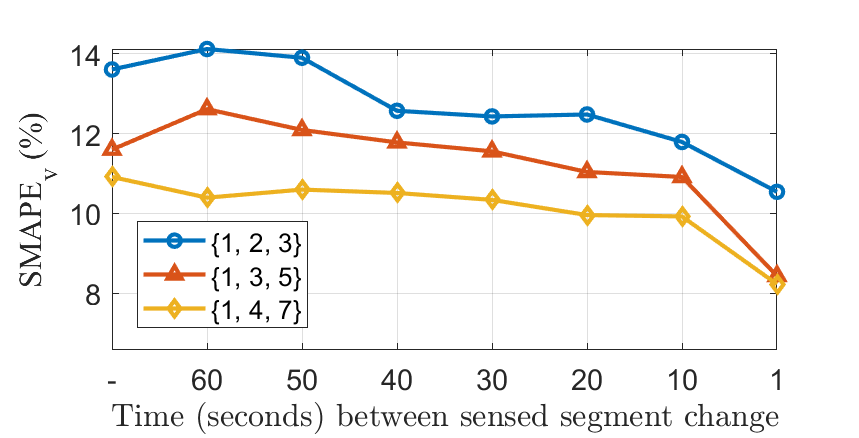}
    \caption{{$\mathrm{RMSE}_{\rho}$ [top left], $\mathrm{RMSE}_{v}$ [top right], $\mathrm{SMAPE}_{\rho}$ [bottom left], and $\mathrm{SMAPE}_{v}$ [bottom right] with different starting configurations of sensors and varying duration between changes in sensor positions using MHE. The symbol `-' at the beginning of the x-axis represents the scenario with fixed sensor locations for the starting configurations in the legend.}}
    \label{f:sensor_placement}
\end{figure}

A limited budget leads to limited bandwidth for data transfer causing traffic measurement data to be available only for a few segments on the road at a time. Therefore, it is important to determine which segments to query for data to obtain the best state estimates. In this section, we test the impact of querying the same number of sensors placed with different spacing while changing the position of sensors. We consider 3 additional segments with data apart from the fixed sensors on the last mainline segment and all ramp segments. For this study, we only consider MHE as it is observed to perform the best with 3 additional sensors. We consider three scenarios with different spacing between measured segments such that their starting sensor positions are $\{1, 2, 3\}$, $\{1, 3, 5\}$, and $\{1, 4, 7\}$. The duration between the change in the position of segments is varied in the same way as in Section \ref{s:speed_impact}. Figure \ref{f:sensor_placement} presents the plots for evaluation metrics for the different starting configurations of sensors (presented in the legend) with varying duration between changes in sensor positions presented on the x-axis.

There is an overall improvement in the estimation performance in terms of all metrics with decreasing duration between changes in sensor position as also observed in Section \ref{s:speed_impact}. Between starting configurations, the configurations with more uniformly spaced measured segments namely $\{1, 3, 5\}$ and $\{1, 4, 7\}$ perform better than consecutive positions $\{1, 2, 3\}$ at all values of time between position changes. Between $\{1, 3, 5\}$ and $\{1, 4, 7\}$, the latter performs better. The difference in performance between the configurations decreases as the duration between position changes is reduced. Compared to the performance of the uniformly spaced starting configurations, the starting placement $\{1, 2, 3\}$ with an x-axis value of above 10 seconds performs worse than when the former is only fixed. This indicates that a more uniformly spaced positioning of sensors is always desirable and can even outperform when the same number of sensors is clustered even if moving to cover more segments over time.

\subsubsection{Impact of {measurement quality}}
\begin{figure}
\subfigure[]{
    \includegraphics[width=0.235\textwidth]{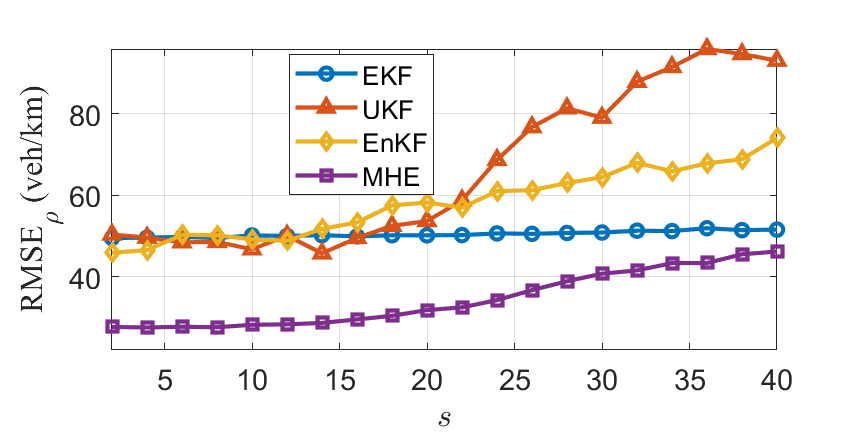}}
    \vspace{-1.5mm}
    \subfigure[]{
    \includegraphics[width=0.235\textwidth]{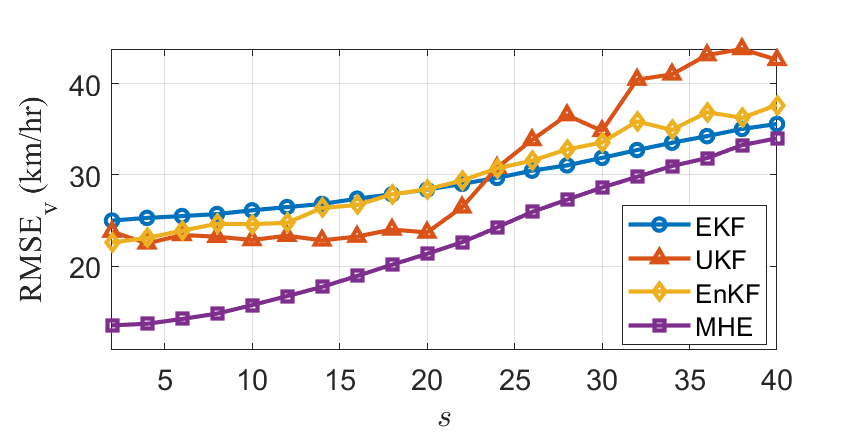}}
    \vspace{-1.5mm}
    \subfigure[]{
    \includegraphics[width=0.235\textwidth]{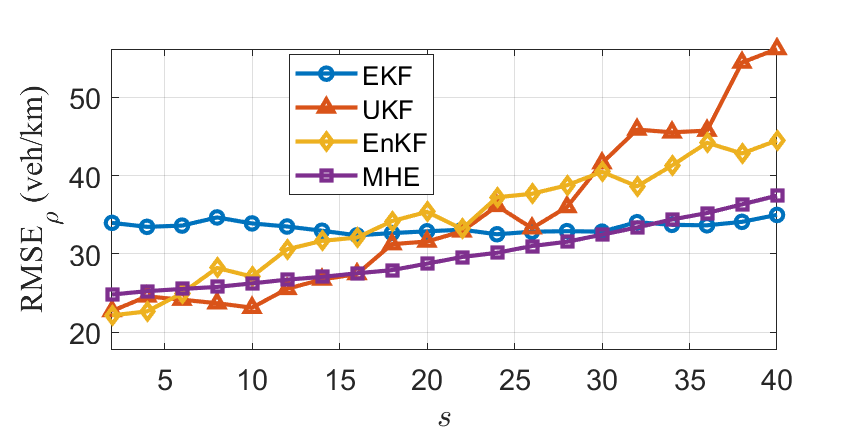}}
    \vspace{-1.5mm}
    \subfigure[]{
    \includegraphics[width=0.235\textwidth]{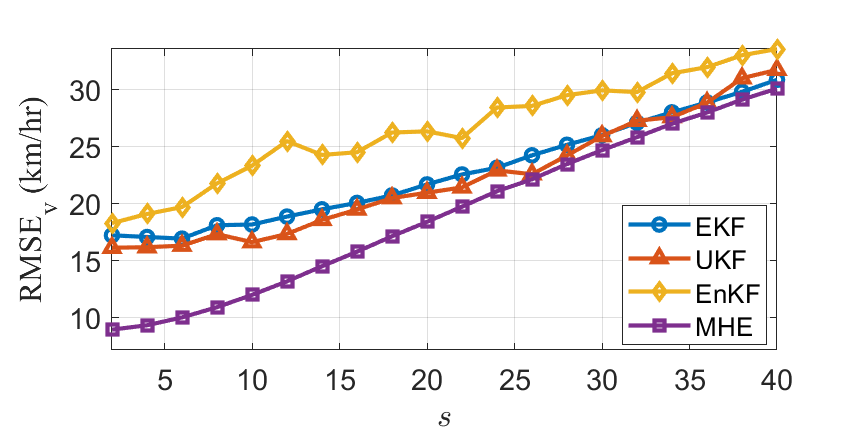}}
    \vspace{-1.5mm}
    \subfigure[]{
    \includegraphics[width=0.235\textwidth]{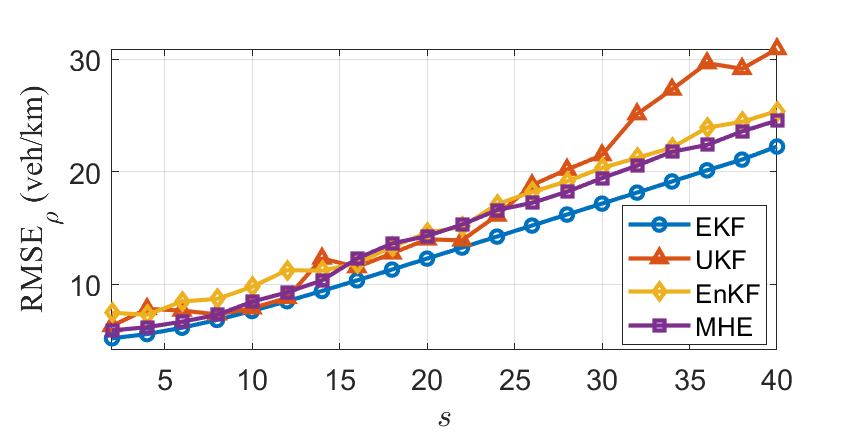}}
    \subfigure[]{
    \includegraphics[width=0.235\textwidth]{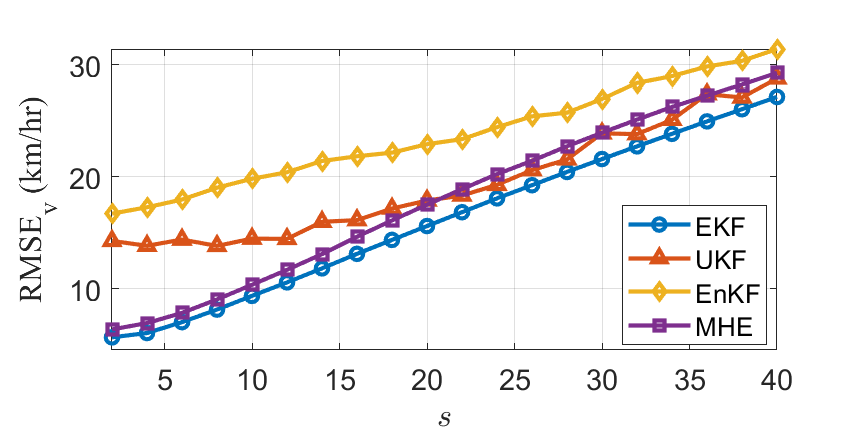}}
    \caption{$\mathrm{RMSE}_{\rho}$ [left] and $\mathrm{RMSE}_{v}$ [right] with changing $s$ (standard deviation of added measurement noise) with 3, 5, and 7 additional fixed sensors in the three rows respectively.} 
    \label{f:noise2}
\end{figure}

As sensors are prone to faults, the sensor noise may change from its manufacturer-specified value for the sensor from time to time. {At the same time, lower penetration rates of CVs in different segments can also result in reduced quality of data measurements.} In such scenarios, a method more robust to measurement errors is considered more reliable. In this section, we check the impact of changing the measurement quality on the estimation performance of the four methods. We further investigate the impact of changing sensor positions on the estimation performance with different levels of error. The goal is to see if the performance improvement offered by moving sensors can offset the deterioration caused by measurement quality. We set the measurement error equal to a random noise drawn from a uniform distribution with bounds $[-1,~1]$, which is standardized and scaled to have zero mean and a standard deviation of $s\in\mathbb{R}$ which is varied to replicate different levels of error. The distribution of noise added to both density and speed measurement is kept the same. In general, the measurement error covariance matrix $R$ for the KF variants is set according to the actual covariance of the noise which in this case is the diagonal matrix with all diagonal elements equal to $s^2$. However, in practice, it is difficult to know the distribution of noise if it is due to varying penetration rates or unexpected sensor faults. Therefore, in this case, we continue to use the value of $R$ defined in Section \ref{s:parameter_tuning}. For MHE, the objective weights and horizon length are also kept the same. Figure \ref{f:noise2} presents the plots of $\text{RMSE}_{\rho}$ and $\text{RMSE}_{v}$ against increasing values of standard deviation $s$ for the four state estimation methods. The plots for SMAPE are omitted for brevity as they show similar relative trends compared to the plots for RMSE and do not contribute to the discussion. The unit of $s$ is the same as the measurements (veh/km for density and km/hr for speed) but is omitted from the plots as it represents the standard deviation for noise in both types of measurements. Three numbers of additional sensors are considered namely 3, 5, and 7 to present the trend in estimation error with increasing measurement noise for different numbers of sensors. For each number of sensors, we take 5 random seed values and average the metrics over the 5 seeds. From the plots, it appears that in all the cases, the estimation error increases with an increase in measurement noise $s$. 

It is observed that the performance of all methods in general deteriorates with increasing noise in the measurements which is expected as the data becomes less reliable. The deterioration in terms of $\text{RMSE}_{\rho}$ is more prominent with the increasing number of sensors. Notice that the performance of the methods in terms of density does not change much up to $s=20$ for the case with three additional sensors. However, the corresponding increase in $\text{RMSE}_v$ shows that the overall estimation performance does indeed deteriorate with increasing noise even at lower levels of noise. The $\text{RMSE}_{\rho}$ does not immediately increase for 3 additional sensors, because the error is already quite large to be sensitive to a small increase in noise. The large increase in the error for UKF is due to large jumps in the state trajectories that reach their bounds from time to time at higher noise possibly due to instability issues with UKF. For MHE, the percentage increase in the error in density and speed between the lowest and highest noise levels are $66.6\%, 50.7\%, 316.2\%$ and $151.8\%, 236.7\%, 360.1\%$ for 3, 5, and 7 additional sensors, respectively. Note that the deterioration in performance becomes more prominent in both density and speed with the increasing number of sensors as the performance becomes more directly associated with available data and thus more sensitive to measurement quality. Overall, UKF appears to be the least reliable in the presence of large measurement errors due to instability issues. Also as observed before, MHE performs better than other methods at smaller numbers of measured segments at lower error levels and while it is affected significantly by measurement error, its performance is still better or comparable to other methods with large noise. Next, we observe the impact of the changing sensor positions on the estimation performance in the presence of noise.

\begin{figure}
    \centering    \includegraphics[width=0.23\textwidth]{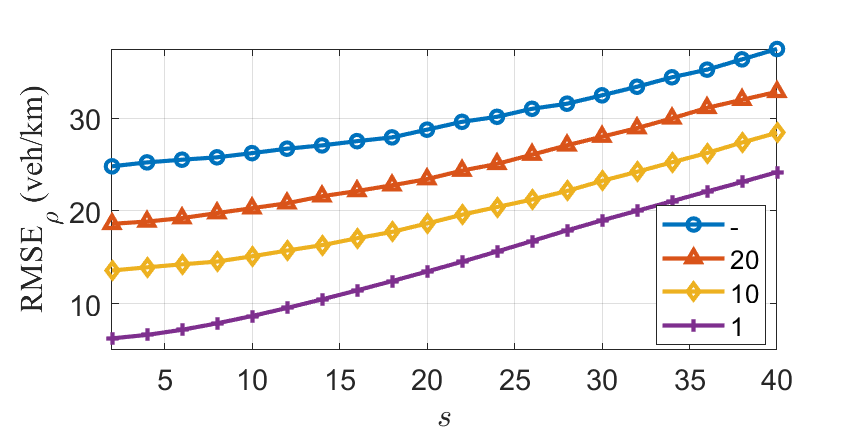}    \includegraphics[width=0.23\textwidth]{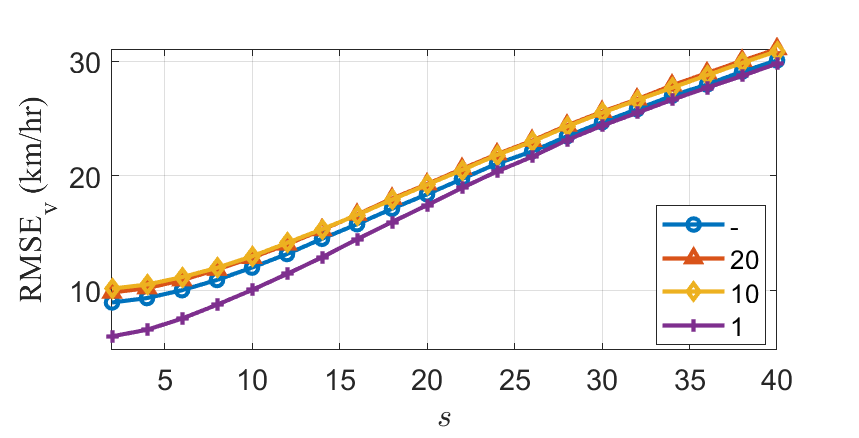}
    \includegraphics[width=0.23\textwidth]{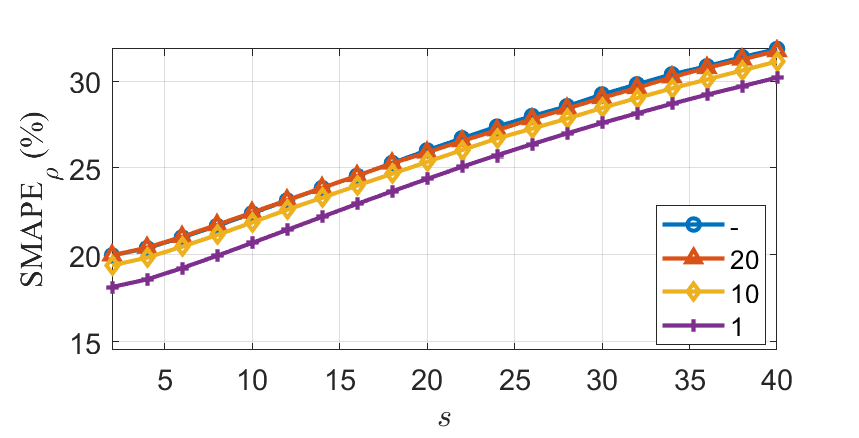}    \includegraphics[width=0.23\textwidth]{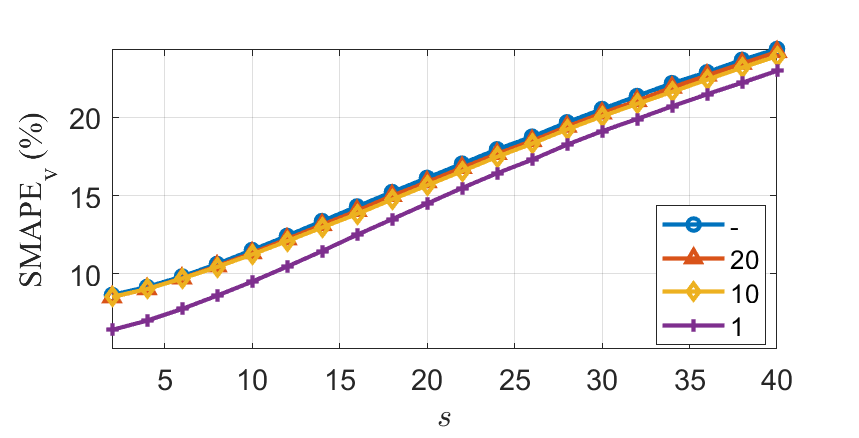}
    \caption{{$\mathrm{RMSE}_{\rho}$ [top left], $\mathrm{RMSE}_{v}$ [top right], $\mathrm{SMAPE}_{\rho}$ [bottom left], and $\mathrm{SMAPE}_{v}$ [bottom right] with changing $s$ (standard deviation of added measurement noise) with 5 additional measured segments with changing positions using MHE.}}
    \label{f:noise_speed_variation}
\end{figure}

Figure \ref{f:noise_speed_variation} presents the variation in error values for estimation using MHE with changing levels of measurement noise in the presence of moving sensors similar to the setting in Section \ref{s:speed_impact}. The legend presents the time (in seconds) between the change in sensor positions using the same position update logic as presented in Section \ref{s:speed_impact}. We consider 5 additional sensors in the configuration presented in Figure \ref{f:sensor_config}. Noise is implemented in the same way as above. The plots for 3 and 7 additional sensors are omitted for brevity. As observed in Section \ref{s:speed_impact}, the estimation performance in terms of $\text{RMSE}_{\rho}$ improves with 20 seconds and further lower duration between changes in sensor positions while $\text{RMSE}_v$ only shows a small improvement at a change duration of 1 second and performs similar to the case with fixed positions with a change duration of even 10 and 20 seconds. While the error at all frequencies of change in sensor position is observed to increase similarly with increasing noise, the $\text{RMSE}_{\rho}$ is consistently smaller for higher frequencies. A change duration of 10 seconds at $s=20$ performs equivalent to a change duration of 20 seconds at $s=0$. In terms of $\text{RMSE}_v$, there is very little change/improvement with a change in position. {Plots for $\text{SMAPE}_{\rho}$ show a smaller difference in magnitude between the curves for the same values on the x-axis compared to $\text{RMSE}_{\rho}$. Qualitative examination of the trajectory plots indicates that a majority of the errors at longer change duration occur at higher densities leading to larger denominator values and hence smaller percentage errors. Also, for certain segments, the scenarios with longer change duration heavily overestimate the densities at lower actual density values which further leads to reduced percentage errors and hence smaller differences in curves. The trajectory plots can be found in Appendix F.} Overall, the evaluation plots reiterate that there is merit in using CVs as moving sensors as the improvement in performance helps offset the deterioration caused by the measurement errors. 

\section{Conclusions and Future Work}\label{sec:summary}
From the previous analysis, we have some preliminary suggestions regarding the questions posed in Section \ref{s:case_study} which are as follows:
\begin{itemize}
    \item \textit{A1:} As expected, the performance of the state estimation methods is improved upon increasing the number of sensors in the system. The performance in terms of density is similar across methods while MHE outperforms other methods in terms of estimated speeds.
    \item \textit{A2:} The performance of all estimation methods in general improves with moving sensors as sensors cover more segments on the highway over time as compared to the case with only fixed sensors. The variation in performance is non-monotonic at lower frequencies of change in sensor positions but a prominent improvement is observed at a duration of 20 seconds or lower. The improvement in density is more profound than the improvement in speeds.
    \item \textit{A3:} More uniformly spaced segments result in a better state estimation performance than the same number of segments placed consecutively in the case of both fixed and moving sensors. Uniformly placed fixed sensors also outperform consecutively placed moving sensors up to a duration of 20 seconds between changes.
    \item \textit{A4:} The performance of all methods worsens with worsening data quality that might result from lower penetration rates of CVs and higher sensor noise. UKF is the least robust out of all methods and shows abrupt increases in error compared to other methods with increasing noise. The impact of quality issues is more prominent in scenarios with more measured segments which show a deterioration in performance at lower levels of measurement noise as compared to scenarios with fewer measured segments. The performance improvement achieved by the use of moving sensors is able to offset the deterioration caused by measurement quality to a fair extent showing the advantage of using moving sensors under adverse conditions.
\end{itemize}
To summarize, we present a state-space formulation for the nonlinear ARZ model while considering junctions in the form of ramp connections. Since the ARZ model is nonlinear, it is not possible to directly apply linear state estimation methods to it which are considered to be more efficient than nonlinear methods. We linearize the ARZ model using Taylor series approximation and use it to implement linear state estimation such as through linear MHE. We present the formulation for linear MHE which has not previously been used for TSE and show that it is a good choice for TSE compared to other popular methods namely EKF, UKF, and EnKF. We also show that the use of moving sensors is better for state estimation compared to fixed sensors and can offset the degradation in performance caused by reduced measurement quality resulting from sensor noise and lower penetration rates of CVs. Various strategies and variations in the selection of segments to obtain CV data are also investigated.

{It is important to note that this study is constrained by a lack of access to real-world data corresponding to the given setting. Given the present theoretical demonstration of the benefits of using mobile CV data sources, a comprehensive investigation is required to assess the real-world capabilities of the approach. In particular, it is important to investigate the technical challenges in ingesting and fusing real-time data streams from both CVs and fixed sensors, examining their impact on estimation performance.}

Future work will also consider the optimal placement of sensors considering CVs for TSE. Besides, while the performance of the ARZ model against the first-order LWR model has been studied in prior research~\cite{Seo2017} which claims the superiority of the former, some of the newer works~\cite{wang2022real} have suggested the possibility of the order of the model being less significant for TSE in the presence of sufficient data. Therefore, it would be interesting to carry out a detailed comparative study between the performance of the ARZ model and a first-order model under different scenarios specifically those depicting non-equilibrium conditions under different sensor placements.

\vspace{-0.144cm}

\bibliographystyle{IEEEtran}	\bibliography{biblio}

\newpage 

\appendices
\section{Traffic System Dynamics}\label{a:model}
This section presents the equations for the traffic demand and supply functions and the traffic flux across junctions according to the ARZ model dynamics. 
\subsection{Demand and supply functions}\label{s:demandupply}
The demand of a segment denotes the traffic flux that wants to leave that segment while the supply of a segment denotes the traffic flux that can enter that segment. Based on these definitions, the demand $D_i[k]$ for Segment $i$ can be written, similar to~\cite{gottlich2021second,kolb2018pareto}, as
\begin{align*}    \hspace{-0.2cm}D_i[k]\hspace{-0.1cm}=\hspace{-0.1cm}
    \begin{cases}
        \rho_i(w_i[k]-p(\rho_i[k])), & \hspace{-0.3cm}\text{if }  \rho_i[k]\le\sigma(w_i[k]),\\
        \sigma(w_i[k])(w_i[k]-p(\sigma(w_i[k]))), & \hspace{-0.3cm}\text{if } \rho_i[k]>\sigma(w_i[k]),
    \end{cases}
\end{align*}
where $\sigma(w_i[k])$ denotes the density that maximizes the demand function and is given as
\begin{align*}
    {\sigma(w_i[k])} = \rho_{m}\left(\frac{w_i[k]}{v_{f}(1+\gamma)}\right)^\frac{1}{\gamma}.
\end{align*}
The supply function $S_i[k]$ on the other hand is given by
\begin{align}
\label{e:supply_function}
S_i[k]\hspace{-0.1cm}=\hspace{-0.1cm}
    \begin{cases}
    \sigma(w_i[k])(w[k]-p(\sigma(w[k]))), & \text{if }  \rho_i[k]\le\sigma(w[k]),\\
    {\rho_i[k]}(w[k]-p(\rho_i[k])), & \text{if } \rho_i[k]>\sigma(w[k]).
    \end{cases}
\end{align}
Notice that $w[k]$ used in \eqref{e:supply_function} does not belong to Segment $i$. Instead, it is calculated from the $\rho[k]$ and $\psi[k]$ of the incoming traffic upstream of Segment $i$. The exact method of calculating this $w[k]$ is given in the following section.
Next, we define the expressions for the flux functions {$q_i[k]$ and $\phi_i[k]$}.

\subsection{Flux formulae at junctions}\label{s:flux_formulae}
This section presents the expressions for the traffic flux {$q_i[k]$} and the relative flux {$\phi_i[k]$} for any Segment $i$, which form the nonlinear part of the state-space model. Development of analytical equations for junction flows in the ARZ and other similar second-order models is an active field of research~\cite{khelifi2017generic,kolb2018pareto,gottlich2021second}, with different papers providing different approaches to model the junction flows, some more complex than the others. However, since state estimation allows for some extent of modeling errors, it is possible to develop a simple state-space formulation for the ARZ model without going into intractable schemes. We consider three types of segment junctions, a one-to-one junction between two mainline segments, a merge junction between two mainline segments and an on-ramp, and a diverge junction between two mainline segments and an off-ramp. In the following discussion, we assume that the mainline segment before the junction has index $i\in\Omega$, the segment after the junction has index $i+1$ and the ramp has index $j\in\hat{\Omega}$ for on-ramp and $j\in\check{\Omega}$ for off-ramp.
\subsubsection{One-to-one junction} The traffic flux leaving Segment $i$ and entering Segment $i+1$ at a one-to-one junction is given as
\begin{align*}
    q_i[k]=\min(D_i[k],S_{i+1}[k])),
\end{align*}
while the relative flux is given as
\begin{align*}
    \phi_i[k] = q_i[k]w_i[k]=q_i[k]\frac{\psi_i[k]}{\rho_i[k]}.
\end{align*}

\subsubsection{Merge junction (on-ramp connection)}
At a merge junction, we have that $\bar{q}_{i+1}[k]=q_i[k]+\hat{q}_j[k]$. We assume that the flow entering Segment $i+1$ from each of the incoming segments is in proportion of their demands, that is, if
\begin{align*}
    \beta_i[k] = \frac{D_i[k]}{D_i[k]+\hat{D}_j[k]},
\end{align*}
then
\begin{subequations}\label{e:merge_flows}
\begin{align}
    q_i[k]&=\beta_i[k]\bar{q}_{i+1}[k],\\  \hat{q}_j[k]&=(1-\beta_i[k])\bar{q}_{i+1}[k].
\end{align}
\end{subequations}
In case of a merge junction, the $w[k]$ used to calculate the supply for the outgoing segment using  \eqref{e:supply_function} is denoted as $\bar{w}[k]$ and is calculated as
\begin{align*}
    \bar{w}[k]=\beta_i[k] w_i[k] + (1-\beta_i[k])\hat{w}_j[k].
\end{align*}

\noindent Then the traffic flux leaving Segment $i$ is given by
\begin{align*}
    q_i[k] = \min(\beta_i[k]S_{i+1}[k],D_i[k],\frac{\beta_i[k]}{1-\beta_i[k]}\hat{D}_j[k]).
\end{align*}
$\bar{q}_{i+1}[k]$ and $\hat{q}_j[k]$ can thereafter be calculated using \eqref{e:merge_flows}.
The relative flux entering Segment $i+1$ is given as
\begin{align*}
    \bar{\phi}_{i+1}[k]=\bar{q}_{i+1}[k]\bar{w}[k],
\end{align*}
and those exiting the incoming segments are given by
\begin{subequations}
\begin{align*}
    \phi_i[k]&=q_i[k]w_i[k],\\
    \hat{\phi}_{j}[k]&=\hat{q}_j[k]\hat{w}_j[k].
\end{align*}
\end{subequations}

\subsubsection{Diverge junction (off-ramp connection)}
At diverge junctions, we have that $q_i[k]=\bar{\check{q}}_j[k]+\bar{q}_{i+1}[k]$. We assume that the proportion of the flow entering the Off-ramp $j$ from Segment $i$ is given by a predefined constant $\alpha_i[k]$, such that
\begin{subequations}\label{e:diverge_flow_relations}
\begin{align*}
    \bar{\check{q}}_j[k]&=\alpha_i[k] q_i[k],\\
    \bar{q}_{i+1}[k]&=(1-\alpha_i[k])q_i[k].
\end{align*}
\end{subequations}
In case of a diverge junction, we use $w_i[k]$ to calculate the supply for both the mainline Segment $i+1$ and the Off-ramp $j$. The flow $q_i[k]$ can then be written as
\begin{align*}
    q_i[k]=\min(D_i[k],\frac{\check{S}_j[k]}{\alpha_i[k]},\frac{S_{i+1}[k]}{(1-\alpha_i[k])}),
\end{align*}
while the relative flux leaving Segment $i$ is given as
\begin{align*}
    \phi_i[k]=q_i[k]w_i[k].
\end{align*}
The relative flux entering the outgoing segments has the same relationship as the flows, that is
\begin{subequations}\label{e:relative_flux_outgoing_diverge}
\begin{align*}
\bar{\phi}_j[k]&=\alpha_i[k]\phi_i[k],\\
    \bar{\phi}_{i+1}[k]&=(1-\alpha_i[k])\phi_i[k].
\end{align*}
\end{subequations}

\section{State-Space Equation Parameters}\label{a:state_space_parameters}
In this section, we present the parameters of the state-space equation \eqref{eq:state_space_gen}. The said parameters are given as follows:
\begin{align}\label{e:matrix_A}
    \m A =&
    \begin{bmatrix}
        1 & 0 & 0 & 0 & 0 & \dots\\
        \frac{v_f}{\tau} & 1-\frac{1}{\tau} & 0 & 0 & 0 & \dots & \\
        0 & 0 & 1 & 0 & 0 & \dots\\
        0 & 0 & \frac{v_f}{\tau} & 1-\frac{1}{\tau} & 0 & \dots \\
        0 & 0 & 0 & 0 & 1 & \dots\\
        \vdots &\vdots &\vdots &\vdots &\vdots &\ddots 
    \end{bmatrix}\\
    \m G=&
    \begin{bmatrix}
        \frac{T}{l} & 0 & 0 & 0 & \dots\\
        0 & \frac{T}{l} & 0 & 0 & \dots\\
        0 & 0 & \frac{T}{l} & 0 & \dots\\
        0 & 0 & 0 & \frac{T}{l} & \dots\\
        \vdots &\vdots &\vdots &\vdots & \ddots 
    \end{bmatrix}\\
    \m f(\m x[k],\m u[k])=&
    \begin{bmatrix}\label{e:f}
        q_{0}[k]-q_1[k]\\
        \phi_{0}[k]-\phi_1[k]\\
        q_{1}[k]-q_2[k]\\
        \phi_{1}[k]-\phi_2[k]\\
        q_{2}[k]-q_3[k]\\        
        \phi_{2}[k]-\phi_3[k]\\
        \vdots
    \end{bmatrix}
\end{align}
where $q_i[k]$ and $\phi_i[k]$ are the traffic flux and relative flux terms for traffic leaving any Segment $i$. As an example of the nonlinear difference terms in \eqref{e:f}, we present the expressions for $q_{i-1}[k]-q_i[k]$ and $\phi_{i-1}[k]-\phi_i[k]$ where Segments $i,i-1,i+1\in\Omega$. There is an On-ramp $j\in\hat{\Omega}$ between Segment $i-1$ and $i$ and no ramp between Segment $i$ and $i+1$. The time parameter is omitted from the notations for all the discrete-time variables for the compactness of the expressions. Also, the state variables are written in terms of the traffic variables that they represent.
\begin{align}
    q_{i-1}-q_i &= \min(\beta_{i-1}S_{i},D_{i-1},\frac{\beta_{i-1}}{1-\beta_{i-1}}\hat{D}_j)-\min(D_i,S_{i+1}),\nonumber\\
    \phi_{i-1}-\phi_{i} &= q_{i-1}\dfrac{\psi_{i-1}}{\rho_{i-1}} - q_i\dfrac{\psi_i}{\rho_i},\nonumber
\end{align}
where the expressions for the various terms are given in Section \ref{s:flux_formulae}. Similarly, if there is an Off-ramp $j\in\check{\Omega}$ between Segment $i$ and $i+1$, and no ramp between Segment $i-1$ and $i$, then the expressions are given as
\begin{align}
    q_{i-1}-q_i &= \min(D_{i-1},S_{i})-\min(D_i,\frac{\check{S}_j}{\alpha_i},\frac{S_{i+1}}{(1-\alpha_i)}),\nonumber\\
    \phi_{i-1}-\phi_{i} &= q_{i-1}\dfrac{\psi_{i-1}}{\rho_{i-1}} - q_i\dfrac{\psi_i}{\rho_i}.\nonumber
\end{align}
Other expressions can also be written in the same manner.

\section{Linear Model Approximation}
This section presents the mathematical expressions for the linear model approximation of the nonlinear state-space formulation of the ARZ model. For the nonlinear function $\m f:\mathbb{R}^{n_x} \times\mathbb{R}^{n_u}\rightarrow \mathbb{R}^{n_x} $, specified in \eqref{eq:state_space_gen}, the first-order Taylor series expansion~\cite{moreno1996vector} about a point $(\m x_0,\m u_0)$ can be written as
\begin{align}\label{e:Taylor_Series_general}
   \m f(\m x,\m u) \hspace{1mm}\approx\hspace{1mm} &\m f(\m x_0,\m u_0)+\nabla\m  f_x(\m x_0,\m u_0)(\m x-\m x_{0})\nonumber\\
    &+\nabla \m f_u(\m x_0,\m u_0)(\m u-\m u_{0}),
\end{align}
where
\begin{align*}
    \nabla \m f_x(\m x_0,\m u_0)=\left[\dfrac{\partial \m f}{\partial x_1}(\m x_0,\m u_0)\hspace{1mm} \cdots \hspace{1mm}\dfrac{\partial \m f}{\partial x_n}(\m x_0,\m u_0)\right]\in\mathbb{R}^{n_x\times n_x},    
\end{align*} 
and
\begin{align*}
    \nabla \m f_u(\m x_0,\m u_0)=\left[\dfrac{\partial \m f}{\partial u_1}(\m x_0,\m u_0)\hspace{1mm} \cdots \hspace{1mm}\dfrac{\partial \m f}{\partial u_m}(\m x_0,\m u_0)\right]\in\mathbb{R}^{n_x\times n_u}.   
\end{align*} 
Here, the operating states $\m x_0$ and operating inputs $\m u_0$ are not fixed for all $k$, instead they are selected as close to the time step $k$ as permitted by the availability of reliable input data and state estimates.
We add the coefficients of $\m x$ from this linearization to the $\m A$ matrix in \eqref{eq:state_space_gen} to get a new coefficient matrix for the approximate model. We obtain the following linear state-space equation
\begin{align*}
    \m x[k+1]\approx\m \tilde{\m A}\m x[k]+\m B\m u[k] + \m c_1,
\end{align*}
where $\tilde{\m A}=\m A + \m G \nabla \m f_x(\m x_0,\m u_0)$, $\m B = \m G \nabla \m f_u(\m x_0,\m u_0),$ and $\m c_1 = \m G (\m f(\m x_0,\m u_0) - \nabla \m f_x(\m x_0,\m u_0)\m x_0 - \nabla \m f_u(\m x_0,\m u_0)\m u_0)$. 
Similarly, we can also linearize the measurement model as follows:
\begin{align*}
    \m y[k] \approx \tilde{\m C}[k]\m x[k] + \m c_2[k],
\end{align*}
where $\tilde{\m C}[k]=\m C[k]\nabla \m h_x(\m x_0)$ and $\m c_2[k] = \m C[k]\m h(\m x_0)-\nabla \m h_x(\m x_0)\m x_0$, where $\nabla \m h_x(\m x_0)$ is the gradient of the measurement function given in \eqref{e:measurement_function} at $\m x_0$.

Since we know the input at every time step, we can always linearize using the current input value. In that case, we do not need the third term in the linearization equation \eqref{e:Taylor_Series_general} as it will always be equal to zero. Besides Taylor series approximation, other methods for linearization such as Carleman linearization~\cite{rauh2009carleman, pruekprasert2020moment} can also be used to obtain a linear approximation to the model. The same was also tested but the results were found to be inferior to the Taylor series approximation and are omitted from this article for brevity.

\section{Moving Horizon Estimation Implementation}\label{a:MHE_implementation}
This section presents the detailed implementation of the MHE algorithm with mathematical expressions for the decision variables, the objective function, and the constraints. It also presents a note on the difference between the current implementation with certain existing approaches for MHE.
\subsection{Decision variables}
The primary decision variables for a single run of the MHE optimization problem at time step $k$ are the state vectors from time step $k-N$ to $k$ denoted by {$\m x_k[\theta]\hspace{1mm}\forall\hspace{1mm} \theta\in\{k-N,k-N+1\dots,k\}$}. These should not be confused with $\hat{\m x}[k-N],\dots,\hat{\m x}[k]$ which are the final state estimates. Out of the decision variables for the optimization at time step $k$, we set the value of the vector $\m x_k[k]$ as the final estimate, that is, $\hat{\m x}[k] = \m x_k[k]$.

\subsection{Objective function}
The objective function for MHE at time step $k\in \mathbb{N}, k\ge N+1$ is denoted by $J[k]$ and is given as
\begin{align}
\hspace{-2mm}
    \label{e:mhe_objective}
    J[k] =& \hspace{1mm}\mu||{\m x}_k[k-N]-\m \bar{\m x}[k-N]||^2\nonumber\\ 
    &+{w_1\hspace{-2mm}\sum^k_{\theta=k-N}\hspace{-2mm}|| \m y[\theta]-( \tilde{\m C}_{\theta}\m x_k[\theta]+\m c_{2\theta}) ||^2}\nonumber\\
    &+{w_2\hspace{-2mm}\sum^{k-1}_{\theta=k-N}\hspace{-2mm}|| \m x_k[\theta+1]\hspace{-0.5mm}-\hspace{-0.5mm} (\tilde{\m A}_{\theta}\m x_k[\theta] + \m B_{\theta}\m u[\theta] + \m c_{1\theta}) ||^2}.
\end{align}

Here, $\bar{\m x}[k-N]$ is a prediction of $\m x[k-N]$ based on a previously obtained state estimate and is expressed as
\begin{align}\label{e:prediction}
    \bar{\m x}[k\hspace{-0.5mm}-\hspace{-0.5mm}N] = \m A \hat{\m x}[k\hspace{-0.5mm}-\hspace{-0.5mm}N\hspace{-0.5mm}-\hspace{-0.5mm}1] + \m G\m f(\hat{\m x}[k\hspace{-0.5mm}-\hspace{-0.5mm}N\hspace{-0.5mm}-\hspace{-0.5mm}1],\m u[k\hspace{-0.5mm}-\hspace{-0.5mm}N\hspace{-0.5mm}-\hspace{-0.5mm}1]).
\end{align}
Some literature such as~\cite{qu2009computation} suggest using another state estimation method like UKF to obtain the predicted states from the previous estimate to better utilize the available measurement data. In this work, we simply use the process model as shown in \eqref{e:prediction}. 
{The notation $\m y[\theta]$ is the measurement vector at time step $\theta\in \{k-N,k-N+1,\dots,k\}$, $\tilde{\m A}_{\theta}, \m B_{\theta}$ and $\m c_{1\theta}$ are parameters of the linearized state-space equation $\hspace{0.5mm}\forall \hspace{0.5mm}\theta \in\{k-N,k-N+1,\dots,k-1\}$, and $\tilde{\m C}_{\theta}$ and $\m c_{2\theta} $ are parameters of the linearized measurement model $\hspace{0.5mm}\forall \hspace{0.5mm}\theta \in \{k-N,k-N+1,\dots,k\}$. Here, $\tilde{\m A}_{\theta}, \m B_{\theta}$ and $\m c_{1\theta}$ are computed at $(\m x_o,\m u[\theta])$ where $\m x_o\hspace{-1mm}=\hspace{-1mm} \sum_{\theta=k-1-N}^{k-1} \m x_{k-1}[\theta]/(N+1)$, and $\tilde{\m C}_{\theta}$ and $\m c_{2\theta} $ are computed at $\m x_o$.}

The first term of the objective known as the \textit{arrival cost} penalizes the error between the current decision state vector at step $k-N$ and its expected values based on past estimates. The second and third terms penalize the measurement error and the process model error. $\mu, w_1$ and $w_2$ represent the weights on the three terms, respectively, and can be adjusted by the modeler. The objective of the problem is to minimize $J[k]$ under the following constraints. 

\subsection{Constraints}

The constraints for the MHE optimization problem consist of the lower and upper bounds on the states, that is, if the bound vectors are $\m x_{\mathrm{min}}$ and $\m x_{\mathrm{max}}\in\mathbb{R}^{n_x}$ respectively, then the constraints are defined as
\begin{align}\label{e:bounds}
    \m x_{\mathrm{min}} \le \m x_k[\theta] \le \m x_{\mathrm{max}}, \forall \hspace{1mm} \theta \in \{k-N,k-N+1, \dots,k\}.
\end{align}
For the problem at hand, we have $\m x_{\mathrm{min}} = \vec{0}$, and $\m x_{\mathrm{max}}=[\rho_m \hspace{2mm} \rho_m v_f \hspace{2mm}\rho_m \hspace{2mm}\rho_m v_f \hspace{2mm}\cdots \hspace{2mm}\rho_m \hspace{2mm}\rho_m v_f]^{\top}$.

\subsection{Optimization problem}
The above objective and constraints are used to write the following optimization problem
\begin{align}\label{e:optimization}
&\underset{\m x_k[k-N],\dots,\m x_k[k]}{\textrm{minimize}} &&\hspace{-20mm}J[k]\nonumber\\
&\hspace{4.8mm}\textrm{subject to}&&\hspace{-20mm}\eqref{e:bounds}.\hspace{20mm}
\end{align}
The objective function $J[k]$ can also be expressed as a sum of quadratic and linear terms of the state vectors as shown in Appendix \ref{a:MHE_optimization}. Defining $\m z_k$ by concatenating the decision variables from \eqref{e:optimization} such that $\m z_k = [\m x_k[k-N]^{\top}\hspace{1mm} \m x_k[k-N+1]^{\top}\hspace{1mm} \cdots\hspace{1mm} \m x_k[k]]^{\top}$, we can write the optimization problem \eqref{e:optimization} in the standard form of a QP defined as
\begin{align}\label{e:QP}
    &\underset{\m z_k}{\textrm{minimize}} &&\hspace{-20mm}\m z_k^{\top}\m H\m z_k+\m q^{\top}\m z_k\nonumber\\
&\textrm{subject to}&&\hspace{-20mm}\m z_{\mathrm{min}}\le \m z_k \le \m z_{\mathrm{max}}.
\end{align}
where $\m H\in\mathbb{R}^{(N+1)n_x\times (N+1)n_x}$ and $\m q\in\mathbb{R}^{(N+1)n_x}$ consist of the coefficients of the quadratic and linear terms in the objective respectively. $\m z_{\mathrm{min}}$ and $\m z_{\mathrm{max}}\in\mathbb{R}^{(N+1)n_x}$ are the lower bound and upper bound vectors of $\m z_k$ obtained by concatenating $\m x_{\mathrm{min}}$ and $\m x_{\mathrm{max}}$ respectively. From Appendix \ref{a:MHE_optimization}, it can be seen that $\m H$ is a positive definite matrix. This makes \eqref{e:QP} a convex program that can be solved efficiently using readily available QP solvers like CPLEX or MATLAB's \texttt{quadprog} function. Algorithm \ref{algorithm1} presents the steps involved in MHE as implemented in this study.

\setlength{\floatsep}{5pt}{
\begin{algorithm}[t]
	\caption{MHE Implementation for TSE}\label{algorithm1}
	\DontPrintSemicolon
	\textbf{input:} total time $t_f$, horizon length $N$, weights $\mu$, $w_1$, and $w_2$, state-space matrices $\m A, \m G$ and function $\m f$, measurements from sensors $\m y[k]\hspace{0.5mm}\forall\hspace{0.5mm}k\in\{1,2,\dots,t_f\}$, inputs $\m u[k]\hspace{0.5mm}\forall\hspace{0.5mm}k\in\{1,2,\dots,t_f\}$, assumed initial state $\hat{\m x}[0]$, and state bounds $\m x_{\mathrm{min}}$ and $\m x_{\mathrm{max}}$\;
	
	\While{$k\hspace{0.5mm}\le \hspace{0.5mm}t_f$}{	
	
	\textbf{set:} operating state $\m x_o=\sum_{\theta=k-1-N}^{k-1} \m x_{k-1}[\theta]/(N+1)$ \;
	\textbf{compute:} predicted state $\bar{\m x}[k-N]$ using \eqref{e:prediction}, linearized state-space equation matrices $\tilde{\m A}_{\theta},\m B_{\theta},$ and $\m c_{1\theta}$ at $(\m x_o,\m u[\theta])$ $\forall\hspace{0.5mm}\theta\in\{k\hspace{-0.5mm}-\hspace{-0.5mm}N,k\hspace{-0.5mm}-\hspace{-0.5mm}N\hspace{-0.5mm}+\hspace{-0.5mm}1,\hspace{-0.5mm}\dots\hspace{-0.5mm},k\hspace{-0.5mm}-\hspace{-0.5mm}1\}$, linearized measurement equation matrices $\tilde{\m C}_{\theta}$ and $\m c_{2\theta}$ at $\m x_o$ $\forall\hspace{0.5mm}\theta\in\{k\hspace{-0.5mm}-\hspace{-0.5mm}N,k\hspace{-0.5mm}-\hspace{-0.5mm}N\hspace{-0.5mm}+\hspace{-0.5mm}1,\dots,k\}$\;
	
	\textbf{set:} coefficient matrices $\m H$ and $\m q$ using $\bar{\m x}[k-N], \tilde{\m A}_{\theta},\m B_{\theta},$ and $\m c_{1\theta}$ $\forall\hspace{0.5mm} \theta \in \{k-N,k-N+1,\dots,k-1\}$, and $\m y[\theta], \tilde{\m C}_{\theta}$ and $\m c_{2\theta}\hspace{0.5mm} \forall \hspace{0.5mm} \theta\in\{k-N,k-N+1,\dots,k\}$, and bound vectors $\m z_{\mathrm{min}}$ and $\m z_{\mathrm{max}}$ using $\m x_{\mathrm{min}}$ and $\m x_{\mathrm{max}}$\;
	
	\textbf{solve:} optimization problem \eqref{e:QP} for $\m z_k$\;
	\textbf{set:} $\hat{\m x}[k] = \m x_k[k]$}
	\textbf{output:} $\hat{\m x}[1],\dots,\hat{\m x}[t_f]$\;
\end{algorithm}}

\subsection{Limitation of other implementation}
The MHE literature presents some other implementations of the optimization problem as well such as the one presented in \cite{alessandri2003receding}. The said approach only considers minimization of the arrival cost and the measurement errors but not the modeling errors that is, the third term in the objective function \eqref{e:mhe_objective} is missing. This results in a problem that is faster to solve. However, since actual traffic states fluctuate more than what is captured by even a second-order traffic model like the ARZ model, there are always some modeling errors that need to be accounted for by considering modeling errors. Additionally, we have some errors due to the linearization of both the process and the measurement models. As a result, not considering modeling error or the third term in \eqref{e:mhe_objective} results in a relatively bad performance of MHE for TSE.

\section{QP formulation for MHE}\label{a:MHE_optimization}
The MHE objective function is given in \eqref{e:mhe_objective} as
\begin{align*}
\hspace{-2mm}
    J[k] =& \hspace{1mm}\mu||{\m x}_k[k-N]-\m \bar{\m x}[k-N]||^2\nonumber\\
    &+w_1\hspace{-2mm}\sum^k_{\theta=k-N}\hspace{-2mm}|| \m y[\theta]-( \tilde{\m C}_{\theta}\m x_k[\theta]+\m c_{2\theta}) ||^2\nonumber\\
    &+w_2\hspace{-2mm}\sum^{k-1}_{\theta=k-N}\hspace{-2mm}|| \m x_k[\theta+1]\hspace{-0.5mm}-\hspace{-0.5mm} (\tilde{\m A}_{\theta}\m x_k[\theta] + \m B_{\theta}\m u[\theta] + \m c_{1\theta}) ||^2.
\end{align*}
The square of the Euclidean norm can be expressed as a product of vectors that can be simplified into quadratic and linear terms in the associated decision variables. For instance, the \textit{arrival cost} term can be expanded as
\begin{align}
    &\mu||{\m x}_k[k-N]-\m \bar{\m x}[k-N]||^2\nonumber\\
    &=({\m x}_k[k-N]-\m \bar{\m x}[k-N])^{\top}\mu \m I_{n_x}({\m x}_k[k-N]-\m \bar{\m x}[k-N])\nonumber\\
    &={\m x}_k[k-N]^{\top} \mu \m I_{n_x} {\m x}_k[k-N] - \m \bar{\m x}[k-N]^{\top} 2\mu \m I_{n_x} {\m x}_k[k-N]\nonumber\\
    &+ \m \bar{\m x}[k-N]^{\top} \mu \m I_{n_x} \m \bar{\m x}[k-N]
\end{align}
Similarly, the second term can be expressed as
\begin{align}
\raggedleft
    &w_1\hspace{-2mm}\sum^k_{\theta=k-N}\hspace{-2mm}|| \m y[\theta]-( \tilde{\m C}_{\theta}\m x_k[\theta]+\m c_{2\theta}) ||^2\nonumber\\
    &=\hspace{-2mm}\sum^k_{\theta=k-N}\hspace{-2mm}{( (\tilde{\m C}_{\theta}\m x_k[\theta])^{\top}} w_1 \m I_{n_p[k]} \tilde{\m C}_{\theta}\m x_k[\theta]\nonumber \\&- (\m y[\theta]-\m c_{2\theta})^{\top} 2w_1 \m I_{n_p[k]} \tilde{\m C}_{\theta}\m x_k[\theta]\nonumber\\
    &+ (\m y[\theta]-\m c_{2\theta})^{\top} w_1 \m I_{n_p[k]} \m (\m y[\theta]-\m c_{2\theta})),
\end{align}
and the third term as
\begin{align}
    &w_2\hspace{-2mm}\sum^{k-1}_{\theta=k-N}\hspace{-2mm}|| \m x_k[\theta+1]\hspace{-0.5mm}-\hspace{-0.5mm} (\tilde{\m A}_{\theta}\m x_k[\theta] + \m B_{\theta}\m u[\theta] + \m c_{1\theta}) ||^2\nonumber\\
    &=\hspace{-2mm}\sum^{k-1}_{\theta=k-N}\hspace{-2mm}( (\m x_k[\theta+1]-\tilde{\m A}_{\theta}\m x_k[\theta])^{\top} w_2 \m I_{n_x} (\m x_k[\theta+1]-\tilde{\m A}_{\theta}\m x_k[\theta])\nonumber\\
    &- (\m B_{\theta}\m u[\theta] + \m c_{1\theta})^{\top} 2w_2 \m I_{n_x} (\m x_k[\theta+1]-\tilde{\m A}_{\theta}\m x_k[\theta])\nonumber\\
    &+ (\m B_{\theta}\m u[\theta] + \m c_{1\theta})^{\top} w_2 \m I_{n_x} \m (\m B_{\theta}\m u[\theta] + \m c_{1\theta})).
\end{align}
Here, the last term in each expansion is a constant and can be removed from the objective function. The sum of the remaining terms can be expressed in terms of the vector $\m z_k = [\m x_k[k-N]^{\top}\hspace{1mm} \m x_k[k-N+1]^{\top}\hspace{1mm} \cdots\hspace{1mm} \m x_k[k]]^{\top}$ as
\begin{align}
    &=\m z_k^{\top} (\m H_1 +\m H_2 + \m H_3) \m z_k + (\m q_1 + \m q_2  + \m q_3)^{\top} \m z_k
\end{align}
where the various matrices and vectors are defined as follows:
\begin{align}
    \m H_1 =
    \begin{bmatrix}
        \mu \m I_{n_x} & \vec{0}\\
        \vec{0} & \vec{0}\\
    \end{bmatrix},
\end{align}

\begin{align}
    \m H_2 =
    \m H_C^{\top}w_1\m I_{N_p}\m H_C
\end{align}
where 
\begin{align*}
\m H_{C}=
\begin{bmatrix}
        \tilde{\m C}_{k-N} & \vec{0} & \vec{0}\\
        \vec{0} & \ddots &\vec{0}\\
        \vec{0} & \vec{0} & \tilde{\m C}_{k}\\
    \end{bmatrix},
\end{align*}
and $N_p={\sum_{\theta=k-N}^kn_p[\theta]}$, and

\begin{align}
    \m H_3 =
    \m H_A^{\top}w_2\m I_{Nn_x}\m H_A
\end{align}
where
\begin{align*}
\m H_A = 
    \begin{bmatrix}
       -\tilde{\m A}_{k-N} & \m I_{n_x} & \vec{0} & \vec{0} & \vec{0}\\
        \vec{0} &  -\tilde{\m A}_{k-N+1} & \m I_{n_x} & \vec{0} & \vec{0}\\        
        \vec{0} & \vec{0} & \ddots &\ddots & \vec{0}\\
        \vec{0} & \vec{0} & \vec{0} & -\tilde{\m A}_{k-1} & \m I_{n_x}\\
    \end{bmatrix}.
\end{align*}

\begin{align}
    \m q_1 = -2
    \left(\begin{bmatrix}
         \bar{\m x}[k-N]\\
         \vec{0}\\
    \end{bmatrix}^{\top}
    \m H_1\right)^{\top},
\end{align}
\begin{align}
    \m q_2 = -2
    \left(\begin{bmatrix}
       \m y[k-N]-\m c_{2 k-N}\\
       \vdots\\
       \m y[k]-\m c_{2 k}\\
    \end{bmatrix}^{\top}w_1\m I_{N_p}\m H_C\right)^{\top},
\end{align}
and
\begin{align}
    \m q_3 = -2
    \left(\begin{bmatrix}
       \m B_{k-N}\m u[k-N]+\m c_{1 k-N}\\
       \vdots\\
       \m B_{k-1}\m u[k-1]+\m c_{1 k-1}\\
    \end{bmatrix}^{\top}w_2\m I_{Nn_x}\m H_A\right)^{\top}.
\end{align}
Replacing $\m H_1 +\m H_2 + \m H_3$ with $\m H$ and $\m q_1 +\m q_2 + \m q_3$ with $\m q$ we get the objective function in \eqref{e:QP}.

\section{Additional estimation results}
\label{a:trajectories}

\subsection{Estimated trajectories using UKF and EnKF with fixed sensors}
\begin{figure}
  \subfigure[]{\includegraphics[width=0.24\textwidth]{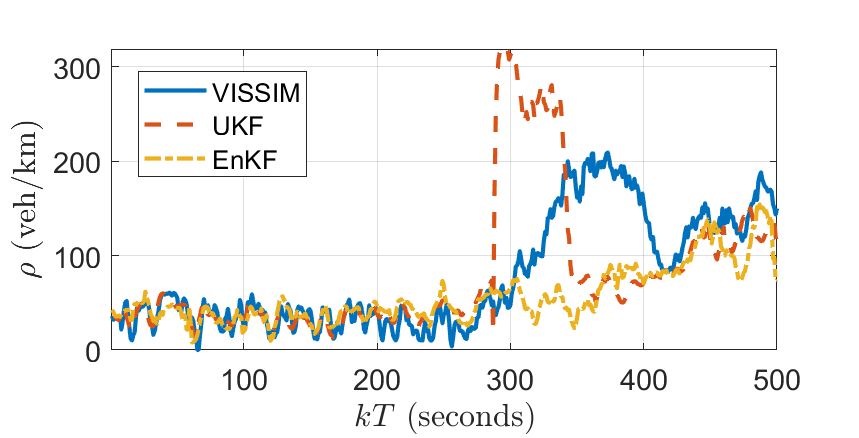}}
  \subfigure[]{\includegraphics[width=0.24\textwidth]{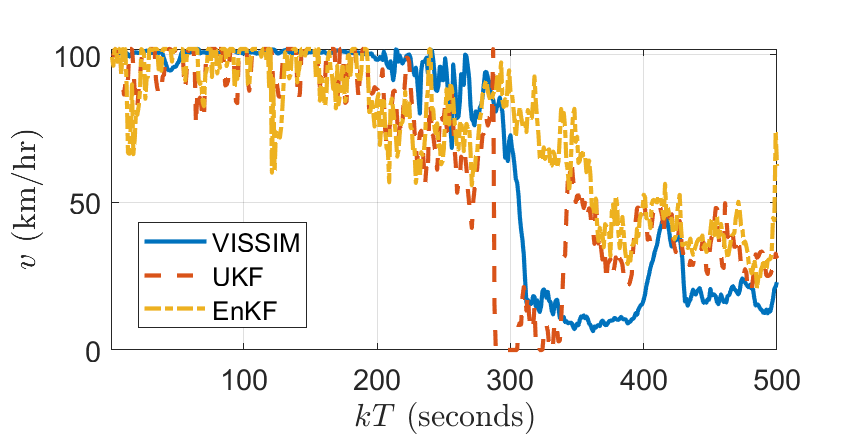}}
    \subfigure[]{\includegraphics[width=0.24\textwidth]{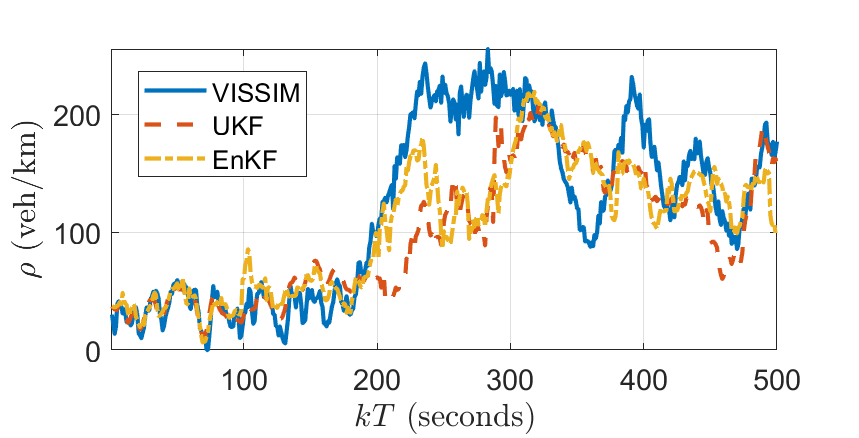}}
    \subfigure[]{\includegraphics[width=0.24\textwidth]{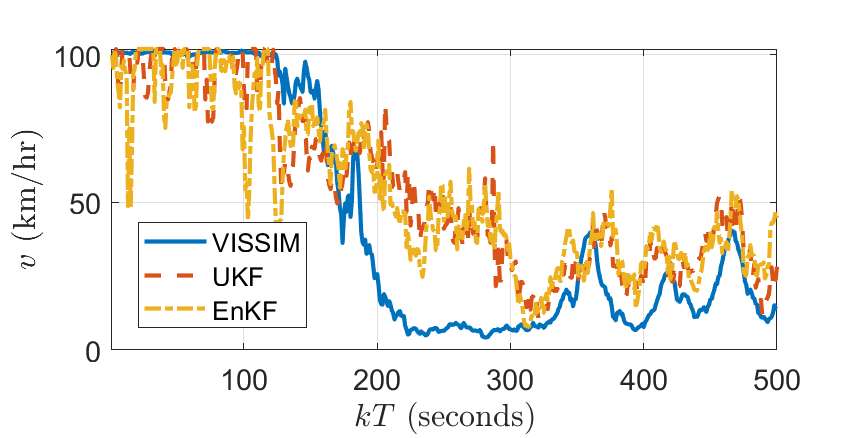}}
      \subfigure[]{\includegraphics[width=0.24\textwidth]{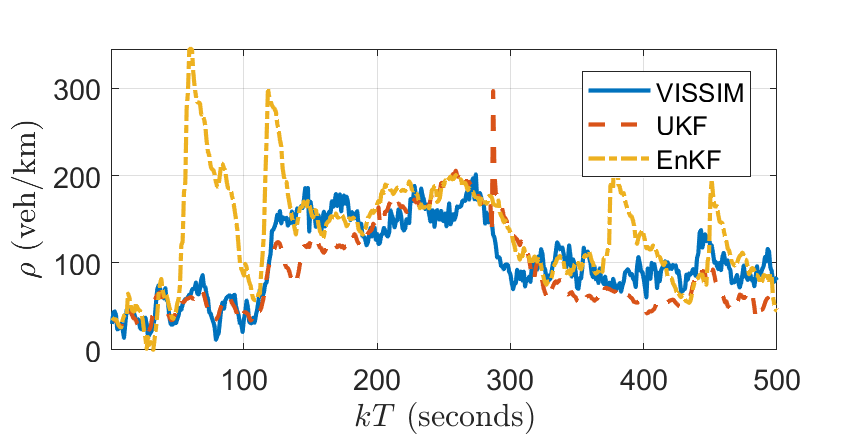}}
    \subfigure[]{\includegraphics[width=0.24\textwidth]{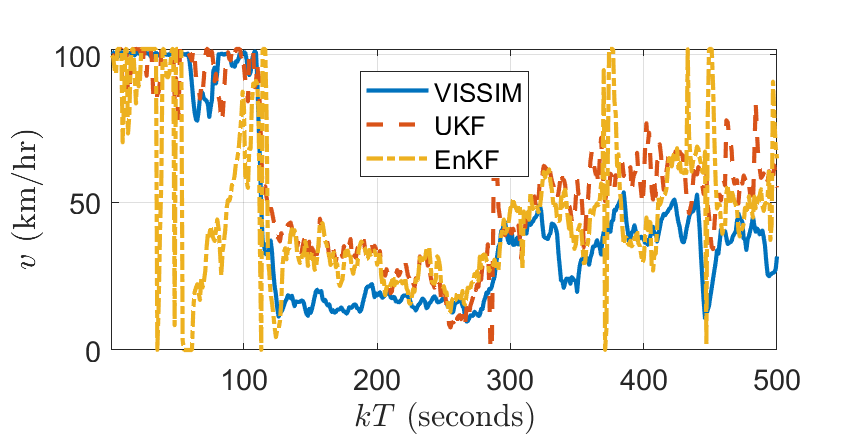}}
  \subfigure[]{\includegraphics[width=0.24\textwidth]{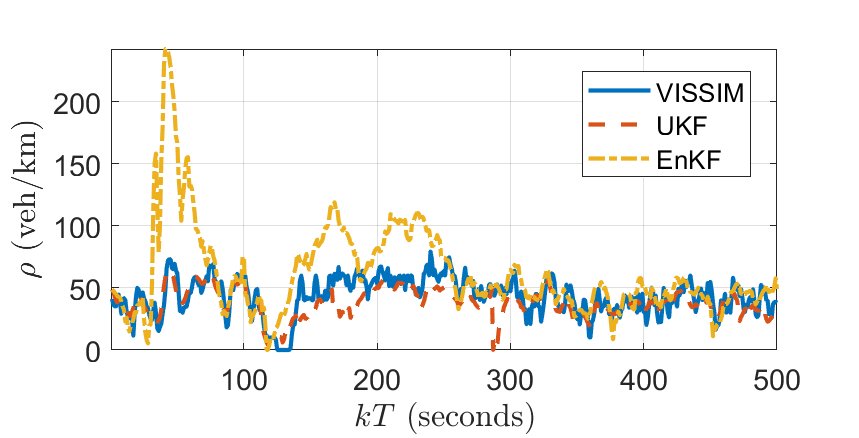}}
    \subfigure[]{\includegraphics[width=0.24\textwidth]{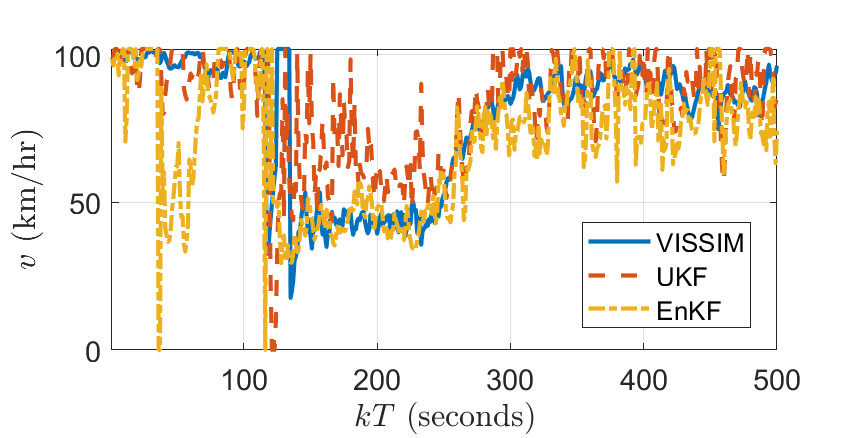}}
  \caption{Plots of simulated and estimated trajectories for densities [left] (a, c, e, g) and speeds [right] (b, d, f, h) in the presence of 4 additional fixed sensors using UKF and EnKF. Rows of figures correspond to the unmeasured Segments 2, 4, 6, and 8 respectively.}\label{f:trajectories_UKF}
\end{figure}

Figure \ref{f:trajectories_UKF} presents the real density and speed for the numerical example presented in Section \ref{s:numerical_study} along with those estimated using UKF and EnKF with 4 additional fixed sensors. It is observed that EnKF tends to overshoot the density estimate on several occasions on Segments 2, 6, and 8. UKF also overshoots density but only on Segment 2 and on one occasion on Segment 6 around 300 seconds. The speed is underestimated at the same time. Besides the overestimated density in the various segments, both UFK and EnKF are able to estimate the congestion in Segment 6 but underestimate the congestion in Segments 2 and 4. EnKF also overestimates the density of Segment 8.

\subsection{Smoothed estimated trajectories with moving sensors}
\begin{figure}
    \subfigure[]{\includegraphics[width=0.24\textwidth]{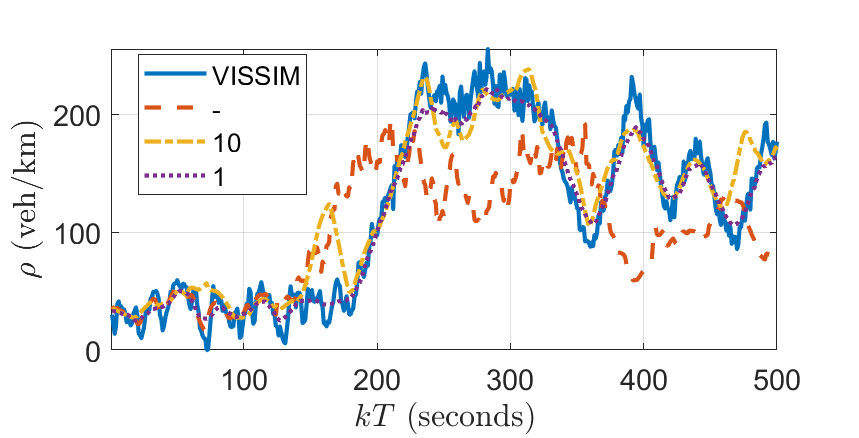}}
    \subfigure[]{\includegraphics[width=0.24\textwidth]{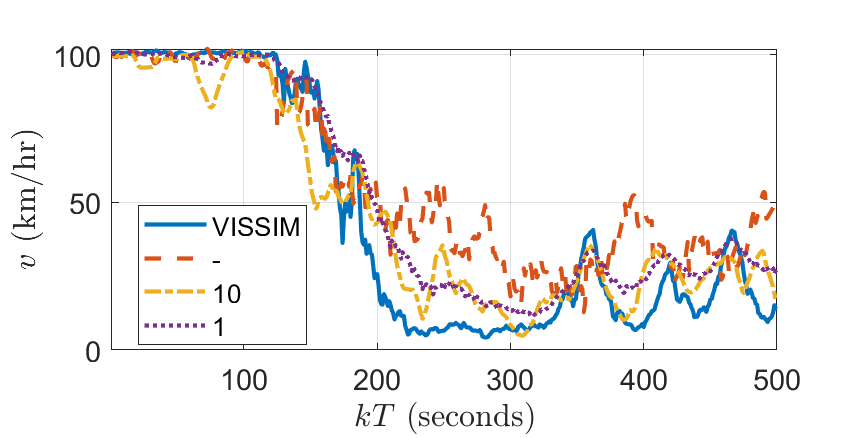}}
    \subfigure[]{\includegraphics[width=0.24\textwidth]{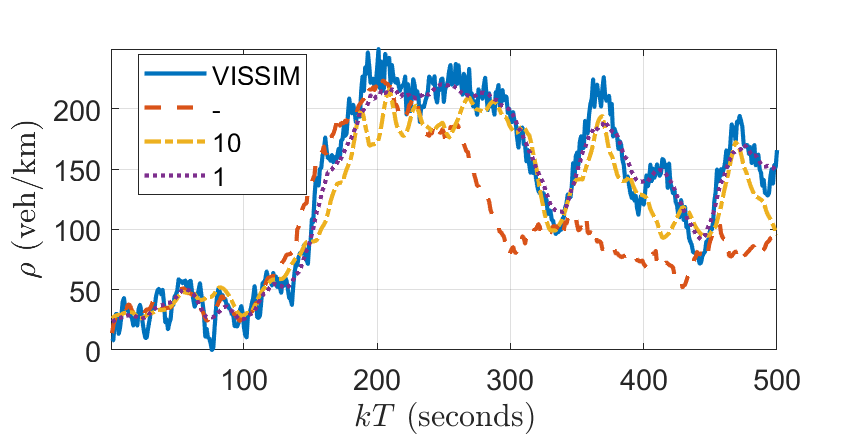}}
    \subfigure[]{\includegraphics[width=0.24\textwidth]{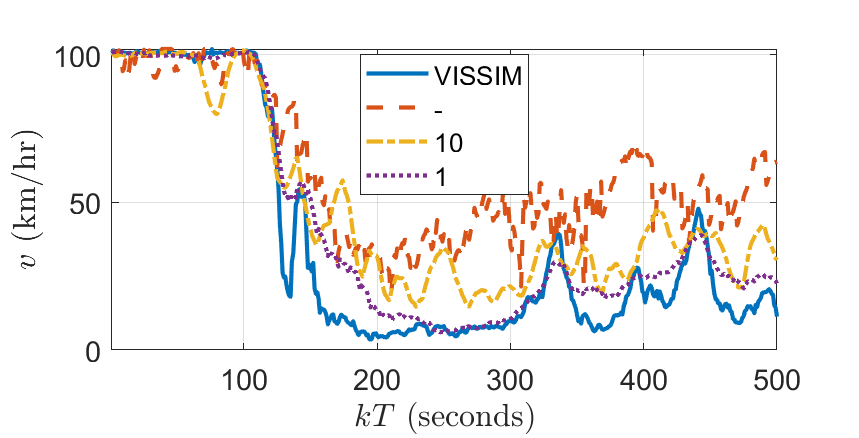}}
  \caption{Plots of simulated and estimated trajectories with smoothing for densities [left] (a, c) and speeds [right] (b, d) in the presence of 3 additional measured segments with changing positions and application of a smoothing filter. Rows of figures correspond to Segments 4 and 5 respectively.}\label{f:trajectories_averaged}
\end{figure}

 Figure \ref{f:trajectories_averaged} presents the simulated and estimated trajectories for density and speed for the numerical example in Section \ref{s:numerical_study} using MHE with 3 additional measured segments whose position changes over the duration mentioned in the legend along with a moving average filter to smooth the oscillations caused by changing sensor positions. As expected, the moving average filter is able to reduce the oscillations in the estimated trajectories while closely following the real density and speed.

\subsection{Estimated density trajectories to explain observed trends in $\text{SMAPE}_{\rho}$ for UKF with moving sensors}
\begin{figure}
    \subfigure[]{\includegraphics[width=0.24\textwidth]{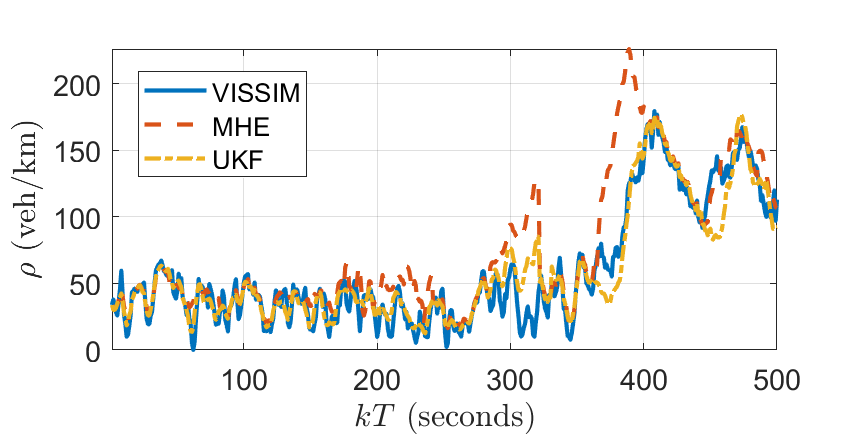}}
    \subfigure[]{\includegraphics[width=0.24\textwidth]{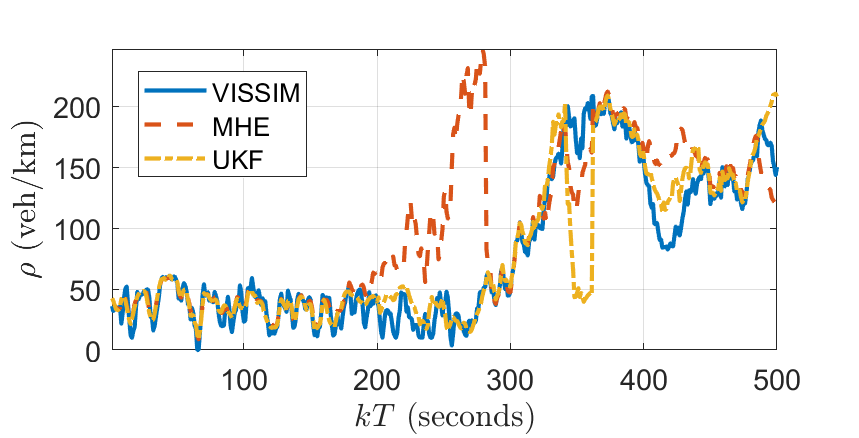}}
    \subfigure[]{\includegraphics[width=0.24\textwidth]{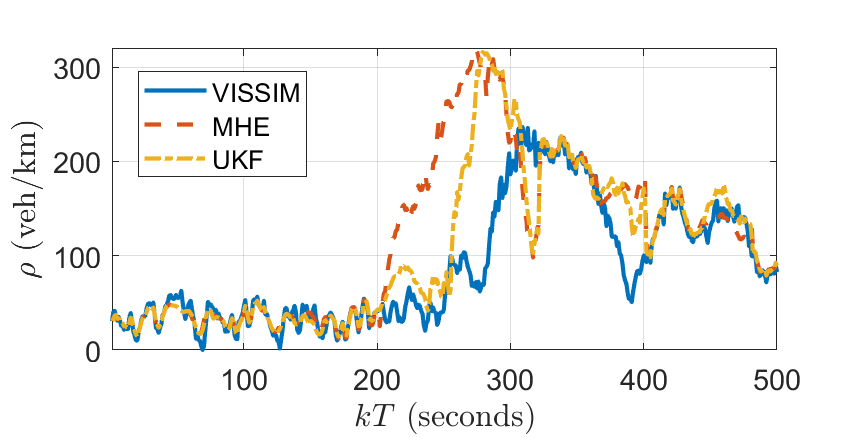}}
    \subfigure[]{\includegraphics[width=0.24\textwidth]{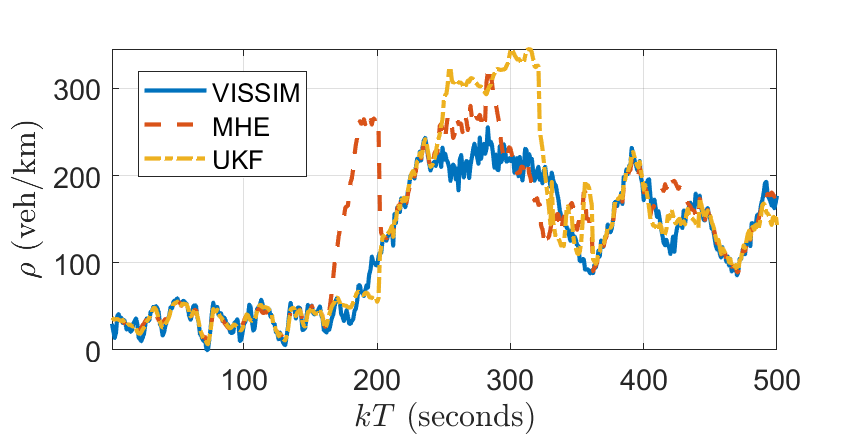}}
    \subfigure[]{\includegraphics[width=0.24\textwidth]{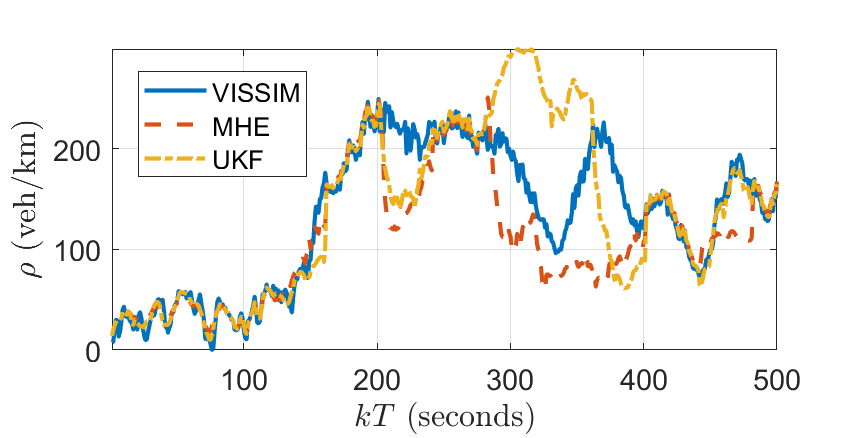}}
    \subfigure[]{\includegraphics[width=0.24\textwidth]{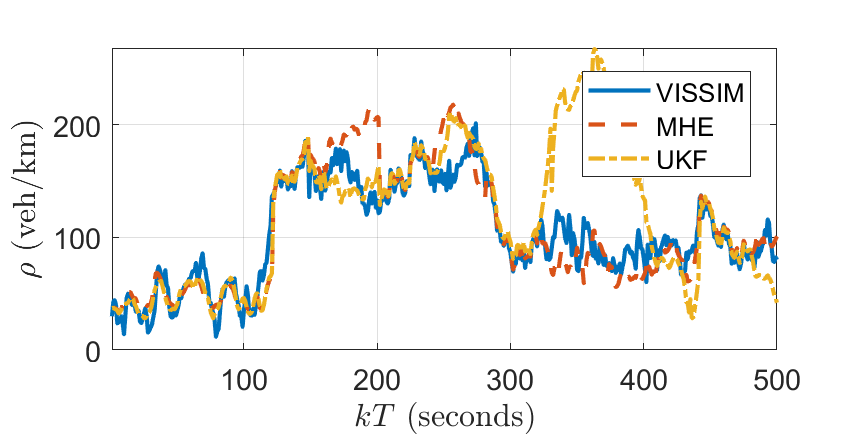}}
    \subfigure[]{\includegraphics[width=0.24\textwidth]{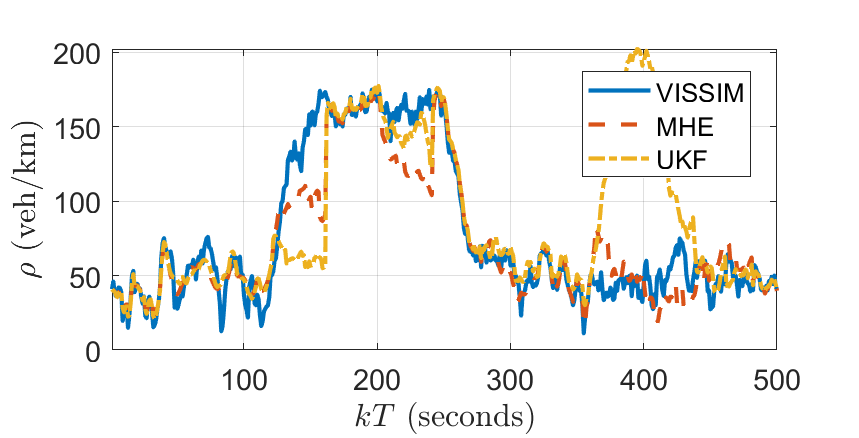}}
    \subfigure[]{\includegraphics[width=0.24\textwidth]{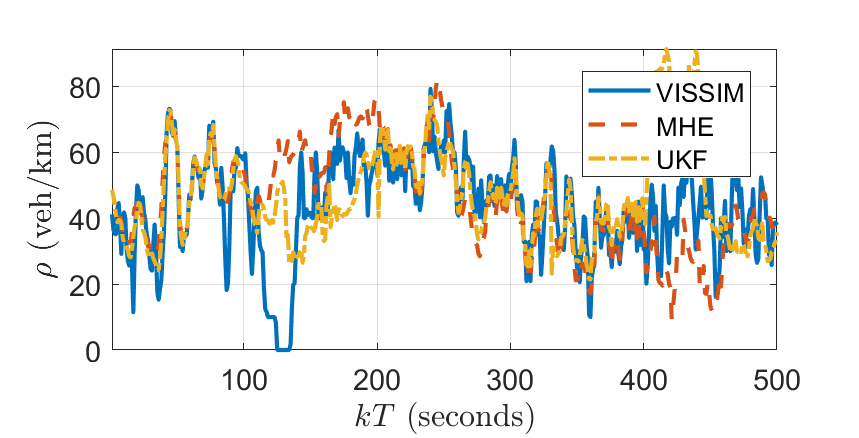}}
  \caption{{Plots of simulated and estimated density trajectories on Segments 1 to 8 in the presence of 3 additional measured segments with changing positions using UKF and MHE. Figures from top to bottom on the left correspond to Segments 1, 3, 5, and 7, and on the right correspond to Segments 2, 4, 6, and 8.}}\label{f:trajectories_ukf}
\end{figure}

{Figure \ref{f:trajectories_ukf} presents the plots of the simulated and estimated trajectories with moving sensors using UKF and MHE. The plots serve to illustrate the reasons for the observed trends with the density evaluation metrics for UKF compared to other methods.}

\subsection{{Estimated density trajectories to explain observed trends in $\textrm{SMAPE}_{\rho}$ with varying measurement quality and moving sensors}}
\begin{figure}
    \subfigure[]{\includegraphics[width=0.24\textwidth]{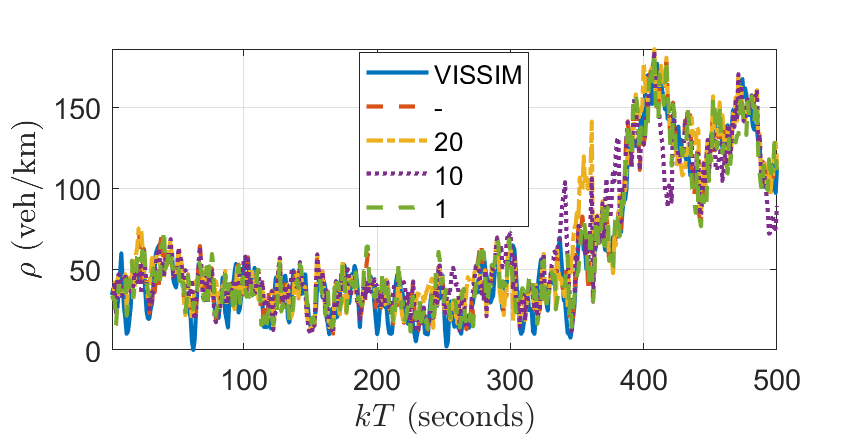}}
    \subfigure[]{\includegraphics[width=0.24\textwidth]{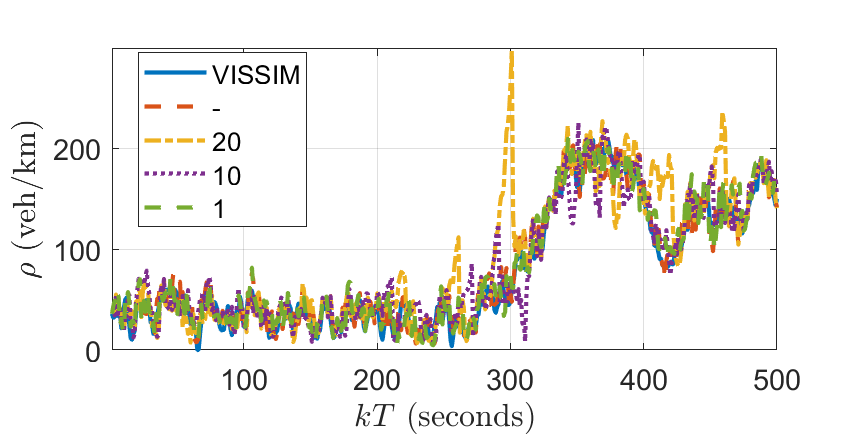}}
    \subfigure[]{\includegraphics[width=0.24\textwidth]{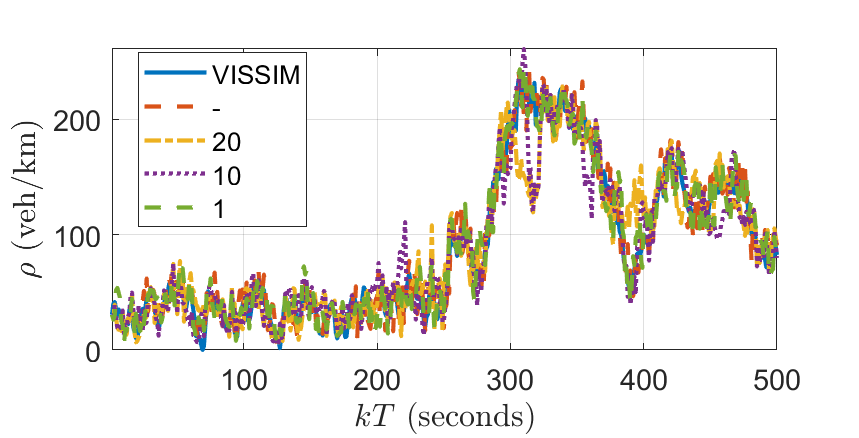}}
    \subfigure[]{\includegraphics[width=0.24\textwidth]{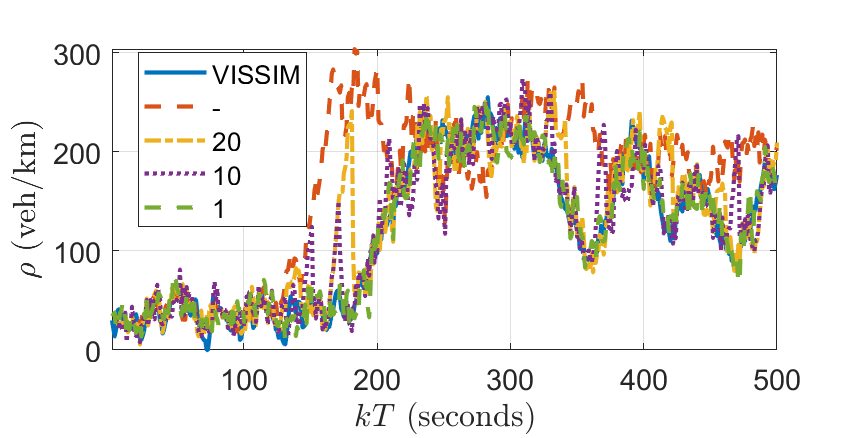}}
    \subfigure[]{\includegraphics[width=0.24\textwidth]{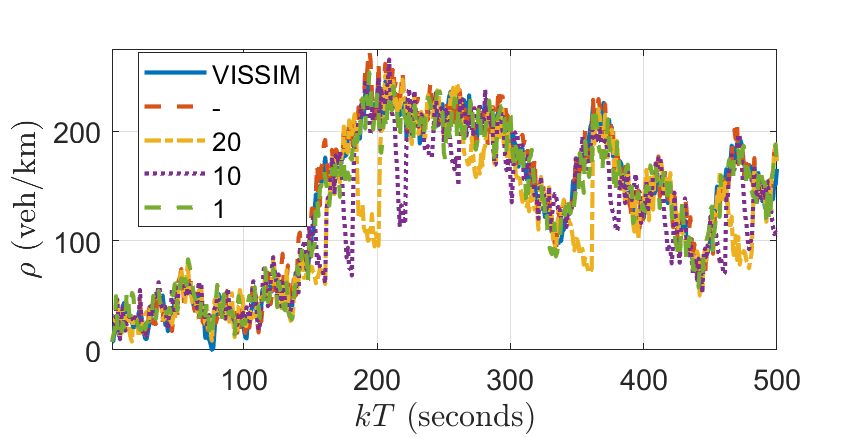}}
    \subfigure[]{\includegraphics[width=0.24\textwidth]{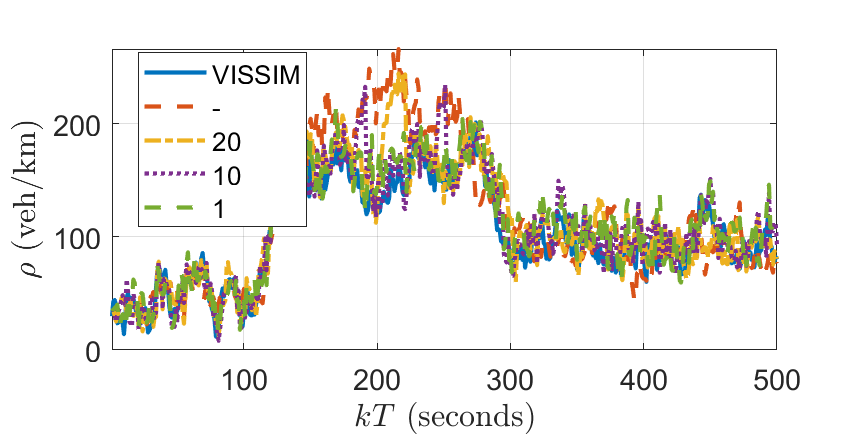}}
    \subfigure[]{\includegraphics[width=0.24\textwidth]{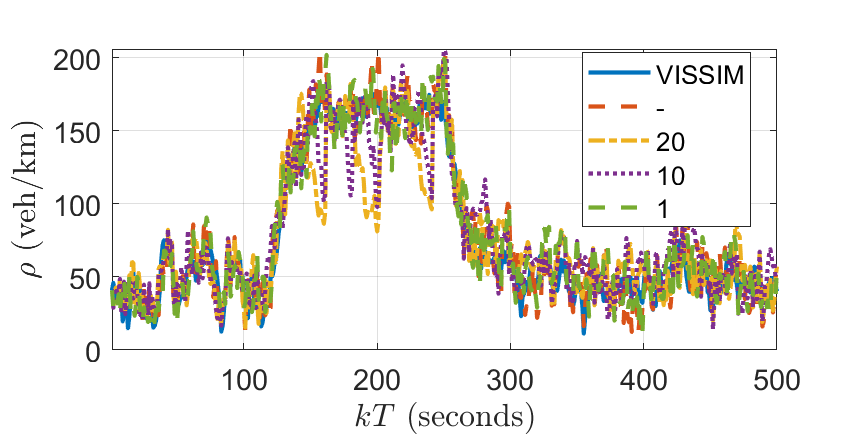}}
    \subfigure[]{\includegraphics[width=0.24\textwidth]{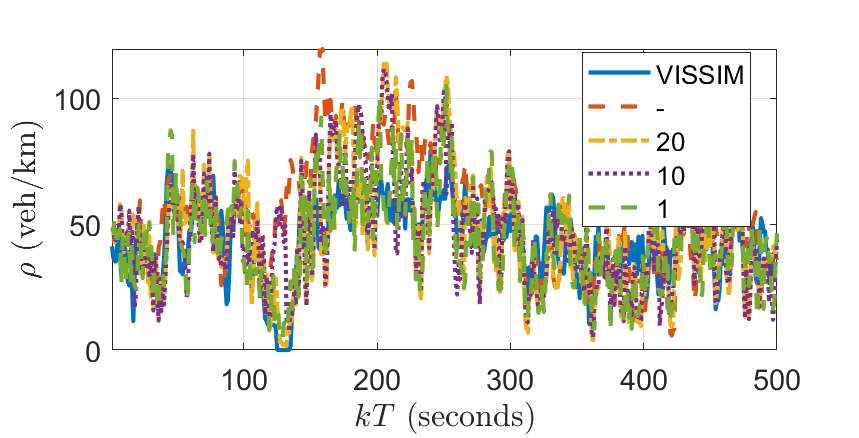}}
  \caption{{Plots of simulated and estimated density trajectories on Segments 1 to 8 in the presence of 5 additional measured segments with changing positions and $s=22$ with MHE. Figures from top to bottom on the left correspond to Segments 1, 3, 5, and 7, and on the right correspond to Segments 2, 4, 6, and 8.}}\label{f:trajectories_noise}
\end{figure}

{Figure \ref{f:trajectories_noise} presents the plots of the simulated and estimated trajectories with moving sensors and noise in measurement data using MHE. The plots serve to illustrate the reasons for the observed trends with the density evaluation metrics under varying measurement quality with moving sensors.}

\end{document}

%% file: preamble_new.tex
\usepackage{epsfig,color,amsmath,cite}
\usepackage{amsthm} 
\usepackage{amsmath}    
\usepackage[T1]{fontenc}
\usepackage[utf8]{inputenc}
\IEEEoverridecommandlockouts
\usepackage{bm}
\usepackage{epstopdf}
\usepackage{amssymb}
\usepackage{url}
\usepackage{enumitem} 
\usepackage{multirow}
\usepackage{hhline}
\usepackage{booktabs}
\usepackage{mathtools}
\usepackage{makecell}
\usepackage{empheq}
\usepackage[linesnumbered,boxed,commentsnumbered,ruled,vlined,longend]{algorithm2e}
\usepackage{comment}
\usepackage{caption}
\usepackage{subfigure}

\makeatother
\DeclareMathAlphabet\mathbfcal{OMS}{cmsy}{b}{n}



\usepackage{stackengine}

\renewcommand{\vec}[1]{\boldsymbol{\mathrm{#1}}}

\newcommand{\m}{\boldsymbol}
\allowdisplaybreaks[4]
\usepackage[colorlinks = true,
linkcolor = blue,
urlcolor  = blue,
citecolor = blue,
anchorcolor = blue]{hyperref}

\newcommand{\mbb}[1]{\mathbb{#1}}

\usepackage[framemethod=TikZ]{mdframed}
\mdfdefinestyle{MyFrame}{%
	linecolor=black,
	outerlinewidth=1.25pt,
	roundcorner=1.25pt,
	innerrightmargin=5pt,
	innerleftmargin=5pt,}
	
\usepackage[noabbrev]{cleveref}

\usepackage{mathtools}

\DeclarePairedDelimiter\abs{\lvert}{\rvert}%
\DeclarePairedDelimiter\norm{\lVert}{\rVert}%

\makeatletter
\let\oldabs\abs
\def\abs{\@ifstar{\oldabs}{\oldabs*}}
\let\oldnorm\norm
\def\norm{\@ifstar{\oldnorm}{\oldnorm*}}
\makeatother

\usepackage[english]{babel}
\usepackage[utf8]{inputenc}
\usepackage[super]{nth}

\usepackage{graphicx}
\usepackage{float}

\usepackage{array}
\usepackage{threeparttable}

\usepackage[english]{babel}
\usepackage[utf8]{inputenc}
\usepackage[super]{nth}

\RequirePackage{filecontents}

\SetKwRepeat{Do}{do}{while}%

\usepackage{mathtools}


%% file: main.bbl
\begin{thebibliography}{10}
\providecommand{\url}[1]{#1}
\csname url@samestyle\endcsname
\providecommand{\newblock}{\relax}
\providecommand{\bibinfo}[2]{#2}
\providecommand{\BIBentrySTDinterwordspacing}{\spaceskip=0pt\relax}
\providecommand{\BIBentryALTinterwordstretchfactor}{4}
\providecommand{\BIBentryALTinterwordspacing}{\spaceskip=\fontdimen2\font plus
\BIBentryALTinterwordstretchfactor\fontdimen3\font minus
  \fontdimen4\font\relax}
\providecommand{\BIBforeignlanguage}[2]{{%
\expandafter\ifx\csname l@#1\endcsname\relax
\typeout{** WARNING: IEEEtran.bst: No hyphenation pattern has been}%
\typeout{** loaded for the language `#1'. Using the pattern for}%
\typeout{** the default language instead.}%
\else
\language=\csname l@#1\endcsname
\fi
#2}}
\providecommand{\BIBdecl}{\relax}
\BIBdecl

\bibitem{gomes2006optimal}
G.~Gomes and R.~Horowitz, ``Optimal freeway ramp metering using the asymmetric
  cell transmission model,'' \emph{Transportation Research Part C: Emerging
  Technologies}, vol.~14, no.~4, pp. 244--262, 2006.

\bibitem{carlson2010optimal}
R.~C. Carlson, I.~Papamichail, M.~Papageorgiou, and A.~Messmer, ``Optimal
  motorway traffic flow control involving variable speed limits and ramp
  metering,'' \emph{Transportation Science}, vol.~44, no.~2, pp. 238--253,
  2010.

\bibitem{han2017resolving}
Y.~Han, A.~Hegyi, Y.~Yuan, S.~Hoogendoorn, M.~Papageorgiou, and C.~Roncoli,
  ``Resolving freeway jam waves by discrete first-order model-based predictive
  control of variable speed limits,'' \emph{Transportation Research Part C:
  Emerging Technologies}, vol.~77, pp. 405--420, 2017.

\bibitem{van2018efficient}
G.~S. van~de Weg, A.~Hegyi, S.~P. Hoogendoorn, and B.~De~Schutter, ``Efficient
  freeway {MPC} by parameterization of {ALINEA} and a speed-limited area,''
  \emph{IEEE Transactions on Intelligent Transportation Systems}, vol.~20,
  no.~1, pp. 16--29, 2018.

\bibitem{lu2014connected}
N.~Lu, N.~Cheng, N.~Zhang, X.~Shen, and J.~W. Mark, ``Connected vehicles:
  Solutions and challenges,'' \emph{IEEE Internet of Things Journal}, vol.~1,
  no.~4, pp. 289--299, 2014.

\bibitem{Lighthill1955b}
M.~J. Lighthill and G.~B. Whitham, ``On kinematic waves {II}. {A} theory of
  traffic flow on long crowded roads,'' \emph{Proc. R. Soc. Lond. A}, vol. 229,
  no. 1178, pp. 317--345, 1955.

\bibitem{Richards1956}
P.~I. Richards, ``Shock waves on the highway,'' \emph{Operations Research},
  vol.~4, no.~1, pp. 42--51, 1956.

\bibitem{HARROU202215}
F.~Harrou, A.~Zeroual, M.~M. Hittawe, and Y.~Sun, ``Chapter 2 - road traffic
  modeling,'' in \emph{Road Traffic Modeling and Management}, F.~Harrou,
  A.~Zeroual, M.~M. Hittawe, and Y.~Sun, Eds.\hskip 1em plus 0.5em minus
  0.4em\relax Elsevier, 2022, pp. 15--63.

\bibitem{daganzo1995requiem}
C.~F. Daganzo, ``Requiem for second-order fluid approximations of traffic
  flow,'' \emph{Transportation Research Part B: Methodological}, vol.~29,
  no.~4, pp. 277--286, 1995.

\bibitem{payne1971model}
H.~J. Payne, ``Model of freeway traffic and control,'' \emph{Mathematical Model
  of Public System}, pp. 51--61, 1971.

\bibitem{whitham1974linear}
G.~B. Whitham, \emph{Linear and nonlinear waves}.\hskip 1em plus 0.5em minus
  0.4em\relax John Wiley \& Sons, 2011, vol.~42.

\bibitem{aw2000resurrection}
A.~Aw and M.~Rascle, ``Resurrection of "second order" models of traffic flow,''
  \emph{SIAM Journal on Applied Mathematics}, vol.~60, no.~3, pp. 916--938,
  2000.

\bibitem{zhang2002non}
H.~M. Zhang, ``A non-equilibrium traffic model devoid of gas-like behavior,''
  \emph{Transportation Research Part B: Methodological}, vol.~36, no.~3, pp.
  275--290, 2002.

\bibitem{leclercq2007lagrangian}
L.~Leclercq, J.~Laval, E.~Chevallier \emph{et~al.}, ``The lagrangian
  coordinates and what it means for first order traffic flow models,''
  \emph{Transportation and traffic theory}, pp. 735--753, 2007.

\bibitem{work2008ensemble}
D.~B. Work, O.-P. Tossavainen, S.~Blandin, A.~M. Bayen, T.~Iwuchukwu, and
  K.~Tracton, ``An ensemble {Kalman} filtering approach to highway traffic
  estimation using {GPS} enabled mobile devices,'' in \emph{2008 47th IEEE
  Conference on Decision and Control}.\hskip 1em plus 0.5em minus 0.4em\relax
  IEEE, 2008, pp. 5062--5068.

\bibitem{wright2016fusing}
M.~Wright and R.~Horowitz, ``Fusing loop and {GPS} probe measurements to
  estimate freeway density,'' \emph{IEEE Transactions on Intelligent
  Transportation Systems}, vol.~17, no.~12, pp. 3577--3590, 2016.

\bibitem{nantes2016real}
A.~Nantes, D.~Ngoduy, A.~Bhaskar, M.~Miska, and E.~Chung, ``Real-time traffic
  state estimation in urban corridors from heterogeneous data,''
  \emph{Transportation Research Part C: Emerging Technologies}, vol.~66, pp.
  99--118, 2016.

\bibitem{seo2017survey}
T.~Seo, A.~M. Bayen, T.~Kusakabe, and Y.~Asakura, ``Traffic state estimation on
  highway: A comprehensive survey,'' \emph{Annual Reviews in Control}, vol.~43,
  pp. 128--151, 2017.

\bibitem{wang2005real}
Y.~Wang and M.~Papageorgiou, ``Real-time freeway traffic state estimation based
  on extended {Kalman} filter: a general approach,'' \emph{Transportation
  Research Part B: Methodological}, vol.~39, no.~2, pp. 141--167, 2005.

\bibitem{zhao2020real}
M.~Zhao, X.~Yu, Y.~Hu, J.~Cao, S.~Hu, L.~Zhang, J.~Guo, and Y.~Wang,
  ``Real-time freeway traffic state estimation with fixed and mobile sensing
  data,'' in \emph{2020 IEEE 23rd International Conference on Intelligent
  Transportation Systems (ITSC)}.\hskip 1em plus 0.5em minus 0.4em\relax IEEE,
  2020, pp. 1--7.

\bibitem{agalliadis2020traffic}
I.~Agalliadis, M.~Makridis, and A.~Kouvelas, ``Traffic estimation by fusing
  static and moving observations in highway networks,'' in \emph{20th Swiss
  Transport Research Conference (STRC 2020)(virtual)}.\hskip 1em plus 0.5em
  minus 0.4em\relax STRC, 2020.

\bibitem{kotsialos2002}
A.~Kotsialos, M.~Papageorgiou, C.~Diakaki, Y.~Pavlis, and F.~Middelham,
  ``Traffic flow modeling of large-scale motorway networks using the
  macroscopic modeling tool {METANET},'' \emph{IEEE Transactions on Intelligent
  Transportation Systems}, vol.~3, no.~4, pp. 282--292, 2002.

\bibitem{messner1990metanet}
A.~Messner and M.~Papageorgiou, ``{METANET}: A macroscopic simulation program
  for motorway networks,'' \emph{Traffic Engineering \& Control}, vol.~31, no.
  8-9, pp. 466--470, 1990.

\bibitem{Seo2017}
T.~Seo and A.~M. Bayen, ``Traffic state estimation method with efficient data
  fusion based on the {Aw-Rascle-Zhang} model,'' in \emph{2017 IEEE 20th
  International Conference on Intelligent Transportation Systems (ITSC)}, 2017,
  pp. 1--6.

\bibitem{yu2019boundary}
H.~Yu, A.~M. Bayen, and M.~Krstic, ``Boundary observer for congested freeway
  traffic state estimation via {Aw-Rascle-Zhang} model,''
  \emph{IFAC-PapersOnLine}, vol.~52, no.~2, pp. 183--188, 2019.

\bibitem{wang2017comparing}
R.~Wang, Y.~Li, and D.~B. Work, ``Comparing traffic state estimators for mixed
  human and automated traffic flows,'' \emph{Transportation Research Part C:
  Emerging Technologies}, vol.~78, pp. 95--110, 2017.

\bibitem{bekiaris2016highway}
N.~Bekiaris-Liberis, C.~Roncoli, and M.~Papageorgiou, ``Highway traffic state
  estimation with mixed connected and conventional vehicles,'' \emph{IEEE
  Transactions on Intelligent Transportation Systems}, vol.~17, no.~12, pp.
  3484--3497, 2016.

\bibitem{xing2022traffic}
J.~Xing, W.~Wu, Q.~Cheng, and R.~Liu, ``Traffic state estimation of urban road
  networks by multi-source data fusion: Review and new insights,''
  \emph{Physica A: Statistical Mechanics and its Applications}, vol. 595, p.
  127079, 2022.

\bibitem{kashinath2021review}
S.~A. Kashinath, S.~A. Mostafa, A.~Mustapha, H.~Mahdin, D.~Lim, M.~A. Mahmoud,
  M.~A. Mohammed, B.~A.~S. Al-Rimy, M.~F.~M. Fudzee, and T.~J. Yang, ``Review
  of data fusion methods for real-time and multi-sensor traffic flow
  analysis,'' \emph{IEEE Access}, vol.~9, pp. 51\,258--51\,276, 2021.

\bibitem{nugroho2019control}
S.~A. Nugroho, A.~F. Taha, and C.~G. Claudel, ``A control-theoretic approach
  for scalable and robust traffic density estimation using convex
  optimization,'' \emph{IEEE Transactions on Intelligent Transportation
  Systems}, vol.~22, no.~1, pp. 64--78, 2019.

\bibitem{vishnoi2020asymmetric}
S.~C. Vishnoi, S.~A. Nugroho, A.~F. Taha, C.~Claudel, and T.~Banerjee,
  ``Asymmetric cell transmission model-based, ramp-connected robust traffic
  density estimation under bounded disturbances,'' in \emph{2020 American
  Control Conference (ACC)}.\hskip 1em plus 0.5em minus 0.4em\relax IEEE, 2020,
  pp. 1197--1202.

\bibitem{rao2001constrained}
C.~V. Rao, J.~B. Rawlings, and J.~H. Lee, ``Constrained linear state
  estimation—a moving horizon approach,'' \emph{Automatica}, vol.~37, no.~10,
  pp. 1619--1628, 2001.

\bibitem{rao2003constrained}
C.~V. Rao, J.~B. Rawlings, and D.~Q. Mayne, ``Constrained state estimation for
  nonlinear discrete-time systems: Stability and moving horizon
  approximations,'' \emph{IEEE Transactions on Automatic Control}, vol.~48,
  no.~2, pp. 246--258, 2003.

\bibitem{wang2017resilient}
Y.~Wang, J.~Yuan, S.~Yu, Y.~Hu, and H.~Chen, ``Resilient moving horizon
  estimation for cyber-physical systems under sensor attacks,'' in \emph{2017
  11th Asian Control Conference (ASCC)}.\hskip 1em plus 0.5em minus 0.4em\relax
  IEEE, 2017, pp. 2274--2279.

\bibitem{alessandri2017fast}
A.~Alessandri and M.~Gaggero, ``Fast moving horizon state estimation for
  discrete-time systems using single and multi iteration descent methods,''
  \emph{IEEE Transactions on Automatic Control}, vol.~62, no.~9, pp.
  4499--4511, 2017.

\bibitem{sirmatel2019nonlinear}
I.~I. Sirmatel and N.~Geroliminis, ``Nonlinear moving horizon estimation for
  large-scale urban road networks,'' \emph{IEEE Transactions on Intelligent
  Transportation Systems}, vol.~21, no.~12, pp. 4983--4994, 2019.

\bibitem{sirmatel2020model}
I.~Sirmatel and N.~Geroliminis, ``Model-based identification, estimation, and
  control for large-scale urban road networks,'' in \emph{2020 European Control
  Conference (ECC)}.\hskip 1em plus 0.5em minus 0.4em\relax IEEE, 2020, pp.
  408--413.

\bibitem{timotheou2015moving}
S.~Timotheou, C.~G. Panayiotou, and M.~M. Polycarpou, ``Moving horizon
  fault-tolerant traffic state estimation for the cell transmission model,'' in
  \emph{2015 54th IEEE Conference on Decision and Control (CDC)}.\hskip 1em
  plus 0.5em minus 0.4em\relax IEEE, 2015, pp. 3451--3456.

\bibitem{huang2020physics}
A.~J. Huang and S.~Agarwal, ``Physics informed deep learning for traffic state
  estimation,'' in \emph{2020 IEEE 23rd International Conference on Intelligent
  Transportation Systems (ITSC)}.\hskip 1em plus 0.5em minus 0.4em\relax IEEE,
  2020, pp. 1--6.

\bibitem{zhao2022integrating}
C.~Zhao and H.~Yu, ``Integrating {PDE} observer with deep learning for traffic
  state estimation,'' in \emph{2022 IEEE 25th International Conference on
  Intelligent Transportation Systems (ITSC)}.\hskip 1em plus 0.5em minus
  0.4em\relax IEEE, 2022, pp. 1964--1969.

\bibitem{shi2021physics}
R.~Shi, Z.~Mo, K.~Huang, X.~Di, and Q.~Du, ``A physics-informed deep learning
  paradigm for traffic state and fundamental diagram estimation,'' \emph{IEEE
  Transactions on Intelligent Transportation Systems}, vol.~23, no.~8, pp.
  11\,688--11\,698, 2021.

\bibitem{di2023physics}
X.~Di, R.~Shi, Z.~Mo, and Y.~Fu, ``Physics-informed deep learning for traffic
  state estimation: A survey and the outlook,'' \emph{Algorithms}, vol.~16,
  no.~6, p. 305, 2023.

\bibitem{Godunov1959}
S.~K. Godunov, ``A difference method for numerical calculation of discontinuous
  solutions of the equations of hydrodynamics,'' \emph{Matematicheskii
  Sbornik}, vol.~89, no.~3, pp. 271--306, 1959.

\bibitem{courant1967partial}
R.~Courant, K.~Friedrichs, and H.~Lewy, ``On the partial difference equations
  of mathematical physics,'' \emph{IBM journal of Research and Development},
  vol.~11, no.~2, pp. 215--234, 1967.

\bibitem{yang2019novel}
H.~Yang and H.~Rakha, ``A novel approach for estimation of dynamic from static
  origin--destination matrices,'' \emph{Transportation Letters}, vol.~11,
  no.~4, pp. 219--228, 2019.

\bibitem{lee2011density}
C.~Lee, J.-B. Lee, and M.~Kim, ``Density measurement algorithm for freeway
  segment using two point detectors,'' \emph{Journal of Advanced
  Transportation}, vol.~45, no.~3, pp. 207--218, 2011.

\bibitem{klein2006traffic}
L.~A. Klein, M.~K. Mills, D.~R. Gibson \emph{et~al.}, ``Traffic detector
  handbook: Volume i,'' Turner-Fairbank Highway Research Center, Tech. Rep.,
  2006.

\bibitem{bekiaris2017highway}
N.~Bekiaris-Liberis, C.~Roncoli, and M.~Papageorgiou, ``Highway traffic state
  estimation per lane in the presence of connected vehicles,''
  \emph{Transportation Research Part B: Methodological}, vol. 106, pp. 1--28,
  2017.

\bibitem{seo2015estimation}
T.~Seo, T.~Kusakabe, and Y.~Asakura, ``Estimation of flow and density using
  probe vehicles with spacing measurement equipment,'' \emph{Transportation
  Research Part C: Emerging Technologies}, vol.~53, pp. 134--150, 2015.

\bibitem{panichpapiboon2008evaluation}
S.~Panichpapiboon and W.~Pattara-atikom, ``Evaluation of a neighbor-based
  vehicle density estimation scheme,'' in \emph{2008 8th International
  Conference on ITS Telecommunications}.\hskip 1em plus 0.5em minus 0.4em\relax
  IEEE, 2008, pp. 294--298.

\bibitem{moreno1996vector}
P.~J. Moreno, B.~Raj, and R.~M. Stern, ``A vector {Taylo}r series approach for
  environment-independent speech recognition,'' in \emph{1996 IEEE
  International Conference on Acoustics, Speech, and Signal Processing
  Conference Proceedings}, vol.~2.\hskip 1em plus 0.5em minus 0.4em\relax IEEE,
  1996, pp. 733--736.

\bibitem{yu2020pde}
H.~Yu, Q.~Gan, A.~Bayen, and M.~Krstic, ``Pde traffic observer validated on
  freeway data,'' \emph{IEEE Transactions on Control Systems Technology},
  vol.~29, no.~3, pp. 1048--1060, 2020.

\bibitem{chen1984linear}
C.-T. Chen, \emph{Linear system theory and design}.\hskip 1em plus 0.5em minus
  0.4em\relax Saunders college publishing, 1984.

\bibitem{9585066}
S.~A. Nugroho, S.~C. Vishnoi, A.~F. Taha, C.~G. Claudel, and T.~Banerjee,
  ``Where should traffic sensors be placed on highways?'' \emph{IEEE
  Transactions on Intelligent Transportation Systems}, pp. 1--14, 2021.

\bibitem{6739119}
L.~Li, X.~Chen, and L.~Zhang, ``Multimodel ensemble for freeway traffic state
  estimations,'' \emph{IEEE Transactions on Intelligent Transportation
  Systems}, vol.~15, no.~3, pp. 1323--1336, 2014.

\bibitem{wan2000unscented}
E.~A. Wan and R.~Van Der~Merwe, ``The unscented {Kalman} filter for nonlinear
  estimation,'' in \emph{Proceedings of the IEEE 2000 Adaptive Systems for
  Signal Processing, Communications, and Control Symposium (Cat. No.
  00EX373)}.\hskip 1em plus 0.5em minus 0.4em\relax Ieee, 2000, pp. 153--158.

\bibitem{evensen2003ensemble}
G.~Evensen, ``The ensemble {Kalman} filter: Theoretical formulation and
  practical implementation,'' \emph{Ocean dynamics}, vol.~53, no.~4, pp.
  343--367, 2003.

\bibitem{simon2010kalman}
D.~Simon, ``Kalman filtering with state constraints: a survey of linear and
  nonlinear algorithms,'' \emph{IET Control Theory \& Applications}, vol.~4,
  no.~8, pp. 1303--1318, 2010.

\bibitem{wang2022real}
Y.~Wang, M.~Zhao, X.~Yu, Y.~Hu, P.~Zheng, W.~Hua, L.~Zhang, S.~Hu, and J.~Guo,
  ``Real-time joint traffic state and model parameter estimation on freeways
  with fixed sensors and connected vehicles: State-of-the-art overview,
  methods, and case studies,'' \emph{Transportation Research Part C: Emerging
  Technologies}, vol. 134, p. 103444, 2022.

\bibitem{gottlich2021second}
S.~Gottlich, M.~Herty, S.~Moutari, and J.~Wei{\ss}en, ``Second-order traffic
  flow models on networks,'' \emph{SIAM Journal on Applied Mathematics},
  vol.~81, no.~1, pp. 258--281, 2021.

\bibitem{kolb2018pareto}
O.~Kolb, G.~Costeseque, P.~Goatin, and S.~Gottlich, ``Pareto-optimal coupling
  conditions for the {Aw--Rascle--Zhang} traffic flow model at junctions,''
  \emph{SIAM Journal on Applied Mathematics}, vol.~78, no.~4, pp. 1981--2002,
  2018.

\bibitem{khelifi2017generic}
A.~Khelifi, H.~Haj-Salem, J.-P. Lebacque, and L.~Nabli, ``Generic macroscopic
  traffic flow models on junctions,'' in \emph{2017 International Conference on
  Control, Automation and Diagnosis (ICCAD)}.\hskip 1em plus 0.5em minus
  0.4em\relax IEEE, 2017, pp. 445--450.

\bibitem{rauh2009carleman}
A.~Rauh, J.~Minisini, and H.~Aschemann, ``Carleman linearization for control
  and for state and disturbance estimation of nonlinear dynamical processes,''
  \emph{IFAC Proceedings Volumes}, vol.~42, no.~13, pp. 455--460, 2009.

\bibitem{pruekprasert2020moment}
S.~Pruekprasert, T.~Takisaka, C.~Eberhart, A.~Cetinkaya, and J.~Dubut, ``Moment
  propagation of discrete-time stochastic polynomial systems using truncated
  carleman linearization,'' \emph{IFAC-PapersOnLine}, vol.~53, no.~2, pp.
  14\,462--14\,469, 2020.

\bibitem{qu2009computation}
C.~C. Qu and J.~Hahn, ``Computation of arrival cost for moving horizon
  estimation via unscented kalman filtering,'' \emph{Journal of Process
  Control}, vol.~19, no.~2, pp. 358--363, 2009.

\bibitem{alessandri2003receding}
A.~Alessandri, M.~Baglietto, and G.~Battistelli, ``Receding-horizon estimation
  for discrete-time linear systems,'' \emph{IEEE Transactions on Automatic
  Control}, vol.~48, no.~3, pp. 473--478, 2003.

\end{thebibliography}
